\documentclass[12pt,tightenlines,prb,nofootinbib,superscriptaddress]{revtex4-1}
\newcommand{\bra}[1]{\langle #1|}
\newcommand{\ket}[1]{|#1\rangle}

\newcommand{\eqnref}[1]{Eq.\ (\ref{#1})}

\usepackage{amsmath}
\usepackage{amssymb}
\usepackage{enumerate}
\usepackage{amscd}
\usepackage{amsthm}
\newtheorem{lemma}{Lemma}

\newcommand{\U}{\mathrm{U}}

\usepackage{tikz}
\pgfrealjobname{paper}
\usetikzlibrary{decorations.markings}
\usetikzlibrary{arrows}

\begin{document}

\title{Classifying symmetry-protected topological phases through the
anomalous action of the symmetry on the edge}

\author{Dominic V. Else}
\affiliation{Department of Physics, University of California, Santa Barbara, CA 93106, USA}
\author{Chetan Nayak}
\affiliation{Department of Physics, University of California, Santa Barbara, CA 93106, USA}
\affiliation{Microsoft Research, Station Q, Elings Hall, University of
California, Santa Barbara, CA 93106, USA}
\begin{abstract}
It is well known that ($1+1$)-D bosonic symmetry-protected topological (SPT) phases with symmetry group $G$ can be identified by the projective
representation of the symmetry at the edge. Here, we generalize this result
to higher dimensions. We assume that the representation of the symmetry on the spatial edge of a
($d+1$)-D SPT is \emph{local} but not necessarily \emph{on-site}, such that
there is an obstruction to its implementation on a region with boundary. We show
that such obstructions are classified by the cohomology group $H^{d+1}(G, U(1))$, in agreement with the
classification of bosonic SPT phases proposed in [Chen et al, Science \textbf{338}, 1604
(2012)]. Our analysis  allows for a straightforward
calculation of the element of $H^{d+1}(G, U(1))$ corresponding to physically meaningful
models such as non-linear sigma models with a theta term in the action. SPT
phases outside the classification of Chen et al are those in which the symmetry
cannot be represented locally on the edge. With some modifications, our
framework can also be applied to fermionic systems in (2+1)-D.

\end{abstract}

\maketitle

The classification of phases of matter in quantum systems at zero temperature
has proven to be much richer than in classical statistical mechanical systems.
For many such phases, the feature which distinguishes them from other phases is
quantum mechanical and not related to spontaneous breaking of a symmetry. One
such family of quantum phases which has been much studied in recent years is the
\emph{symmetry-protected topological} (SPT) phases
\cite{haldane1,*haldane2,aklt1,*aklt2,diverging_prl,hasan_kane_review,moore_birth_2010,qi_zhang_review,gu_wen_2009,wen_lu,chen_gu_wen,schuch,pollmann-arxiv-2009,pollmann-prb-2010,spto_higher,*spto_higher_prb,
cenke_neel_order,vishwanath_senthil,lu_vishwanath,cenke_wave_functions,wen_projective_construction,wen_spin_hall,santos_wang_2014,wang_santos_wen_2014,senthil_iqhe_bosons,boson_top_witten,microscopic_iqhe,
levin_gu,window,supercohomology,kapustin_cobordism,orbifolds,wang_levin,schnyder_classification,kitaev_periodic,tenfold_way,wen_fermions,tang_fermions,fidkowski_fermions,supercohomology,wang_potter_senthil,qi_fermions,yao_fermions,gu_levin_fermions,cheng_gu_fermions,thomale,kapustin_fermions}
A system with a symmetry is considered to lie in an SPT phase if (a) the symmetry
is not spontaneously broken; and (b) the system can be connected to one whose ground
state is a trivial product state without a phase transition, but \emph{only} if
we allow the symmetry to be broken explicitly. In some sense, SPT phases are ``trivial'' in the bulk,
but boundaries between different
SPT phases are non-trivial and must either be gapless, break the
symmetry (explicitly or spontaneously), or be topologically ordered.

The central problem in the study of SPT phases is classifying the different
phases that can occur for a given symmetry. In bosonic systems with an internal
symmetry group $G$, an early result was that in (1+1)-D systems, the possible
SPT phases are classified \cite{chen_gu_wen,schuch} by the \emph{second cohomology group} $H^2(G, \U(1))$.
This result has a natural interpretation \cite{pollmann-arxiv-2009} in terms of the symmetry transformation
properties of an edge between the SPT and vacuum (or equivalently, of
the entanglement spectrum \cite{pollmann-prb-2010}). Such
an edge will in general transform \emph{projectively} under the symmetry. The
second cohomology group arises naturally from a consideration of these
projective representations.

It has been argued \cite{spto_higher,*spto_higher_prb} that, more generally,
the SPT phases in $d$ spatial dimensions are classified by the cohomology
group $H^{d+1}(G, \U(1))$. This result was based on an explicit construction of
field theories in discrete space-time which are believed to be representative of each SPT phase. However, making a definitive identification between these
lattice field theories and other, more physically motivated, descriptions of the
corresponding SPT phases
\cite{cenke_neel_order,vishwanath_senthil,lu_vishwanath,senthil_iqhe_bosons} has proved difficult. In this paper, therefore, we
propose to recast the cohomological classification in a different, hopefully
more intuitive viewpoint,
inspired by the original (1+1)-D treatment. The central idea is that, just as in
the (1+1)-D case, the symmetry transformation on the edge of a ($d$+1)-D system
will be, in some sense, anomalous
\cite{spto_2d,Wen13,kapustin_anomalies,kapustin_anomalies_2}. Specifically, if we have a system defined on
a $d$-dimensional spatial manifold $M_{\mathrm{bulk}}$ with a boundary, the edge symmetry acts on the boundary
$\partial M_{\mathrm{bulk}}$, which itself has no boundary [$\partial (\partial
M_{\mathrm{bulk}}) = 0$]. Therefore, there might be an
obstruction to implementing the edge symmetry in a consistent way on a $(d-1)$-dimensional manifold
$M$ \emph{with} boundary $\partial M \neq 0$. We will argue that this obstruction is indeed
classified by the cohomology group $H^{d+1}(G, \U(1))$. [For (2+1)-D systems, our
  approach is related to, though more general than, that of Ref.~\onlinecite{spto_2d}, which was bsed on a
tensor-network representation for the edge symmetry.] In fact, in (2+1)-D our
approach also leads to a classification of SPT phases in interacting fermion
systems,
as we will show.
%

The remainder of this paper is organized as follows. In Section \ref{sec_general}, we
give the general demonstration that the obstruction is classified by $H^{d+1}(G,
\U(1))$. For (2+1)-D SPT's, this argument can be given in full generality
(assuming only that the symmetry acts \emph{locally} on the edge), but in higher
dimensions we will need to make additional assumptions about the form of the
symmetry. 
In Section \ref{sec_chiral_example},
we discuss by way of illustration a simple example of an anomalous symmetry that appears on the edge of
a (2+1)-D SPT. In Section \ref{separation_proof}, we use the ideas of this paper
to prove that (2+1)-D SPT phases characterized by different elements of
$H^3(G, \U(1))$ are necessarily separated by a phase transition unless the symmetry is
broken explicitly. In Section \ref{sec_nlsm}, we show how to use our approach to derive the
element of the cohomology group corresponding to non-linear sigma models
containing a topological term. In Section \ref{sec_discrete}, we make explicit the connection between
our work and the original classification of Ref.
\onlinecite{spto_higher,*spto_higher_prb}. In Section \ref{sec_beyond}, we explain why, in
the presence of anti-unitary symmetries, there exist bosonic SPT phases not captured by our
arguments. In Section \ref{sec_fermions}, we show how our ideas can be
applied also to fermionic systems in (2+1)-D.

\section{The general formalism}
\label{sec_general}
Consider a system in a bosonic SPT phase. By definition, this means it is gapped and
non-degenerate in the bulk, and (disregarding symmetry considerations) can be
continuously connected to a product state without a phase transition.
However, in a system with boundary, we can define an
effective low-energy theory for the boundary, which may be gapless 
notwithstanding the gap in the bulk. A key property of SPT phases is that
the boundary theory of an SPT phase in $d$ spatial dimensions can always be
realized at the microscopic level in a
strictly $(d-1)$-dimensional system (see Appendix \ref{sec_edge_construction} for a careful proof of this
well-known fact.)  This is in contrast to, for example, integer quantum Hall
states in which the boundary is chiral and cannot be realized as a stand-alone
system \cite{wen_gapless}. For SPT phases, the anomalous nature of the edge arises not from the
boundary theory itself but from the way it is acted upon by the symmetry.


We assume that the symmetry in the bulk is unitary and on-site, that is, for a
lattice system with $N$ sites, the symmetry group $G$ is
is represented as a unitary tensor-product $U(g) = [u(g)]^{\otimes N}$ of
operators acting on each site. (We may need to group several sites together into
a single effective site in order to satisfy this condition.)
We now consider the low-energy Hilbert space of states with energies below
some cutoff that is less than the bulk gap; these states are edge excitations.
Projecting the unitary representation of the symmetry group onto this low-energy Hilbert space,
we obtain a unitary representation, acting only
on the boundary degrees of freedom, that may not be on-site.
On the contrary, it appears to be a characteristic of non-trivial SPT phases that
the symmetry is realized on the boundary in a
fundamentally non-on-site way \cite{spto_higher,*spto_higher_prb,spto_2d,Wen13}. Nevertheless, the key assumption that
we make in this paper is that the boundary symmetry, albeit not on-site, is
nevertheless still \emph{local} in the sense of Ref.~\onlinecite{wen_lu} (e.g.\ it can be
represented as a finite-depth quantum circuit.) This seems a natural assumption,
but we expect it to be violated by SPT phases not captured by the cohomological
classification (see Section \ref{sec_beyond} for further discussion).

For a non-on-site symmetry, there is the possibility that there is an
obstruction to implementing the symmetry on a manifold with boundary in a
consistent way. We intend to show that, by classifying these obstructions, one
recovers the cohomological classification of SPT phases.
 A simple example of this idea is the well-known connection
between (1+1)-D SPT's and the projective symmetry transformation of the edge
\cite{pollmann-arxiv-2009,chen_gu_wen,schuch},
which we shall now review.

\begin{figure}
  \includegraphics{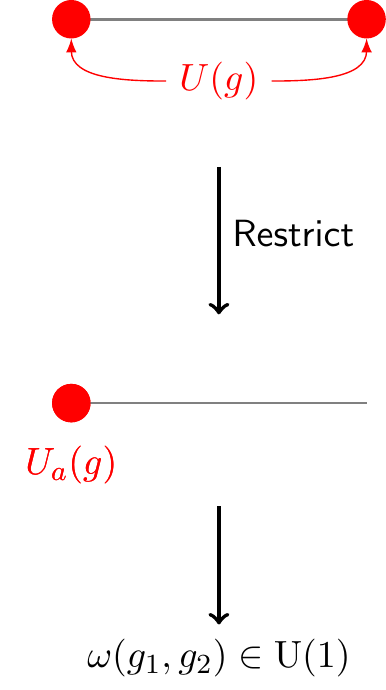}
%
%
%
%
\caption{Obtaining a 2-cocycle on the (0+1)-D edge of a (1+1)-D SPT.}
\end{figure}
\subsection{(1+1)-D SPT's}
The boundary of a one-dimensional system simply comprises a pair of points $a$
and $b$. Let $U(g)$ be the representation of the symmetry group
$G$ on this boundary.
Assuming that we chose the system size such that the end-points $a$ and $b$ are
well-separated (i.e.\ by a distance large compared to all intrinsic length
scales), locality of $U(g)$ simply implies that it
must act on $a$ and $b$ separately; that is, it must be a tensor product $U(g) =
U_a(g) \otimes U_b(g)$. We can think of $U_a(g)$ as the \emph{restriction} of
$U(g)$ to the point $a$. Importantly, however, this restriction is uniquely
defined only modulo phase factors. Indeed, $U(g)$ is left invariant under
$U_a(g) \to \beta(g) U_a(g), U_b(g) \to \beta(g)^{-1} U_b(g)$ for any
$\mathrm{U}(1)$-valued function $\beta(g)$. Thus, while $U(g)$ is always a representation of the
symmetry group $G$, that is $U(g_1) U(g_2) = U(g_1 g_2)$, the non-uniqueness of the restriction procedure implies that
$U_a(g)$ need only be a \emph{projective} representation of $G$, which is to say
that
$U_a(g_1) U_a(g_2) = \omega(g_1, g_2) U_a(g_1 g_2)$ for some
$\mathrm{U}(1)$-valued function $\omega(g_1, g_2)$. The function $\omega$
describes the obstruction to consistently (i.e.\ non-projectively) implementing
the symmetry on the point $a$. 

Since multiplication of the $U_a$'s must be
associative, one can derive a consistency condition on $\omega$ by
evaluating $U_a(g_1) U_a(g_2) U_a(g_3)$ in two different ways, namely
\begin{equation}
  \label{omega_associativity_comm}
  \omega(g_1, g_2) \omega(g_1 g_2, g_3) = \omega(g_2, g_3) \omega(g_1, g_2 g_3).
\end{equation}
A function $\omega$ satisfying \eqnref{omega_associativity_comm} is known as a
2-cocycle. Furthermore, due to the fact that $U_a(g)$ is only defined up to a
$g$-dependent phase factor $\beta(g)$, it follows that we have an equivalence
relation on 2-cocycles:
\begin{equation}
  \omega(g_1, g_2) \sim \omega(g_1, g_2) \beta(g_1) \beta(g_2)
  \beta(g_1 g_2)^{-1}.
\end{equation}
The group of $2$-cocycles quotiented by the above equivalence relation is, by
definition, the second cohomology group $H^2(G, \U(1))$. One can then show that
two models are in the same SPT phase if and only if they correspond to the same
element of $H^2(G, \U(1))$. Therefore, SPT phases in (1+1)-D are classified by
$H^2(G, \U(1))$.


\begin{figure}
  \includegraphics{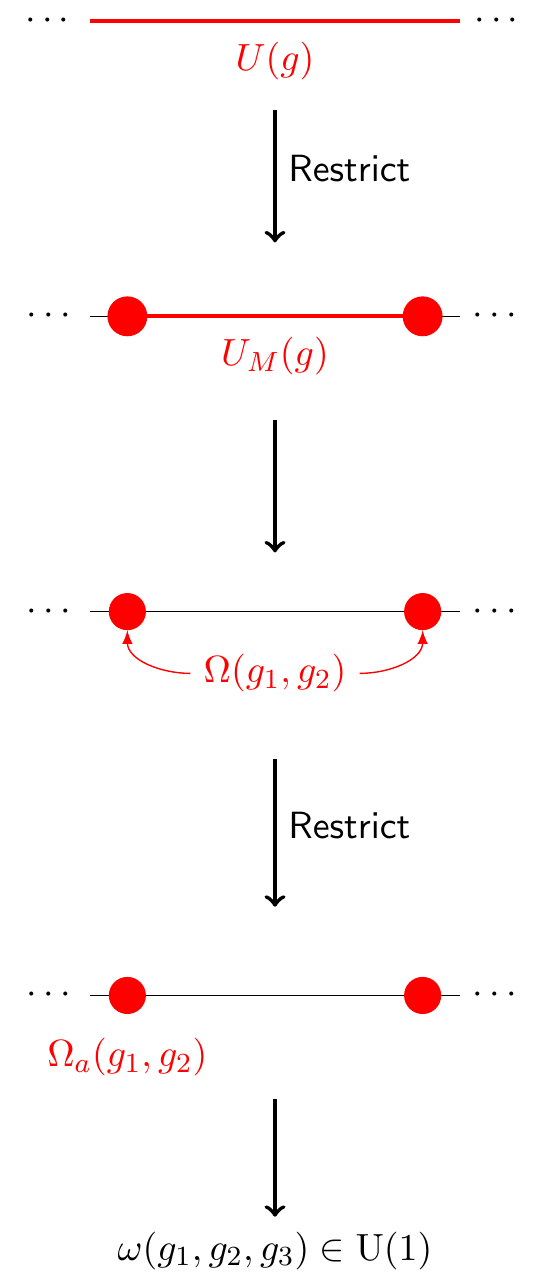}
\caption{Obtaining a 3-cocycle on the (1+1)-D edge of a (2+1)-D system.}
\end{figure}

\subsection{(2+1)-D SPT's}
\label{sec_2p1d}

When presented as it was above, the (1+1)-D case suggests an obvious
generalization to higher dimensions: we consider the symmetry $U(g)$ acting on
the boundary $C$, then \emph{restrict} it to a subregion $M$, which in general will
be a manifold with boundary ($C$ itself has no boundary as it is the
boundary of a higher-dimensional manifold), to see if the symmetry is 
implemented consistently or not.

First, we need to give a more general definition of what it
means to restrict a local unitary $U$ acting on a spatial manifold $C$ to a
sub-manifold $M$, which for the case discussed above was obvious due to the
tensor-product structure. Specifically, we say that a local unitary $U_M$ acting
on the region $M$ is the restriction of
$U$ to the region $M$ if it acts the same as $U$ in the \emph{interior} of $M$, well
away from the boundary $\partial M$. We observe two properties about this
restriction:
\begin{enumerate}[(a)]
  \item It always exists for any local unitary. This can easily be seen from,
    for example, the quantum circuit description.
  \item It is defined modulo local unitaries acting in the vicinity of the
    boundary $\partial M$.
\end{enumerate}
The second property is the higher-dimensional generalization of the restriction
being defined only up to phase factors. Thus, in general, if $U(g)$ is a
representation of the symmetry group $G$, then $U_M(g)$ need only satisfy
\begin{equation}
  \label{first_coboundary_nonabelian}
  U_M(g_1) U_M(g_2) = \Omega(g_1, g_2) U_M(g_1 g_2)
\end{equation}
where $\Omega(g_1, g_2)$ is a local unitary acting in the vicinity of $\partial
M$ which represents the obstruction to a consistent representation on $M$ due to
the fact that it is a manifold with boundary. Thus, we have reduced the problem
of classifying local unitary representations $U(g)$ on a $d$-dimensional manifold to that
of classifying local unitary obstructions $\Omega(g_1, g_2)$ on a $(d-1)$-dimensional
manifold. The idea now is to perform more such reductions, each time reducing
by 1 the dimensionality of the manifold acted upon, until we get down to the
simplest case of 0 dimensions (i.e.\ points).

For $(2+1)$-D SPT's, this reduction can be completed as follows. In this case
the boundary has only one spatial dimension, and so $\Omega(g_1, g_2)$ as constructed
above already acts on just a pair of points $a$ and $b$. We observe that
\eqnref{first_coboundary_nonabelian}, together with the associativity of the
operators $U_M(g)$, implies that $\Omega$ must satisfy
\begin{equation}
  \label{omega_associativity}
  \Omega(g_1, g_2) \Omega(g_1 g_2, g_3) = \; ^{U_M({g_1})} \Omega(g_2, g_3)
\Omega({g_1},{g_2}{g_3}),
\end{equation}
which is a non-abelian analogue of \eqnref{omega_associativity_comm}, and where we
have introduced the conjugation notation $^x y = xyx^{-1}$.  Now we perform a
second restriction, from $\partial M = \{ a, b \}$ to the single point
$a$. The restriction
$\Omega \to \Omega_a$ is defined only up to phase factors, and so we conclude
that $\Omega_a$ satisfies \eqnref{omega_associativity} only up to phase factors:
\begin{equation}
  \label{hacker}
  \Omega_a(g_1, g_2) \Omega_a (g_1 g_2, g_3) = \omega(g_1, g_2, g_3) \;
  ^{U_M({g_1})} \Omega_a(g_2, g_3) \Omega_a(g_1,
  g_2 g_3),
\end{equation}
where $\omega(g_1, g_2, g_3) \in \U(1)$. We show in Appendix
\ref{sec_crossed_module} that
$\omega$ must satisfy the 3-cocycle condition
\begin{equation}
  \label{three_cocycle_condition}
  \omega(g_1, g_2, g_3) \omega(g_1 g_2, g_3, g_4)^{-1} \omega(g_1, g_2 g_3,
  g_4) \omega(g_1, g_2, g_3 g_4)^{-1} \omega(g_2, g_3, g_4) = 1.
\end{equation}
Furthermore, as $\Omega_a(g, g^{\prime})$ is only defined up to phase factors
$\beta(g, g^{\prime})$, we
must identify
\begin{equation}
  \label{three_coboundary}
  \omega(g_1, g_2, g_3) \sim \omega(g_1, g_2, g_3) \; \beta(g_1, g_2) \beta(g_1
  g_2, g_3) \beta(g_2, g_3)^{-1} \beta(g_1, g_2 g_3)^{-1}.
\end{equation}
We show in Appendix \ref{sec_crossed_module} that, up to equivalence, the choice of restriction
$U(g) \to U_M(g)$ does not affect the 3-cocycle. 
The group of 3-cocycles quotiented by the equivalence relation
\eqnref{three_coboundary} is, by
definition, the third cohomology group $H^3(G, \U(1))$. Hence, we recover the
cohomological classification of (2+1)-D SPT's.

\subsection{Higher dimensions}
\label{sec_higher_dims}
In higher dimensions it is not clear whether we can still do the reduction
procedure in complete generality as in the (2+1)-D case. Nevertheless, we can still
perform the reduction if we make some simplifying assumptions about
the action of the symmetry on the boundary. (The non-linear sigma models
discussed in Section \ref{sec_nlsm} are a non-trivial example in which the symmetry on the
edge takes the required form.) Specifically, we consider a symmetry group $G$ acting on a Hilbert space equipped with a set of
basis states labeled by the variables $\alpha(x)$ associated with each
spatial location in a closed $(d-1)$-dimensional space $C_1$. We can take the spatial coordinate
$x$ to be either discrete (i.e.,
a lattice) or continuous. The class of symmetry actions that we
consider are those that can be written in the form
\begin{equation}
  \label{simple_symmetry}
  U(g) = N(g) S(g),
\end{equation}
such that:
\begin{enumerate}[(a)]
\item $S(g)$ is the on-site part of the symmetry which can be written
  in the form 
  \begin{equation}
    S(g) = \sum_{ \alpha } \ket{ g \alpha  }\bra{ \alpha },
  \end{equation}
  where $\alpha \to g \alpha$ is some on-site action of the symmetry on the classical
  labels $\alpha$; and
\item in the same basis, the non-on-site part $N(g)$ is diagonal, namely
  \begin{equation}
    N(g) = \sum_{\alpha} e^{i \mathcal{N}^{(1)}(g)[\alpha]} \ket{\alpha}\bra{\alpha}
  \end{equation}
  where $\mathcal{N}(g)$ are functionals of the configuration
  $\alpha$. We require these functionals to be sufficiently local that $N(g)$, and hence
  $U(g)$, are local unitaries.
\end{enumerate}
The requirement that $U(g)$ be a representation, $U(g_1) U(g_2) = U(g_1 g_2)$,
can be written in terms of the functionals $\mathcal{N}(g)$ as
\begin{equation}
  \label{first_coboundary}
  g_1 \mathcal{N}^{(1)}(g_2) + \mathcal{N}^{(1)}(g_1) - \mathcal{N}^{(1)}(g_1 g_2) = 0 \quad
  \mbox{(mod $2\pi$)},
\end{equation}
where we have defined the action of group elements on functionals in the obvious
way: $(g \mathcal{F})[\alpha] = \mathcal{F}[g^{-1} \alpha]$. Henceforth we will take
the (mod $2\pi$) to be implied, or in other words we consider the functionals to take values in
$\mathbb{R}/(2\pi\mathbb{Z})$.

Now as before, we can restrict $U(g)$ to a subregion $M_1$ with boundary, which
(since $S(g)$ can be trivially restricted)
amounts to restricting the functionals $\mathcal{N}^{(1)}(g)$. Then
\eqnref{first_coboundary} need be satisfied by the restricted functionals
$\widetilde{\mathcal{N}}^{(1)}(g)$ only up to boundary terms,
\begin{equation}
  \label{first_coboundary_boundary}
  g_1 \widetilde{\mathcal{N}}^{(1)}(g_2) + \widetilde{\mathcal{N}}^{(1)}(g_1) -
  \widetilde{\mathcal{N}}^{(1)}(g_1 g_2) =
  \mathcal{N}^{(2)}(g_1, g_2)
\end{equation}
where the $\mathcal{N}^{(2)}(g_1, g_2)$ are functionals which depend only on the
value of $\alpha$ near the boundary $\partial M_1$ and describe the obstruction.
This corresponds to \eqnref{first_coboundary_nonabelian}. 

In order to continue the reduction process, we find it useful to define the
group coboundary operators $\delta_k$ which map functionals depending on $k$
group elements into functionals depending on $k+1$ group elements, as follows:
\begin{multline}
  \label{coboundary_operator}
  (\delta_k \mathcal{N}^{(k)})(g_1, \cdots, g_{k+1})
  = g_1 \mathcal{N}^{(k)}(g_2, \cdots g_n) + (-1)^{k+1} \mathcal{N}^{(k)}(g_1,
  \cdots, g_k) \\+ \sum_{i=1}^k (-1)^i \mathcal{N}^{(k)}(g_1, \cdots, g_{i-1}, g_i
  g_{i+1}, g_{i+2}, \cdots, g_{k+1}).
\end{multline}
In particular, $(\delta_1 \mathcal{N}^{(1)})(g_1, g_2)$ corresponds to the left-hand side of
\eqnref{first_coboundary}. The important property which the coboundary operators
satisfy is that they form a chain complex, i.e. $\delta_{k+1} \circ \delta_k =
0$.

\begin{figure}
  \includegraphics{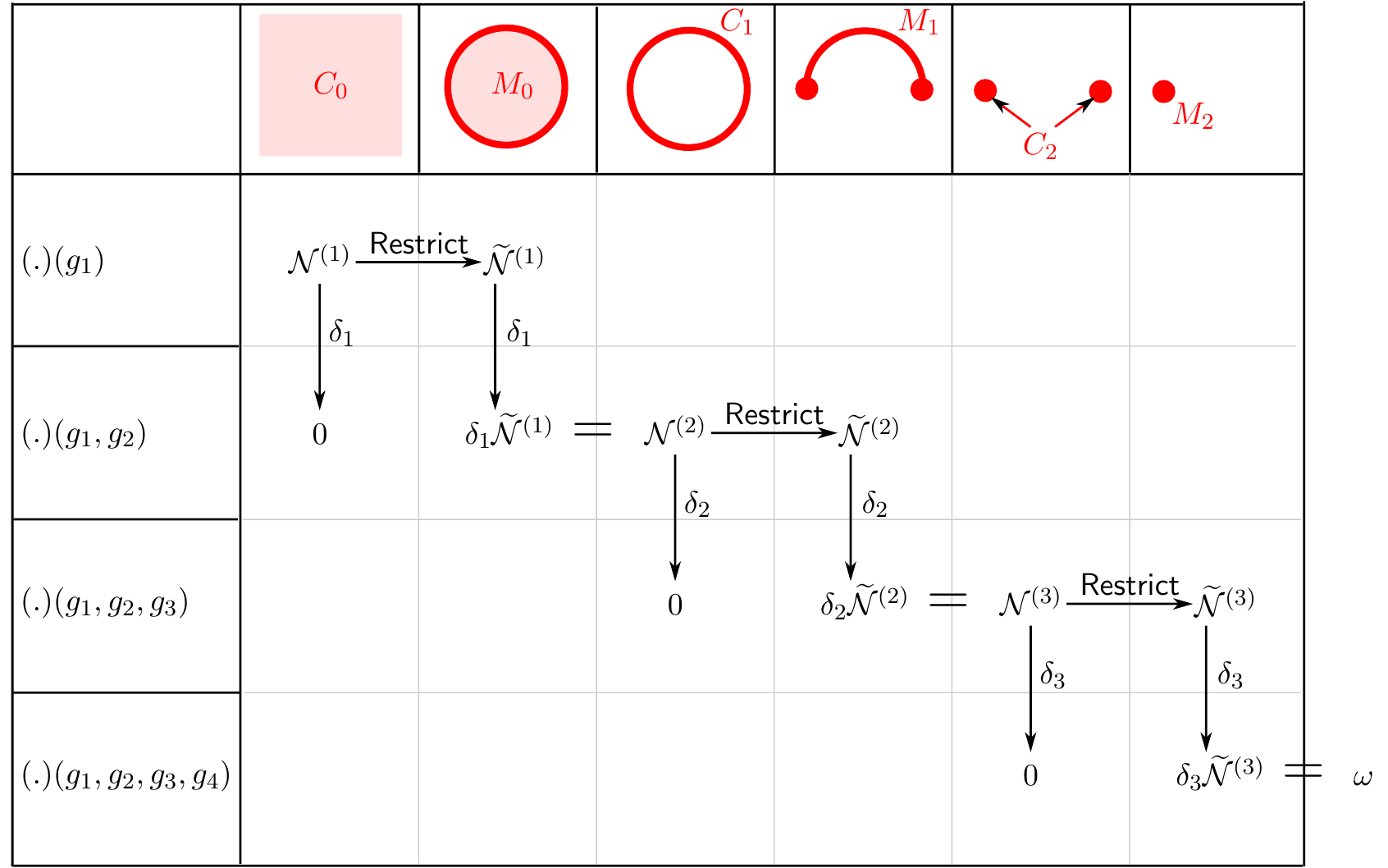}
  \caption{The reduction process to obtain a 4-cocycle $\omega$ on the (2+1)-D
    edge of a (3+1)-D system, assuming a symmetry representation on the edge of
  the form \eqnref{simple_symmetry}.}
\end{figure}

We can now formulate the reduction process for symmetries acting on a manifold
of arbitrary spatial dimension $d$. At the $k$-th step of the process, we have
a set of functionals $\mathcal{N}^{(k)}$ acting on a closed $d-k$-dimensional
manifold $C_k$ and indexed by $k$ group elements, satisfying $\delta_k
\mathcal{N}^{(k)} = 0$. We then consider restrictions
$\widetilde{\mathcal{N}}^{(k)}$ of these functionals onto the manifold $M_k$,
where $M_k$ is a submanifold of $C_k$ with boundary. As
$\widetilde{\mathcal{N}}^{(k)}$ must act the same as $\mathcal{N}^{(k)}$ in the
\emph{interior} of $M_k$, it follows that $\mathcal{N}^{(k+1)} \equiv \delta_k
\widetilde{\mathcal{N}}^{(k)}$ acts on 
the boundary $\partial M_k \equiv C_{k+1}$.  Furthermore, as $\delta_{k+1} \circ
\delta_k = 0$, it follows that $\delta_{k+1} \mathcal{N}^{(k+1)} = 0$. Thus, we
just iterate these reduction steps, terminating when we reach $\omega = \mathcal{N}^{(d+1)}$, which is simply a
mapping from $d+1$ group elements to $\mathrm{U}(1)$ satisfying
$\delta_{d+1} \mathcal{N}^{(d+1)}$; this the definition of a $\U(1)$
$(d+1)$-cocycle. Due to the ambiguity in the choice of
restrictions, it follows that $\omega$ is only defined up to
\begin{equation}
  \label{general_equivalence_relation}
  \omega \sim \omega + \delta_{d+1} \lambda
\end{equation}
where $\lambda$ is some element of $\U(1)$ depending on $d+1$ group elements. The
group of $(d+1)$-cocycles quotiented by the equivalence relation
\eqnref{general_equivalence_relation} is, by definition, the cohomology group $H^{d+1}(G, \U(1))$. Thus, we recover the cohomological
classification of SPT phases in arbitrary dimensions.

Finally, let us discuss the case of symmetry
groups that contain anti-unitary operations. It is perhaps unclear in general what is meant
by restriction of an anti-unitary operation (although see
Ref.~\onlinecite{gauging_time_reversal}). Nevertheless, if we consider only
symmetries that can be represented as a suitable generalization of \eqnref{simple_symmetry},
the same arguments as above can be applied with only minor modifications.
Specifically, we consider symmetries of the form
\begin{equation}
  U(g) = N(g) S(g) K^{n(g)},
\end{equation}
where $N(g)$ and $S(g)$ are as before, $K$ is complex conjugation in the $\{
\ket{\alpha} \}$ basis, and $n(g)$ is $0$ for unitary elements of $G$ and $1$ for
anti-unitary elements. If we define the action of $G$ on functionals as
$g \mathcal{F}[\alpha] = (-1)^{n(g)} \mathcal{F}[g^{-1} \alpha]$, all of the steps in
the above derivation can be carried through without change, except that there is a residual non-trivial action of $G$ on $\U(1)$.
Thus, the classification is
$H^{d+1}(G, \U(1))$, but with $\U(1)$ considered
as a non-trivial $G$-module, with anti-unitary elements acting by complex
conjugation.

\section{Example: ``Chiral'' symmetry on the edge of a (2+1)-D SPT}
\label{sec_chiral_example}
It was shown in Ref. \onlinecite{chen_wen_chiral} that the action of the symmetry on the gapless
edge of some non-trivial (2+1)-D SPT's is ``chiral'', as expressed (for example)
in the fact that it acts differently on the left- and right-moving fields. Let
us show how this corresponds to a local but not on-site symmetry and calculate
the corresponding 3-cocycle. We will focus on the simplest case where the
symmetry is just $\mathbb{Z}_2$, but similar arguments can be made for $\mathbb{Z}_n$
or $\U(1)$ symmetries.

We assume the low-energy theory of the (1+1)-D edge is described by a massless
boson field $\varphi$ with compactification radius $2\pi$, i.e. a bosonic Luttinger liquid,
with Lagrangian density
\begin{equation}
  \mathcal{L} = \frac{g}{2\pi}\left[\frac{1}{v}(\partial_t \varphi)^2 - v (\partial_x
  \varphi)^2\right].
\end{equation}
We introduce the dual boson field $\theta$ according to $\partial_x \theta
= 2\pi \Pi$, where $\Pi$ is the canonical momentum conjugate to $\varphi$.
The commutation relation for $\theta$ and $\varphi$ is, therefore,
\begin{equation}
[ \varphi(x),\theta(x')] = -2\pi i \Theta(x-x')
\label{eqn:LL-comm-rel}
\end{equation}
where $\Theta(x)$ is the unit step function.
Note that this definition, together with the fact that total angular
momentum is quantized to integers, implies that $\theta$ is also an angular
variable defined modulo $2\pi$. 

Now, suppose that the fields $\varphi$ and $\theta$ transform under $\mathbb{Z}_2$
according to
\begin{align}
  \varphi &\to \varphi + n \pi, \nonumber \\
\theta &\to \theta + m \pi, \label{various}
\end{align}
Here $(n,m) = (1,0)$ corresponds to a normal on-site
$\pi$ rotation of the boson field. On the other hand, as we shall see, $(n,m) = (1,1)$ is the
non-on-site symmetry that we would expect at the edge of a non-trivial
$\mathbb{Z}_2$ SPT. Also, $m \neq 0$ corresponds to a
superficially ``chiral'' symmetry in the sense that the left- and right-moving
fields $\phi_{L,R} = \varphi \pm \theta$ transform differently under
$\mathbb{Z}_2$, but in the $\mathbb{Z}_2$ case [though not for $\mathbb{Z}_n$ or
$\U(1)$] this chirality is not physically meaningful because $\theta \sim
\theta + 2\pi$ so $m$ is actually only defined modulo 2.

From the commutation relations (\ref{eqn:LL-comm-rel}),
one can show that \eqnref{various} is effected by the unitary
operator $U = (-1)^{nL + mW} = N^m S^n$, where
where $L$ is the total angular momentum and $W$ is the total winding number, and
we define 
\begin{align}
  N &= \exp\left(-\frac{i}{2} \int \partial_x \varphi \, dx\right) \\
  S &= \exp\left(-\frac{i}{2} \int \partial_x \theta dx \right). 
\end{align}
We now define the restriction $U_{[a,b]} = N_{[a,b]}^m
S_{[a,b]}^n$ to a finite interval $[a,b]$, where
\begin{align}
  N_{[a,b]} &= \exp\left(-\frac{i}{2} \int_a^b \partial_x \varphi \, dx\right) \\
  S_{[a,b]} &= \exp\left(-\frac{i}{2} \int_{a-\epsilon}^{b+\epsilon} \partial_x \theta \, dx \right),
\end{align}
where we have made use of our freedom to redefine the restriction near the
boundary of $[a,b]$ to shift the endpoints of the second integral by some
small $\epsilon > 0$. This ensures that $N_{[a,b]}$ and $S_{[a,b]}$ commute.
Hence, we find that $U_{[a,b]}^2 = N_{[a,b]}^{2m} S_{[a,b]}^{2n}$, where
\begin{align}
  N_{[a,b]}^2 &= \exp\left(-i \int_a^b \partial_x \varphi dx\right) \\
            &= e^{i \varphi(a)} e^{-i \varphi(b)} \\
S_{[a,b]}^2 &= \exp\left(- i \int_{a-\epsilon}^{b+\epsilon} \partial_x \theta dx\right) \\
            &=  e^{i \theta(a-\epsilon)} e^{- i \theta(b+\epsilon)}.
\end{align}
Thus, as expected, we find that
$\Omega \equiv U_{[a,b]}^2 = [e^{in \theta(a-\epsilon)}
e^{i m\varphi(a)}] 
[e^{- in \theta(b+\epsilon)} e^{-im\varphi(b)} ] \equiv \Omega_a \Omega_b$ still acts non-trivially at the endpoints $a$
and $b$ even though $U^2 = 1$.

In the present example, \eqnref{omega_associativity} takes the form
\begin{equation}
  U_{[a,b]} \Omega U_{[a,b]}^{-1} = \Omega,
\end{equation}
and this equality can readily be verified directly from the forms of $U_{[a,b]}$ and $\Omega$ given
above. On the other hand, the restriction $\Omega_a$
satisfies this equation in general only up to a phase factor. Indeed, we find
\begin{equation}
  U_{[a,b]} \Omega_a U_{[a,b]}^{-1} = e^{- i n \theta(a-\epsilon)} e^{-im \varphi(a) - imn \pi} = (-1)^{mn} \Omega_a.
\end{equation}
Hence, we find that the $3$-cocycle associated with the realization of $\mathbb{Z}_2$
is given by $\omega(X,X,X) = (-1)^{mn}$ and $\omega(g_1, g_2, g_3) = 1$ for $(g_1, g_2, g_3)
\neq (X,X,X)$, where $X$ is the generator of $\mathbb{Z}_2$. For $m = n = 1$
this corresponds to a non-trivial 3-cocycle, and the corresponding
representation of $\mathbb{Z}_2$ would appear at the boundary of a
non-trivial (2+1)-D $\mathbb{Z}_2$ SPT.

\section{Proof of separation of phases in (2+1)-D.}
\label{separation_proof}
In this section, we will outline how one can use the ideas given above to prove for
(2+1)-D systems that systems characterized by different elements of the
cohomology group $H^3(G, \U(1))$ must be separated by a bulk phase transition;
the details are left to the appendices.
(Unfortunately, the proof cannot be applied in higher dimensions due to the
lack of a completely general characterization of anomalous symmetry.)

First, as we want to make statements about bulk properties, we need to
reformulate the ideas of Section \ref{sec_general} in a slightly different way, in terms of
properties of the ground state in the bulk rather than the low-energy physics at
the edge. 
We
show in Appendix \ref{sec_completing} that, given a general ground state $\ket{\Psi}$ in some
SPT phase in $d$ spatial dimensions ($d \leq 2$), and a region $A$ in the bulk, one can find a
representation $V_{\partial A}(g)$ of the symmetry group, which acts \emph{inside} $A$, but only near the
boundary $\partial A$, such that $U_A(g) \ket{\Psi} = V_{\partial A}(g)
\ket{\Psi}$. Here
$U_{A}(g)$ is the restriction of the symmetry onto the region $A$ (which can
be defined consistently since we are assuming the symmetry is represented
on-site in the bulk.) The physical interpretation of this result is simply that,
as $\ket{\Psi}$ is invariant under $U(g)$, therefore $U_A(g) \ket{\Psi}$ can
differ from $\ket{\Psi}$ only near the boundary $\partial A$. This representation
$V_{\partial A}(g)$ can be anomalous in the same way as the representation of
the symmetry on a physical edge, and the anomaly can be classified using the
method of Section \ref{sec_general}.

The final result that we need is that the element of $H^{3}(G, \U(1))$ is
independent of the choice of region $A$, \emph{even} in the presence of
spatial inhomogeneity; this is also proved in Appendix \ref{sec_completing}. (Actually, as
discussed in that appendix, we only prove this for certain regions $A$, but that
is sufficient for the following discussion.) This allows us to prove that two
systems $\mathcal{S}$ and $\mathcal{S}^{\prime}$ characterized by different elements of $H^{d+1}(G, \U(1))$ must be
separated by a phase transition \cite{top_ent_entr}. Indeed, consider two systems connected
without a phase transition. Then, without closing the gap, one can create an
interpolated system that looks like $\mathcal{S}$ on some region $A$ and like
$\mathcal{S}^{\prime}$ on another region $A^{\prime}$ (see Appendix
\ref{sec_completing} for a
careful proof of this fact.) It therefore follows that the same element of
$H^{d+1}(G, \U(1))$ must be obtained in both cases. By a similar argument, one
also finds that a spatial boundary between two different SPT phases must either
be gapless or break the symmetry.

\section{Non-linear sigma models}
\label{sec_nlsm}
It has been found \cite{boson_iqhe,cenke_neel_order,vishwanath_senthil,adr_classification} that a quite general way to reproduce the essential features of various SPT phases
is through the field theory of a quantum non-linear sigma model (NL$\sigma$M), where topological
properties of the SPT phase arise out of the bulk theta term included in the
action. Here, we will show in such models, the presence of the theta term indeed
leads to an obstruction to on-site representation of the symmetry on a spatial
edge, in such a way as to
allow a straightforward calculation of the corresponding element of the
cohomology group.

For example, consider in $D$ space-time dimensions (i.e.\ $D = d+1$) a NL$\sigma$M for
the $(D+1)$-component vector field $\mathbf{n}$, constrained to have unit norm,
i.e.\ $\mathbf{n}$ lies on a unit $D$-sphere.
The (Euclidean) action can be written as the sum of a dynamical
contribution $S_{\mathrm{dyn}}$ and a topological contribution
$S_{\mathrm{top}}$:
\begin{align}
  S^{\mathrm{bulk}} &= S_{\mathrm{dyn}}^{\mathrm{bulk}} +
  S_{\mathrm{top}}^{\mathrm{bulk}} \label{S_bulk}, \\
  S_{\mathrm{dyn}}^{\mathrm{bulk}} &=\frac{1}{g}\int d^D x \; \partial_\mu \mathbf{n} \cdot
  \partial_\mu \mathbf{n} \label{S_dyn} \\
  S_{\mathrm{top}}^{\mathrm{bulk}} &= i\Theta \frac{1}{V_D} \int
  \mathbf{n}^{*}(\omega_V)
  \label{S_top},
\end{align}
where $V_{D}$ is the volume of the unit $D$-sphere, and $\mathbf{n}^{*}(\omega_V)$ is the
pullback through the map $\mathbf{n}$ of the volume form on the unit $D$-sphere.
Written componentwise, this
amounts to
\begin{equation}
  S_{\mathrm{top}}^{\mathrm{bulk}} = i\Theta \frac{1}{V_D} \int d^{D} x \; \epsilon^{a_1, \cdots, a_{D+2}} 
  n^{a_1} \partial_0 n^{a_2} \partial_1 n^{a_2} \cdots \partial_{D-1} n^{a_{D}},
\end{equation}
where $\epsilon^{a_1, \cdots, a_{D+1}}$ is the $(D+1)$-dimensional Levi-Civita symbol.
The theta term
$S_{\mathrm{top}}^{\mathrm{bulk}}$ measures a topologically invariant ``generalized
winding number'' in $\pi_D(S^D) \cong \mathbb{Z}$, and for spacetimes without boundary is quantized to integer
multiples of $i\Theta$. Hence, we implement the requirement that SPT phases be
trivial in the bulk by setting $\Theta$ to be an integer multiple of $2\pi$,
thus ensuring that $S_{\mathrm{top}}^{\mathrm{bulk}}$ makes no contribution to the partition
function $\int \mathcal{D}[\mathbf{n}] e^{-S}$. In fact, although we have given
a specific form of $S_{\mathrm{dyn}}^{\mathrm{bulk}}$ for concreteness, it will not be important
for our analysis as the topological features of the system are entirely captured
by $S_{\mathrm{top}}^{\mathrm{bulk}}$.


Although the inclusion of $S^{\mathrm{bulk}}_{\mathrm{top}}$ has no effect on the partition
function in the bulk, it does play a crucial role once we introduce a spatial
edge. In that case $S_{\mathrm{top}}^{\mathrm{bulk}}$ depends (mod $2\pi i$) only on the values of
$\mathbf{n}$ on the boundary (to see this, note that any two extensions into the
bulk can be connected at the boundary to give a closed surface, on which
$e^{-S_{\mathrm{top}}} = 1$); the action on the boundary is referred to as the
\emph{Wess-Zumino-Witten} action $S_{WZW}$.
Thus, we can integrate out the gapped bulk to
give an effective
action for the low-energy excitations on the edge of the form
\begin{equation}
  \exp(-S^{\mathrm{edge}}) = \exp(-S^{\mathrm{edge}}_{\mathrm{dyn}} -
  S_{\mathrm{WZW}}),
\end{equation}
where $S^{\mathrm{edge}}_{\mathrm{dyn}} = \int d^d x
\mathcal{L}^{\mathrm{edge}}_{\mathrm{dyn}}$ is some unimportant dynamical term
derived from $S^{\mathrm{bulk}}_{\mathrm{dyn}}$.
Note that one can then write 
$S_{\mathrm{WZW}} = \int d^d x \mathcal{L}_{\mathrm{WZW}}$ for some local
Lagrangian density $\mathcal{L}_{\mathrm{WZW}}$ defined on the edge. 
However, there is no canonical way to
do so.

Now let us consider the symmetry group $G$ in the bulk corresponding to some
invertible action $\mathbf{n} \to
g\mathbf{n}$ for $g \in G$. We demand that $S_{\mathrm{dyn}}^{\mathrm{bulk}}$ and
$S_{\mathrm{top}}$ be \emph{locally}
invariant under the symmetry, i.e.\ that the integrands in Eqs. (\ref{S_dyn}) and
(\ref{S_top}) must be invariant, not just the integral. Then we expect that
$S^{\mathrm{edge}}_{\mathrm{dyn}}$ is also locally invariant under the symmetry.
  $S_{WZW}$ must also be globally invariant (at least,
  modulo $2\pi i$) but in general we do not expect it to be locally
  invariant. Indeed, because
there is no canonical choice for $\mathcal{L}_{WZW}$, one expects
that the symmetry will transform $\mathcal{L}_{WZW}$ to a different Lagrangian
that nevertheless integrates to the same
action (modulo $2\pi i$) in a spacetime without boundary.

%
We will now show that, after quantization, the lack of local invariance of
$S_{WZW}$ implies the non-on-site
nature of the unitary
representation of the symmetry on the edge.
We assume that after quantization the Hilbert space is spanned by a basis of
states labeled by spatial configurations of $\mathbf{n}$ at a fixed time. We
can calculate the imaginary-time propagator $e^{-\beta H}$ (or equivalently, the
Hamiltonian $H$) by a path integral
\begin{equation}
  \label{propagator}
  \bra{ \mathbf{n}^{\prime} } e^{-\beta H} \ket{ \mathbf{n} } = \int
  \mathcal{D}[\mathbf{n}(\tau)] e^{-S^{\mathrm{edge}}\{0,\beta\}},
\end{equation}
where
\begin{equation}
  S^{\mathrm{edge}}\{0,\beta\} =
  \int d^{D-2} x \int_0^\beta d\tau \;
  (\mathcal{L}^{\mathrm{edge}}_{\mathrm{dyn}} + \mathcal{L}_{WZW})
\end{equation}
is the action evaluated on a spacetime with temporal boundaries at $\tau=0$ and
$\tau=\beta$. Now, so far we only know that $S_{\mathrm{WZW}}$ is globally
invariant (modulo $2\pi i$) on a space-time manifold without boundary. Since
$S_{\mathrm{WZW}}$ is not \emph{locally} invariant, in the presence of a
temporal boundary we can only conclude 
that it will transform as
$S_{\mathrm{WZW}}\{0,\beta\}
\to g S_{\mathrm{WZW}}\{0,\beta\}$ ($g \in G$), where the
difference can be expressed in terms of the field configurations at the
temporal boundaries:
\begin{equation}
  \label{symdiff}
  g S_{\mathrm{WZW}}\{0,\beta\} - S_{\mathrm{WZW}}\{0,\beta\} =
  i\mathcal{N}(g)[\mathbf{n}(\tau)] - i\mathcal{N}(g)[\mathbf{n}(0)] \quad (\mbox{mod $2\pi i$}),
\end{equation}
where $\mathcal{N}(g)$ is a functional of the field configuration at a fixed time. 

\eqnref{symdiff} implies that the edge Hamiltonian is not invariant under the naive
on-site implementation of the symmetry, $S(g) = \int \mathcal{D}[\mathbf{n}] \,
\ket{g \mathbf{n}} \bra{\mathbf{n}}$. Indeed, combined with \eqnref{propagator},
we find
\begin{align}
  \bra{\mathbf{n}^{\prime}} S(g)^{\dagger} e^{-\beta H} S(g) \ket{\mathbf{n}} &= e^{i
\mathcal{N}(g)[\mathbf{n}^{\prime}] - i \mathcal{N}(g)[\mathbf{n}]} \bra{\mathbf{n}^{\prime}}
  e^{-\beta H} \ket{\mathbf{n}} \\
  &= \bra{\mathbf{n}^{\prime}} N(g)^{\dagger} e^{-\beta H} N(g) \ket{\mathbf{n}}.
\end{align}
where
\begin{equation}
  \label{Ng_nlsm}
  N(g) = \int \mathcal{D}[\mathbf{n}] e^{i \mathcal{N}(g)[\mathbf{n}]}
  \ket{\mathbf{n}} \bra{\mathbf{n}}.
\end{equation}
Hence, we see that the correct implementation of the symmetry on the edge, which does
commute with the Hamiltonian, is $U(g) = N(g) S(g)$. In general,
there is no reason to expect $N(g)$ to be on-site, as we shall see.
However, as we show in Section \ref{nlsm_cochains}, it is necessarily local.
Thus, the symmetry on the edge is a local but non-on-site symmetry precisely of
the form considered in Section \ref{sec_higher_dims}, and we can calculate the appropriate element
of the cohomology group using the reduction procedure of that section.

We can also consider anti-unitary symmetries by a straightforward extension of
the above considerations. Specifically, an anti-unitary symmetry is 
implemented in the action by $\mathbf{n} \to g \mathbf{n}$, $i \to (-1)^{n(g)}i$.
Then we find that the representation of the symmetry on the edge is $U(g) = N(g)
S(g) K^{n(g)}$, with $N(g)$ and $S(g)$ as before and $K$ complex conjugation in
the $\mathbf{n}$ basis.

\subsection{Calculating the cocycle in nonlinear sigma models using $\U(1)$
cochains on the target manifold}
\label{nlsm_cochains}
A particularly compact and elegant way of calculating the cocycle for
NL$\sigma$Ms
is by
interpreting the theta term in terms of a $\U(1)$ cochain defined on the target
manifold $T = S^{D}$. First we need to state some defintions. 
We refer to $k$-dimensional oriented integration domains on a
manifold $T$ as $k$-chains. Given a $k$-chain $A$, we denote the opposite
orientation by $-A$, and we can also define a sum operation on $k$-chains in the
natural way, so that the $k$-chains can be viewed as an additive group. (If one
wanted to be rigorous, one would define $k$-chains as formal linear combinations
of oriented $k$-simplices with integer coefficients.) A
$\U(1)$ $k$-\emph{cochain} is a linear mapping from $k$-chains to $\U(1)$ [which
we here write additively as $\mathbb{R}/(2\pi\mathbb{Z})$]. (Note that we are
here referring to \emph{topological} cochains on a manifold; these should
be distinguished from the \emph{group} cochains that are used to construct the group cohomology of
some group $G$.) In particular, each
differential $k$-form $\omega$ induces a $\U(1)$ $k$-cochain by integration,
\begin{equation}
  \omega(A) = \left( \int_A \omega \right) \operatorname{mod} 2\pi.
\end{equation}
where in an abuse of notation we will denote the $k$-form and the $\U(1)$ $k$-cochain
by the same symbol.
Any $\U(1)$ $k$-cochain $\omega$ on the target manifold $T$ can be used to define a
local  $\U(1)$-valued functional $F_\omega$ for a $T$-valued field
$\mathbf{n}$ on
a $k$-dimensional space(-time) manifold $M$ via
\begin{equation}
  \label{F_omega}
  F_\omega[\mathbf{n}] = \omega(\mathbf{n}(M)),
\end{equation}
where $\mathbf{n}(M)$ is the image of $M$, viewed as a chain, under the mapping
$\mathbf{n}$. If $\omega$ is derived from a differential $k$-form, this is equivalent to
defining $F_\omega$ as the integral of the pullback, $F_\omega[\mathbf{n}] =
\left(\int_M \mathbf{n}^{*}(\omega)\right) \operatorname{mod} 2\pi$. In
particular, the
topological theta term action of \eqnref{S_top} is a special case of \eqnref{F_omega}.

We define the coboundary operator $d$ which maps $k$-cochains to
$(k+1)$-cochains according to
\begin{equation}
  (d\omega)(A) = \omega(\partial A),
\end{equation}
where $\partial A$ is the boundary of $A$.
We call a $k$-cochain $\omega$ \emph{exact} if it can be written as $\omega =
d\kappa$ for some $(k-1)$-cochain $\kappa$. Our central tool is the following
result.
  \begin{lemma}
    \label{lemma:central}
  A $\U(1)$ $k$-cochain $\omega$ on a manifold $T$ is exact if and only if
  $\omega(C) = 0$ for all closed (i.e.\ boundaryless) $k$-chains $C$.
  \begin{proof}
    See Appendix \ref{appendix_universal}.
  \end{proof}
  \end{lemma}
The property that $\omega(C) = 0$ for closed $C$ in turn is equivalent to requiring of the induced functional
$F_{\omega}$ that it vanish on all closed space-time manifolds. If this is satisfied, then
one expects that for a space-time manifold $M$ with boundary, $F_{\omega}[\mathbf{n}]$ should depend
only on the values of $\mathbf{n}$ on the boundary $\partial M$. Indeed, given
$\omega = d\kappa$, one finds that
\begin{align}
  F_\omega[\mathbf{n}] &= (d\kappa)(\mathbf{n}(M)) \\
                       &= \kappa(\partial \mathbf{n}(M))  \\
                       &= \kappa(\mathbf{n}(\partial M)) \\
                       &\equiv F_{\kappa}[\mathbf{n}(\partial M)].
\end{align}
Given the above considerations, one can show that the procedure for obtaining the edge symmetry from the theta
term, and then the cocycle from the edge symmetry, can be reduced to a
simple prescription in terms of the $\U(1)$ cochains defined on the target manifold, with no reference to
the space-time manifold at all, which we now describe.

\begin{table}
  \begin{tabular}{c|c|ccccc|}
    \multicolumn{6}{r}{$\leftarrow d$} \\
    \cline{2-7}
    & & 3              & 2              &   1            & 0 & \\
    \cline{2-7}
  $\delta$ & 0 & $\omega^{(0)}$ & $\kappa^{(0)}$ &              &              & \\
  $\downarrow$ &  1 & 0          & $\omega^{(1)}$ & $\kappa^{(1)}$ &              & \\
               & 2 &            & 0              & $\omega^{(2)}$ & $\kappa^{(2)}$ & \\
               & 3 &            &                & 0              & $\omega^{(3)}$ & \\
               & 4 &            &                &                & 0 & \\
    \cline{2-7}
  \end{tabular}
  \caption{A tabular representation of the reduction process to extract a
    $\U(1)$ group 3-cocycle $\nu = \omega^{(3)}$ starting from a symmetric topological term in (2+1)-D represented by a
  topological $\U(1)$ cochain $\omega^{(0)}$. Each cell in the table is specified by a row
  label $l$ and a column label $k$, and corresponds to a set of $k$-cochains
  labeled by $l$ group elements. Going left in the table corresponds to applying
  the \emph{topological} coboundary operator $d$, whereas going down
corresponds to applying the \emph{group} coboundary operator $\delta$ defined by
\eqnref{coboundary_operator_cochains}. These two operations commute, so the table can be
interpreted as a commutative diagram.}
\end{table}

We start from a topological action $S_{\mathrm{top}}$ on a spacetime-manifold
$M$ with $d$-dimensional target manifold $T$, written as $S_{\mathrm{top}}[\mathbf{n}] =
F_{\omega^{(0)}}[\mathbf{n}]
= \omega^{(0)}(\mathbf{n}(M))$, where $\omega^{(0)}$ is an exact $\U(1)$ $d$-cochain on
$T$ which is invariant under the action
of the symmetry, $g \omega^{(0)} = \omega^{(0)}$. Here we defined the action of the
symmetry on a cochain by $g \omega(A) = (-1)^{n(g)} \omega(gA)$, where $n(g)$ is
1 for anti-unitary elements and 0 for unitary elements, and the action of $g$ on
chains is derived from its action on $\mathbf{n}$. Hence, we have $\delta_0
\omega^{(0)}
= g\omega^{(0)} - \omega^{(0)} = 0$, where we have introduced the \emph{group} coboundary operators
$\delta_k$ (\emph{not} the same as the topological coboundary operator $d$ defined above) in
the same way as
\eqnref{coboundary_operator} above, namely:
\begin{multline}
  \label{coboundary_operator_cochains}
  (\delta_k \omega^{(k)})(g_1, \cdots, g_{k+1})
  = g_1 \omega^{(k)}(g_2, \cdots g_n) + (-1)^{k+1} \omega^{(k)}(g_1,
  \cdots, g_k) \\+ \sum_{i=1}^k (-1)^i \omega^{(k)}(g_1, \cdots, g_{i-1}, g_i
  g_{i+1}, g_{i+2}, \cdots, g_{k+1}),
\end{multline}

Given a set of exact $(d-k)$-cochains $\omega^{(k)}$ indexed by $k$ group
  elements which satisfy $\delta_k \omega^{(k)} = 0$, we
  can write $\omega^{(k)} = d\kappa^{(k)}$ for some set of $(d-k-1)$-cochains
  $\kappa^{(k)}$. Now, $\delta_k \omega^{(k)} = 0$ implies that, for
  closed chains $C$, $(\delta_k \kappa^{(k)})(C) = (\delta_k
\omega^{(k)})(\partial C) = 0$. Hence, we can define $\omega^{(k+1)} = \delta_k
\kappa^{(k)}$ which is exact and satisfies $\delta_{k+1} \omega^{(k+1)} = 0$.
The sequence terminates when we reach $\omega^{(D)}$, which is a set of
  $0$-cochains indexed by $k$ group elements. Now a $0$-cochain is essentially
  just a scalar $\U(1)$ function defined on the target manifold $T$. But the fact
  that $\omega^{(D)}$ evaluates to zero for the closed $0$-chain $a - b$ (where
  $a$ and $b$ are any two points) implies that the $\omega^{(D)}$ are
  \emph{constant} $U(1)$ functions. Thus, $\omega^{(D)}$ defines a mapping from
  $D$ group elements to $\U(1)$ satisfying $\delta_D \omega^{(D)} = 0$, which
  defines an element of the group cohomology group $H^{D}(G, \U(1))$.

\subsection{Examples}
The possible symmetry transformations that leave the Lagrangian of
\eqnref{S_bulk}
invariant in space-time dimensions $D = 2,3,4$ were constructed in
Ref.~\onlinecite{adr_classification} 
for a variety of different symmetry groups. Our framework allows in
principle for the element of the cohomology group $H^D(G, U(1))$ to be calculated in all of these
cases. Let us consider a few examples.

\subsubsection{$Z_2^T$ in (1+1)-D}
\label{Z2t_1p1}
We write the symmetry group as $Z_2^T = \{ 1, \mathbb{T} \}$.
The target manifold is $S^2$ and we work in spherical coordinates $\mathbf{n} =
(\cos \theta, \sin \theta \cos \varphi, \sin \theta \sin \varphi)$. The action
of $\mathbb{T}$ on $\mathbf{n}$ is $\mathbb{T} \mathbf{n} = - \mathbf{n}$, or in terms
of the spherical coordinates $\theta \to \pi - \theta, \varphi \to \varphi + \pi$. The initial
$\U(1)$ cochain can be written in terms of a 2-form 
\begin{equation}
  \label{a_omega_0}
  \omega^{(0)} = \Theta \frac{1}{4\pi} \sin \theta  (d\theta \wedge d \varphi).
\end{equation}
As $\omega^{(0)}$ integrates to 0 (mod $2\pi$) over the whole $2$-sphere, it
follows that it can written as $\omega^{(0)} = d\kappa^{(0)}$ for some $\U(1)$
$1$-cochain $\kappa^{(0)}$. We can write
$\kappa^{(0)}$ explicitly as
\begin{equation}
  \kappa^{(0)} = \Theta \frac{1}{4\pi} (1-\cos \theta) d\varphi
\end{equation}
Treating $\kappa^{(0)}$ as a differential 1-form and taking the exterior
derivative, one recovers \eqnref{a_omega_0}. When written as a $1$-form,
$\kappa^{(0)}$ appears to have a singularity at $\theta = \pi$. To show that, as
a $\U(1)$ $1$-cochain, $\kappa^{(0)}$ is actually well-defined and satisfies
$d\kappa^{(0)} = \omega^{(0)}$ globally, it is sufficient to check that
$\int_C \kappa^{(0)} = 0$ (mod $2\pi$) for an loop $C$ of infinitesimal size encircling the
apparent singularity at $\theta = \pi$, which is indeed the case.

Now, following the general prescription of Section \ref{nlsm_cochains}, we define $\omega^{(1)} =
\delta_0 \kappa^{(0)}$. The only non-trivial component is
\begin{align}
  \omega^{(1)}(\mathbb{T}) &= \mathbb{T} \kappa^{(0)} - \kappa^{(0)} \\
                           &= -\frac{\Theta}{4\pi} (1 + \cos \theta) - \frac{\Theta}{4\pi}(1-\cos \theta) \\
&= -\frac{\Theta}{2\pi} d\varphi,
\end{align}
from which we immediately read off that $\omega^{(1)} = d\kappa^{(1)}$, where
$\kappa^{(1)} = -\frac{\Theta}{2\pi}\varphi$ (which is well-defined as a $U(1)$ $0$-cochain
because $\varphi$ is defined modulo $2\pi$). Thus, we can define the cocycle
$\nu = \delta_1 \kappa^{(1)}$, and the only non-zero component is
\begin{align}
  \nu(\mathbb{T},\mathbb{T}) &= \mathbb{T} \kappa^{(1)} + \kappa^{(1)} \\
                                      &= \frac{\Theta}{2\pi} \{\varphi + \pi - \varphi \}\\
                                      &= \frac{\Theta}{2}.
\end{align}
Thus, if $\Theta$ is an odd multiple of $2\pi$, this $2$-cocycle corresponds to
a non-trivial SPT phase, with the zero-dimensional boundary transforming
projectively under the symmetry, i.e.\ as a Kramers doublet with $\mathbb{T}^2 =
-1$. On the other hand, if $\Theta$ is an even multiple of $2\pi$, we have a
trivial SPT phase with the edge transforming as $\mathbb{T}^2 = 1$. Thus, by
different choices of $\Theta$ one recovers both elements of the cohomology group
$H^2(Z_2^T, \U(1)) \cong Z_2$.

\subsubsection{$Z_2$ in (2+1)-D}
We write the symmetry group as $Z_2 = \{ 1, X \}$. 
The target manifold is $S^3$ and we work in generalized spherical coordinates
$\mathbf{n} = (\cos \theta, \sin \theta \mathbf{n}_2)$, where $\mathbf{n}_2 \in S^2$. 
 The action of $X$ on $\mathbf{n}$ is $X \mathbf{n} = - \mathbf{n}$, or in terms
 of the generalized spherical coordinates $\theta \to \pi - \theta, \mathbf{n}_2 \to -\mathbf{n}_2$. The initial
$U(1)$ cochain is 
\begin{equation}
  \omega^{(0)} = \Theta \frac{1}{V_3} \sin^2 \theta  (d\theta \wedge \omega_{V,2})
\end{equation}
where $V_3$ is the volume of the $3$-sphere, and $\omega_{V,2}$ is the volume
form for $\mathbf{n}_2$. We then find that $\omega^{(0)} = d\kappa^{(0)}$, where
\begin{equation}
  \kappa^{(0)} = \Theta \frac{1}{V_3} \left(\int_0^\theta \sin^2 x \, dx\right) \omega_{V,2}.
\end{equation}
We observe that $V_3$ can be expressed as ($V_2 = 4\pi$)
\begin{equation}
  \label{relation}
V_3 = V_2 \int_0^\pi \sin^2 \theta \, d\theta.
\end{equation}
From this one can show that $\kappa^{(0)}$ is well-defined despite the apparent
singularity at $\theta = \pi$. Now the only non-trivial element of $\omega^{(1)}
= d\kappa^{(1)}$ is
\begin{align}
  \omega^{(1)}(X) &= X \kappa^{(0)} - \kappa^{(0)} \\
  &= \Theta \frac{1}{V_3} \omega_{V,2} \int_0^{\pi} \sin^2 x dx \\
  &= \Theta \frac{1}{4\pi} \omega_{V,2}
\end{align}
(here we used \eqnref{relation} and the fact that $\omega_{V,2}$ is odd under $\mathbf{n}_2 \to
-\mathbf{n}_2$.) In fact, this is identical to \eqnref{a_omega_0}. The reduction process
then proceeds nearly identically to that in Section \ref{Z2t_1p1} above and one finds
that the only non-zero component of
the $3$-cocycle is
\begin{equation}
  \nu(X, X, X) = \frac{\Theta}{2}.
\end{equation}
Thus, one recovers both elements of $H^3(Z_2, U(1)) \cong Z_2$ for $\Theta$ an
odd or even multiple of $2\pi$ respectively.

\section{Lattice models of SPT phases}
\label{sec_discrete}
In Ref.~\onlinecite{spto_higher,*spto_higher_prb}, the classification of SPT phases in $d$ spatial dimensions was
based on an explicit construction of a field theory for a $d+1$-dimensional
discrete spacetime for each element of the cohomology group $H^{d+1}(G, \U(1))$.
Although a discrete spacetime is perhaps hard to interpret physically, the
construction of Ref.~\onlinecite{spto_higher,*spto_higher_prb} can also be used to derive a ground-state
wavefunction on a spatial lattice; a gapped Hamiltonian with this wavefunction
as its ground state constitutes an (albeit
unrealistic) lattice Hamiltonian realizing the SPT phase. Hence, it is worthwhile to show
that the symmetry on the edge of such of a lattice model is indeed classified
under our scheme by the same element of the cohomology group that was
used to construct the wavefunction. We do this in Appendix \ref{sec_app_discrete}. In
particular, this shows that every element of the cohomology group $H^{d+1}(G,
\U(1))$ can be realized in an explicit lattice model.

\section{Beyond the cohomological classification}
\label{sec_beyond}
It is now well established
\cite{vishwanath_senthil,beyond_cohomology_walker_wang,window,kapustin_cobordism,thorngren_cobordism}
that in (3+1)-D there exists an SPT phase with
respect to time-reversal symmetry that is beyond the standard
cohomological classification. The reason why this phase is outside the
cohomological classification can be readily understood, as follows. Deriving the cohomological classification using
arguments such as those presented in this paper requires at the very least the
assumption that the symmetry can be implemented locally on a standalone
realization of the edge. We will now argue that the beyond-cohomology phase
violates this assumption.

Indeed, one possible surface termination for the beyond-cohomology phase is a gapped
``three-fermion'' topological phase $\mathcal{F}$ in which all three non-trivial particle
sectors are fermions. Any purely (2+1)-D realization of this phase is
necessarily chiral; that is, its conjugate $\overline{\mathcal{F}}$ under time
reversal cannot be connected to $\mathcal{F}$ without a phase transition. (One way to
see this is to note that $\mathcal{F}$ and $\overline{\mathcal{F}}$ have opposite edge chiral central charges
$c_{-} = \pm 4$ and hence a spatial boundary between them must be gapless. If we
make the spatial variation from $\mathcal{F}$ to
$\overline{\mathcal{F}}$ sufficiently slow, this gapless spatial boundary must be interpreted as a
bulk phase transition \cite{Chen2013}.) Suppose that a state $\ket{\Psi}$ within the
phase $\mathcal{F}$ could be invariant under a local anti-unitary operation $T$.
Then one can always write $T = U \mathbb{T}$, where $\mathbb{T}$ is the normal
on-site representation of time-reversal, and $U$ is a local unitary. But then,
since $\mathbb{T} \ket{\Psi}$ is in the conjugate phase
$\overline{\mathcal{F}}$, we see that $U$ is a local unitary connecting
$\mathcal{F}$ and $\overline{\mathcal{F}}$, which is a contradiction.

\section{Fermionic systems}
\label{sec_fermions}
The restriction arguments given in Section \ref{sec_general} are quite general and
therefore can be equally well applied to fermionic systems, at least in (2+1)-D.
(Generalizing to higher dimensions would require one to find an appropriate
fermionic equivalent of the special form of the symmetry considered in Section
\ref{sec_higher_dims}.) Here we will discuss in general terms the issues arising
which result in the fermionic classification differing from the bosonic one,
with reference to a particular example of a Fermion SPT protected by a $Z_2$ symmetry.
As the general classification is somewhat complicated, we
we leave the details to Appendix \ref{appendix_fermionic}. It would be
interesting to see whether it can be related to the ``supercohomology''
classification proposed in Ref.~\onlinecite{supercohomology}. We will consider
only cases in which the symmetry is unitary and on-site; Thus, our
classification will not include the well-known cases of topological insulators
and superconductors\cite{hasan_kane_review,moore_birth_2010,qi_zhang_review}, which are protected by non-unitary symmetries.

The first issue that needs to be considered is the privileged role of fermion parity. Any local fermionic system must be invariant under the fermion parity $(-1)^F$,
where $F$ is the total fermion number. Therefore, the fermionic symmetry
group $G_f$ characterizing a fermion SPT always contains fermion parity. This
must commute with all the other elements of $G_f$ if they describe local
symmetries. If we now consider the (1+1)-D edge of a (2+1)-D SPT, by assumption
it is realizable as a strictly (1+1)-D local fermion system. As this (1+1)-D system must
always be invariant under the fermion parity of the edge, we expect that, in the
realization of $G_f$ on the (1+1)-D edge, the parity element is represented as
the actual fermion parity of the edge. (This can be verified by using the
techniques of Appendix \ref{sec_edge_construction} to construct the edge representation.) That is, by
contrast to the bosonic case, the fermionic symmetry group contains an element
that is \emph{always realized on-site on the boundary}. Furthermore, even when we restrict and
consider the action of the symmetry on a finite interval, \emph{the restricted
operations must be local, and therefore must still commute with the fermion
parity} (whereas there is no analogous requirement in the bosonic case.)

The other main difference from the fermionic case occurs when defining the
restriction of the obstruction operator $\Omega(g_1, g_2)$, which acts on a pair
of points $a$ and $b$, to a single point $a$.
At this point, one encounters a subtlety that was glossed over in the bosonic
treatment.  $\Omega(g_1, g_2)$ is clearly local in the sense (``locality
preserving'') that it
maps local operators (including fermion creation and
annihilation operators) to local operators under Heisenberg evolution. (We can
deduce this from the fact that it is true for the $U_M(g)$'s and that the
locality preserving property is invariant under multiplication.)
This does not necessarily imply \cite{gross_index,hastings_torus_trick} that it is a local unitary in
the sense (``locally generated'') that it can be written as the time evolution of a local fermionic
Hamiltonian in a domain
containing only the two points $a$ and $b$. In other words, we might not be able
to write $\Omega(g_1, g_2) = \Omega_a(g_1, g_2) \Omega_b(g_1, g_2)$, where
$\Omega_a$ and $\Omega_b$ are fermionic local unitaries acting only near the
points $a$ and $b$. For example, the following unitary is locality preserving
but not locally generated:
\begin{equation}
  \label{paradise_lost}
  \Omega = (c_a + c_a^{\dagger}) (c_b + c_b^{\dagger}),
\end{equation}
where $c_{a,b}$ are the annihilation operators for fermions at points $a$ and
$b$ respectively. If $\Omega$ is not locally generated, this presents an
obstacle for defining the restriction $\Omega \to \Omega_a$. This problem was
not present in the bosonic case because for bosonic systems locality preserving
actually does imply locally generated for unitaries acting on a pair of points (for
unitaries acting on higher-dimensional regions this is no longer the case
\cite{gross_index,hastings_torus_trick}.)
Nevertheless, it is clear that, for the $\Omega$ given by \eqnref{paradise_lost} there is still a
natural definition of ``restriction'' $\Omega_a = (c_a + c_a^{\dagger})$, even
though $\Omega_a$ is not really local. More generally, it is always true that
$\Omega(g_1, g_2) = \Omega_a(g_1, g_2) \Omega_b(g_1, g_2)$, where $\Omega_a(g_1,
g_2)$ is
either a fermionic local unitary acting near the point $a$, or it is such
a local unitary multiplied by $c_a + c_a^{\dagger}$ (and similarly for
$\Omega_b$). In the latter case, however, the \emph{restricted operations
$\Omega_a(g_1, g_2)$ and $\Omega_b(g_1, g_2)$ will anti-commute rather than commute.} This
anti-commutation leads to fermionic corrections to the 3-cocycle condition
[\eqnref{three_cocycle_condition}], to the the equivalence relation
[\eqnref{three_coboundary}], and to the product rule for ``stacking'' SPT phases; see Appendix \ref{appendix_fermionic} for more details.

\subsection{Example: Fermionic SPT with $Z_2$ symmetry}
In order to illustrate the ideas discussed above, let us consider a (1+1)-D
field theory which we expect to describe the edge of a
(2+1)-D fermionic SPT protected by a $Z_2$ symmetry. (This $Z_2$ is in addition
to the always-present fermion parity; thus the full fermionic symmetry group is
$G_f = Z_2 \times Z_2^f$.) This theory is the fermionic analogue of
the bosonic edge we considered in Section \ref{sec_chiral_example}. The low-energy physics is 
described by a gapless Dirac point (which can emerge, for example, from a microscopic
 lattice model of non-interacting electrons with a Fermi surface.) Thus, we define
the fermionic fields $\Psi_R(x)$ and $\Psi_L(x)$ corresponding to left- and
right-moving fermions (in terms of the original lattice operators, these will be
local on a length scale set by the energy cutoff.) We can define the
corresponding number operators $N_{L,R} = \int \Psi_{L,R}^{\dagger} \Psi_{L,R} dx$. The Hamiltonian is
\begin{equation}
  H = J (N_L + N_R).
\end{equation}
The fermion parity is $(-1)^{N_R + N_L}$ and sends $\Psi_L \to -\Psi_L, \Psi_R
\to -\Psi_R$. We assume that the additional $Z_2$ symmetry is given by $U = (-1)^{N_R}$;
thus, it acts only on the right-movers and sends $\Psi_L \to \Psi_L, \Psi_R \to
-\Psi_R$. This forbids perturbations like $\Psi_L^{\dagger} \Psi_R$ which would open up
a gap, suggesting that the gapless edge is protected by the symmetry. Indeed,
we will show that the symmetry corresponds to a non-trivial fermionic cocycle.

We can define the restriction of the $Z_2$ symmetry to a finite interval $[a,b]$
according to
\begin{equation}
  U_{[a,b]} = \exp\left(-i \pi \int_a ^b \Psi_R^{\dagger} \Psi_R dx\right).
  \label{cslewis}
\end{equation}
If we invoke the bosonization correspondences $\Psi_R^{\dagger} \Psi_R \sim
\partial_x \phi_R(x)/(2\pi)$, $\Psi_R(x) \sim e^{i \phi_R}$, we see that $\Omega
\equiv U_{ [a,b]
}^2 \sim \Psi_R(a) \Psi_R^{\dagger}(b) \equiv \Omega_a \Omega_b$. Thus,
$U_{ [a,b] }^2$ acts only
on the endpoints as expected. 
However, this is an example of the possibility
discussed above, of the operators $\Omega_a$ carrying non-trivial fermion
parity.

The parity of $\Omega_a$, which we call $\sigma$ ($\sigma = -1$ in the current
calculation) constitutes one aspect of the non-trivial fermionic
cocycle. The other aspect comes from the relation $U_{[a,b]} \Omega
U_{[a,b]}^{\dagger} = \Omega$. The restricted operations $\Omega_a$ satisfy this
relation only up to a phase factor $\omega$. To calculate this phase factor we need to
regularize the integral \eqnref{cslewis} by introducing a soft cutoff; that is,
we replace \eqnref{cslewis} by
\begin{equation}
  U_{[a,b]} = \exp\left(-i\pi \int_{-\infty}^{\infty} f(x) \Psi_R^{\dagger}
  \Psi_R dx
  \right),
\end{equation}
where $f$ is a smooth function such that $f(x) = 1$ for $x \in
[a+\epsilon,b-\epsilon]$, and $f(x) = 0$ for $x < (a - \epsilon)$ or $x > (b +
\epsilon)$. Using the bosonization correspondence to express $U_{[a,b]}^2$ in
terms of $\partial_x \phi_R$ and integrating by parts gives
\begin{equation}
\Omega_a \sim \exp\left(i \int_{a-\epsilon}^{a+\epsilon} f'(x) \phi_R dx\right).
\end{equation}
Using the fact that $\phi_R \to \phi_R + f(x) \pi$ under $U_{[a,b]}$ gives
\begin{align}
  \omega \equiv U_{[a,b]} \Omega_a U_{[a,b]}^{\dagger} \Omega_a^{\dagger} 
  &= \exp\left(i\pi \int_{a-\epsilon}^{a + \epsilon}
  f'(x) f(x) dx\right)  \\
  &= \exp\left(\frac{i\pi}{2} \int_{a-\epsilon}^{a+\epsilon} \frac{d}{dx} [f(x)]^2
  \right) \\
  &= i.
\end{align}

The numbers $(\omega,\sigma) = (i,-1)$ constitute the fermionic 3-cocycle for the $Z_2$
symmetry. We see that taking four copies of the same edge leads to a trivial
fermionic 3-cocycle (in agreement with the results of
Ref.~\onlinecite{lu_vishwanath} showing that four
copies of the theory under consideration can be gapped out without breaking the
symmetry.) Furthermore, if one applies the fermionic 3-cocycle condition
[see Appendix \ref{appendix_fermionic}] one sees that the only allowable values of the fermionic
3-cocycle are the ones obtained by taking copies in this way, namely $(1,1),
(i,-1), (-1,1)$ and $(-i,-1)$. Thus, we have recovered all the elements of a $\mathbb{Z}_4$ classification
for fermionic SPT's with $G_f = Z_2 \times Z_2^f$ (which is the same result
obtained from supercohomology\cite{supercohomology}). By contrast,
Refs.~\onlinecite{ryu_zhang_fermions,qi_fermions,yao_fermions,gu_levin_fermions}
obtained a $\mathbb{Z}_8$ classification for the same $G_f$. The odd-numbered
phases in this classification have an odd number of gapless Majorana modes at
the edge, each
of which is ``half'' of the gapless Dirac mode considered here. The explanation for the
discrepancy in the classification is that the
symmetry in these odd-numbered phases does not act locally on the edge, and hence
they are not captured by our approach.

\section{Conclusions}

Suppose we have a system whose bulk ground state is invariant
under a group $G$ of symmetries that commute with the Hamiltonian.
Let us further suppose that there is an energy gap to all bulk excitations and a concomitant finite correlation length
and that we can solve the Hamiltonian (with a sufficiently powerful computer, for instance)
for systems much larger than the correlation length. Armed with this information,
we wish to determine if the system is in a symmetry-protected topological phase and, if so, which one.
In a 1D system on a finite interval, we can identify an SPT phase by the presence
of gapless excitations at the ends of the system that transform under a
projective representation of the symmetry (or, alternatively, the presence of such states in
the bipartite entanglement spectrum) \cite{pollmann-arxiv-2009,chen_gu_wen,schuch,fidkowski_fermions}.
But how do we identify an SPT phase in higher dimensions?
One approach is to gauge the symmetry $G$.\cite{Etingof10,levin_gu,Chen14,wang_levin}
In 2D, the resulting theory has anyonic excitations in the bulk \cite{levin_gu}.
By determining the statistics of these excitations,
one can deduce the SPT phase of the ungauged system. In 3D, the gauged theory has anyonic excitations
on its surface \cite{Chen14} and extended excitations (e.g. vortex lines) in its bulk \cite{wang_levin}.
The topological properties of surface and bulk excitations of the gauged model
can be used to identify the underlying SPT phase.
But this approach involves modifying the system drastically, and it cannot be used if
all that we are given are the low-energy eigenstates of the original Hamiltonian. Moreover, it may be more difficult,
as a practical matter, to solve the gauged model and deduce its quasiparticles' topological
properties than it is to solve the original Hamiltonian.

Here, we take a different approach, which identifies an SPT directly from the realization
of the symmetry group $G$ on its boundary states. We consider $d$-dimensional SPTs
for which the restriction of $G$ to the low-energy Hilbert space has a local
action on the $(d-1)$-dimensional boundary of the system.
In such a phase, there may be an obstruction to restricting the action of the
symmetry to a $(d-1)$-dimensional proper submanifold of the boundary. To analyze such
an obstruction, we construct a new functional of two group elements by taking
a suitably defined coboundary of the restriction. This localizes the obstruction to the $(d-2)$-dimensional boundary of
the $(d-1)$-dimensional proper submanifold of the boundary. We then continue in the same fashion, either
restricting a functional of $k$ group elements on a closed $(d-k)$-dimensional manifold
to a $(d-k)$-dimensional submanifold with boundary or
constructing the coboundary of a functional of
$k$ group elements on a $(d-k)$-dimensional submanifold with boundary, thereby obtaining
a functional of $k+1$ group elements on a $(d-k-1)$-dimensional closed manifold.
These functionals are operators that act on the local Hilbert spaces of the corresponding
submanifolds.
The resulting sequence of maps between functionals terminates after we reach functionals
of $d$ group elements acting on a single point; the coboundary of such a functional must be an ordinary phase.
Equivalence classes of such sequences are classified by the
cohomology group $H^{d+1}(G,U(1))$ in $d=1,2$ and, with an additional assumption, in $d\geq 3$.
Consequently, given the low-energy states of the boundary (or large eigenvalue eigenstates
of the reduced density matrix for a bipartition of a system without a real boundary),
we can, in principle, determine the corresponding element of $H^{d+1}(G,U(1))$.
The Hamiltonian need not take any special form -- in fact, it is not even necessary to know
the Hamiltonian. As we have shown, this procedure gives the expected results when applied to
discrete \cite{spto_higher,*spto_higher_prb} and continuous \cite{boson_iqhe,cenke_neel_order,vishwanath_senthil,adr_classification}
non-linear sigma models.

The obstructions classified by these arguments prevent a model from being continuously deformed into a model
in which the symmetry is realized on the boundary in an on-site manner. (By assumption, the symmetry
can be realized in an on-site manner in the full bulk theory -- by grouping multiple sites into a single site, for instance.)
As a result of the incorrigibly non-on-site nature of the symmetry, if we try to gauge it, the resulting gauge
theory will be
anomalous\cite{Ryu12,Wen13,for_dummies,kapustin_cobordism,kapustin_anomalies,kapustin_anomalies_2}.
Only the action of the symmetry on the whole system, bulk and edge together,
can be gauged in an anomaly-free fashion. A simple example is a $2+1$-dimensional
$U(1)$ SPT, which is very similar to the $\mathbb{Z}_2$ case discussed in Section \ref{sec_chiral_example}.
Such a state is a bosonic integer quantum Hall state \cite{boson_iqhe}. If the theory is gauged,
the edge effective Lagrangian takes the form ${\cal L}_{\rm edge} = \frac{g}{2\pi}(\partial_\mu \varphi - n A_\mu)^2
+ \frac{m}{2\pi} A_\mu \epsilon_{\mu\nu} \partial_\nu \varphi$.
Charge is no longer conserved at the edge since an electric field along the edge will cause charge to flow
from the bulk to the edge. Following Laughlin \cite{Laughlin81},
we can understand this in an annular geometry. By adiabatically
increasing the flux through the center of the annulus by $2\pi$, the charge at the outer edge is
increased by $2nm$, the integer (necessarily even in a bosonic SPT)
that characterizes the Hall conductance. The $3$-cocycle obtained by our construction reflects this
charge pumped to the edge, as may be seen by noting that
a $U(1)$ transformation applied to a finite interval along the edge is equivalent to applying equal and opposite
gauge fields at the ends of the interval.\footnote{By restricting the rotation to an interval, we cannot introduce a net
twist, but we can separate equal and opposite twists and focus on the vicinity of just one of them.}
Since they are equal and opposite, such gauge fields cannot increase
the total charge on the edge, but if we focus on the charge to the left of an arbitrary point in the middle of the interval,
then this increases by $2nm$ when the gauge field winds by $2\pi$. The restriction $\Omega_a$ 
defined in Eq. \ref{first_coboundary_nonabelian} measures such a
charge\footnote{
  The precise relation is slightly subtle, however, because $\Omega_a$ measures
  the charge to the left of the endpoint $a$, the very place at which
  charge is accumulating due to the Hall effect. The precise manner in which $\Omega_a$ takes into
account the charge located at the endpoint itself depends on the restriction $U \to
U_M$. However, as long as we use the same restriction both to define $\Omega_a$
and to implement the winding, the value of the measured ``charge'' is independent of
the choice of restriction. Perhaps the easiest case to interpret is where $U_M$ 
is defined so that the winding of the phase interpolates linearly over a transition region
between no winding outside the interval
and the desired winding in the interior of the interval. Then $\Omega_a$ measures the charge to the left
of a point $x$, averaged over all $x$ in the transition region. This is $nm$, rather than the charge $2nm$
measured to the left of a point in the interior of the interval.
}.
Meanwhile, $U_M$ applies gauge fields at the ends of the interval.
Then, according to the definition (\ref{hacker}), the cocycle measures accumulated charge
in response to this change in gauge field.
We expect that similar reasoning can relate our constructions to anomalies in higher dimensions and for
discrete symmetries.

In this paper, we have confined our attention to ``internal'' symmetries. It would be interesting
to extend them to space group symmetries. States of free fermions protected by
inversion symmetry \cite{Turner09,Hughes11};
time-reversal symmetry combined with a point group symmetry \cite{Fu11}; or a rotational symmetry alone
have been classified \cite{Jadaun13}. With the methods described here, it might be possible
to extend these ideas to interacting fermion systems and to bosonic systems in which
a space group symmetry, projected to the low-energy
boundary theory, maps sites to sites and then has an additional ``internal'' action that is non-on-site.
However, care must be taken to consider a boundary that respects
the space group symmetry and to consider a sequence of submanifolds (which are, presumably, not connected
manifolds) that also respect the symmetry.

We have given a prescription that, in principle, allows one to identify an SPT phase, given
its ground state wavefunction, and we have shown how to apply it to some long-wavelength effective
field theories and exactly soluble lattice models. But how useful can this prescription be in practice,
given an arbitrary -- perhaps experimentally-motivated model? This remains to be
seen. However, ground state wavefunctions with tensor network descriptions are
natural candidates for the reduction procedure\cite{spto_2d}.
A numerical implementation would open an important
avenue for future research.

Conversely, we have shown in Section \ref{sec_discrete}
and Appendix \ref{sec_app_discrete} that each SPT phase in the cohomological classification has
a representative wavefunction which is the ground state of some lattice Hamiltonian. However, these Hamiltonians are certainly
not expected to describe any experimentally realizable systems; finding more realistic
Hamiltonians giving rise to SPT phases is an
important open problem.

As noted above, our construction leads to $H^{d+1}(G,U(1))$ in $d\geq 3$ provided we make
an additional assumption: there exists a local basis for the Hilbert space of the $(d-1)$-dimensional boundary
in which the symmetry acts on the boundary in an on-site manner, except for a diagonal part
which cannot be made on-site.
This assumption holds in a system that is described by a $d$-dimensional non-linear sigma
model with $\theta$-term at long wavelengths
\cite{boson_iqhe,cenke_neel_order,vishwanath_senthil,adr_classification}
since the symmetry acts in an on-site manner
on all gradient energy terms in
the $(d-1)$-dimensional boundary effective action and non-on-site only on the Wess-Zumino term,
which only enters the phase of the ground state wavefunction.
However, it remains an interesting open question whether there
are SPT phases in three dimensions that violate this assumption and, consequently,
realize the aforementioned non-trivial sequence but in a
manner that is not classified by group cohomology. Such an exception to a cohomological classification,
if it exists, would be distinct from the so-called ``beyond cohomology'' SPT phases
\cite{vishwanath_senthil}, which occur due to the violation of
a different assumption -- that the symmetry is realized locally (but not necessarily on-site)
at the boundary of the system. In ``beyond cohomology'' SPT phases,
the symmetry is realized in an inherently non-local manner at the boundary of the system.
Our methods do not enable us to classify such phases;
once the condition of locality is relaxed, a very different approach may be necessary.

This comment also applies to the most famous SPT phase, the 3D time-reversal-invariant
topological insulator \cite{Moore07,Fu07,Roy09},
where time-reversal acts in an inherently non-local manner at a 2D surface. However, there
are fermionic SPT phases in which the symmetry is realized locally on the boundary, and these can be classified
along the lines discussed in Section \ref{sec_fermions}. Carrying out this classification to completion and
relating it to the notion of ``supercohomology''\cite{supercohomology} is an
important goal for future work. We remark that dimensional reductions
have previously been employed in the classification of fermionic
SPT phases \cite{topfieldtheory_insulators,tenfold_way}, and it would be interesting if a connection could be drawn with the
reduction procedure described here.

Finally, we note that symmmetry-enriched topological (SET) phases \cite{Etingof10,Mesaros13,Hung13,Lu13,Xu13,Essin13}
generalize SPT phases to systems with topological order.
In SET phases, symmetry realization interacts non-trivially with the fusion and braiding properties of anyons,
as already occurs in topological phases at the (2+1)-D boundary of a (3+1)-D SPT. 
The possible symmetry fractionalization patterns in (2+1)-D correspond to different
projective representations of the anyons and are classified by ${H^2}(G,A)$, where $A$ is
the group of Abelian anyons. It is possible that an extension of our methods can
also be applied to the analysis of symmetry fractionalization in (3+1)-D SET
phases which have topological excitations occupying closed loops.

\begin{acknowledgments}
We thank M.~Cheng, C.~Xu, A.D.~Rasmussen, and B.~Ware for helpful discussions.
D.~Else was supported by a gift from the Microsoft Corporation.
\end{acknowledgments}

\appendix

\section{Explicit construction of the edge theory}
\label{sec_edge_construction}
In this Appendix we will give an explicit proof of the property that the edge theory
of an SPT in $d$ spatial dimensions can always be realized in a strictly
$(d-1)$-dimensional system and show given certain assumptions how to construct
the representation of the symmetry in this realization.

By definition, the
ground state of an SPT phase is gapped and not topologically ordered, which means it
can be connected to a product state by a local unitary. Indeed,
let $\mathcal{D}$ be the local unitary which turns the bulk ground state
$\ket{\Psi_{\mathrm{gr}}}$ on a boundaryless spatial region product state $\ket{\phi}^{\otimes N}$, and
let $\widetilde{\mathcal{D}}$ be the restriction of $\mathcal{D}$ to a spatial region with
boundary.
Any low-energy state $\ket{\Psi}$ associated with the boundary must be identical to
$\ket{\Psi_{\mathrm{gr}}}$ far from the boundary. It follows that
$\widetilde{\mathcal{D}} \ket{\Psi}$ must be identical to $\ket{\phi}^{\otimes N}$ far from the boundary.
Hence, $\widetilde{\mathcal{D}} \ket{\Psi}$ is simply a direct product with
copies of $\ket{\phi}$ in the bulk of some state
$\ket{\Psi}_B$ defined on a strip $B$ near the boundary:
\begin{equation}
  \label{decoupling}
  \widetilde{\mathcal{D}} \ket{\Psi} = \ket{\Psi}_{B} \otimes \ket{\Phi}_{B^c}
\end{equation}
where $B^c$ is the complement of $B$ and $\ket{\Phi}_{B^c}$ is a product state
of $\ket{\phi}$ on every site in $B^c$.
Thus, the states $\{\ket{\Psi}_{B} : \ket{\Psi} \mbox{ a low-energy boundary
state } \}$ constitute a $(d-1)$-dimensional realization of the boundary theory.
One can also apply the mapping $\widetilde{\mathcal{D}}$ to the original Hamiltonian for the
system with boundary in order to obtain a Hamiltonian for this realization of
the boundary theory.

Now suppose that the bulk ground state is invariant under an on-site
representation $U(g)$ of the symmetry. As the local unitary $\mathcal{D}$ is not required to have any particular
properties with respect to the symmetry, in general it might not be easy to
determine how the symmetry acts on the boundary theory. However, the task
becomes easier if we make the following simplifying assumption: we assume
that $\mathcal{D}$ can be chosen to
commute with $U(g)$ in the absence of boundary. (We emphasize that this does
\emph{not} necessarily imply that we are considering a trivial SPT phase. That would be only
be true if we made the stronger assumption that $\mathcal{D}$ can be
continuously connected to the identity by a path which everywhere commutes with
the symmetry.).  In fact, this assumption is always true in any SPT phase described by the
cohomological classification, because, in particular, it is true for the ground
states constructed via the discrete topological term construction of
Ref.~\onlinecite{spto_higher,*spto_higher_prb} (see Section \ref{sec_discrete} above.) 
This
implies that it is also true for any other ground state in the same SPT phase,
since, by definition, any two ground states in the same SPT phase can be related
by a symmetry-respecting local unitary.

Given this assumption, one can explicitly construct the realization of the
symmetry on the edge, as follows. That $U(g)$ and $\mathcal{D}$ commute in the
absence of boundary implies that their restrictions $\widetilde{U}(g)$ and
$\widetilde{\mathcal{D}}$ to a region with boundary must commute up to boundary terms.
Thus, $\widetilde{\mathcal{D}} \widetilde{U}(g) \widetilde{\mathcal{D}}^{\dagger} =
W_B(g) U_{B^c}(g)$ where $W_B(g)$ acts only in the
strip $B$, and $U_{B^c}(g)$ is simply the restriction of $U(g)$ into the
complement $B^c$. Hence, we find that
\begin{align}
  \widetilde{\mathcal{D}} \{ \widetilde{U}(g) \ket{\Psi} \} &= 
  \widetilde{\mathcal{D}} \widetilde{U}(g) \widetilde{\mathcal{D}}^{\dagger}
  \widetilde{\mathcal{D}} \ket{\Psi} \\
  &= W_{B}(g)
  U_{B^c}(g) (\ket{\Psi}_{B} \otimes \ket{\Phi}_{B^c}) \\
  &= 
  \{ W_{B}(g)
  \ket{\Psi}_{B}\} \otimes \ket{\Phi}_{B^c}.
  \label{symact}
\end{align}
To get to the last line, we used the fact that $U_{B^c}(g) \ket{\Phi}_{B^c} =
\ket{\Phi}_{B^c}$. 
This follows from the fact that, \emph{without}
boundary, $\ket{\phi}^{\otimes N}$ is invariant under $U(g)$, since it is
obtained from $\ket{\Psi_{\mathrm{gr}}}$ [which is certainly invariant under
$U(g)$] by $\mathcal{D}$ which by assumption commutes with $U(g)$.
Comparing \eqnref{symact} with \eqnref{decoupling},
we see that $W_{B}(g)$
is the representation of the symmetry on the stand-alone realization of the
boundary.

\section{The (2+1)-D reduction procedure}
\label{sec_crossed_module}
Here we will prove the two key properties of $\omega(g_1, g_2, g_3)$ defined by
\eqnref{hacker} in Section \ref{sec_2p1d}; firstly, that it must be a 3-cocycle, and secondly,
that up to equivalence it is independent of the choice of restriction $U(g) \to
U_M(g)$.

We first make a general remark: the structure described in Section
\ref{sec_2p1d} is known in the mathematics
literature as a \emph{crossed module extension}.
Recall that a projective representation of a group $G$ corresponds
to a \emph{central extension}, which is an exact sequence
\begin{equation}
  1 \to \U(1) \to H \to G \to 1
\end{equation}
such that the image of $\U(1)$ is in the center of $H$. Similarly, a crossed
module extension is an exact sequence
\begin{equation}
  1 \to \U(1) \to K \xrightarrow{\varphi} H \to G \to 1
\end{equation}
along with a left-action of $H$ on $K$, represented by $k \mapsto \, ^h k$, 
such that $^{\varphi(k)} k^{\prime} = k k^{\prime} k^{-1}$ for all $k,
k^{\prime} \in K$. 
It is a well-known theorem in the mathematics literature
\cite{holt_cohomology,huebschmann_cohomology,brown_cohomology,crossed_module_extensions}
that the crossed module
extensions of $G$ by $\U(1)$ are classified by $H^3(G, \U(1))$. The procedure
described in Section \ref{sec_2p1d} for obtaining the 3-cocycle $\omega(g_1, g_2, g_3)$,
as well as the proofs of the properties of $\omega$ given below, are adapted from the proof of
this classification theorem given in
Ref.~\onlinecite{crossed_module_extensions}; however, the reader does not need
to understand the connection to crossed module extensions in order to follow
these proofs.

To prove that $\omega$ is a 3-cocycle, we calculate $\Omega_a(g_1, g_2)
\Omega_a(g_1 g_2, g_3) \Omega_a(g_1 g_2 g_3, g_4)$ in two different ways.
Firstly,
\begin{align}
  & \Omega_a(g_1, g_2) \Omega_a(g_1 g_2, g_3) \Omega_a(g_1 g_2 g_3, g_4) \\
  &= \omega(g_1 g_2, g_3, g_4) \times \Omega_a(g_1, g_2) ^{U_M(g_1 g_2)} \Omega_a(g_3, g_4) \Omega_a(g_1 g_2, g_3 g_4)
   \\
  &= \omega(g_1 g_2, g_3, g_4) \times \, ^{\Omega_a (g_1, g_2) U_M(g_1 g_2)} \Omega_a(g_3, g_4) \Omega_a(g_1, g_2)
   \Omega_a(g_1 g_2, g_3 g_4) \label{asdf1} \\
  &= \omega(g_1 g_2, g_3 g_4) \times \, ^{\Omega (g_1, g_2) U_M(g_1 g_2)} \Omega_a(g_3, g_4) \Omega_a(g_1, g_2)
   \Omega_a(g_1 g_2, g_3 g_4) \label{asdf2} \\
  &= \omega(g_1 g_2, g_3, g_4) \times \, ^{U_M(g_1) U_M(g_2)} \Omega_a(g_3, g_4)
   \Omega_a(g_1, g_2) \Omega_a(g_1 g_2, g_3 g_4)  \label{asdf3} \\
  &= \omega(g_1 g_2, g_3, g_4) \omega(g_1, g_2, g_3 g_4) \times \, ^{U_M(g_1)
U_M(g_2)} \Omega_a(g_3, g_4) ^{U_M(g_1)} \Omega_a(g_2, g_3 g_4) \Omega_a(g_1,
g_2 g_3 g_4) \label{asdf5},
\end{align}
where we applied \eqnref{hacker} twice. To get from \eqnref{asdf2} to \eqnref{asdf3} we used
\eqnref{first_coboundary_nonabelian}. To get from \eqnref{asdf1} to \eqnref{asdf2},
we used the fact that $\Omega(g, g^{\prime})$ can be written as a product of a
contributions near $a$ and contributions near $b$, which commute; it follows
that for any operator $X_a$ localized near $a$, 
\begin{equation}
  \label{xavier}
^{\Omega(g_1, g_2)} X_a = \, ^{\Omega_a(g_1, g_2)} X_a.
\end{equation}
Proceeding in a different way, we also have
\begin{align}
  &  \Omega_a(g_1, g_2) \Omega_a(g_1 g_2, g_3) \Omega_a(g_1 g_2 g_3, g_4) \\
  &= \omega(g_1, g_2, g_3) \times \, ^{U_M(g_1)} \Omega_a(g_2, g_3) \Omega_a(g_1, g_2 g_3) \Omega_a(g_1 g_2 g_3, g_4) \\
  &= \omega(g_1, g_2, g_3) \omega(g_1, g_2 g_3, g_4) \times \, ^{U_M(g_1)}
  \Omega_a(g_2, g_3) ^{U_M(g_1)}  \Omega_a(g_2 g_3, g_4) \Omega_a(g_1, g_2 g_3
  g_4) \\
  &= \omega(g_1, g_2, g_3) \omega(g_1, g_2 g_3, g_4) ^{U_M(g_1)} \{ \Omega_a(g_2, g_3) \Omega_a(g_2 g_3, g_4) \} \Omega_a(g_1,
  g_2 g_3 g_4) \\
  &= \omega(g_1, g_2, g_3) \omega(g_1, g_2 g_3, g_4) \omega(g_2, g_3, g_4)
  \times \, ^{U_M(g_1)} \{ ^{U_M(g_2)} \Omega_a(g_3, g_4) \Omega_a(g_2, g_3 g_4)
  \} \Omega_a(g_1, g_2 g_3 g_4) \\
&= \omega(g_1, g_2, g_3) \omega(g_1, g_2 g_3, g_4) \omega(g_2, g_3, g_4) \times
  \, ^{U_M(g_1) U_M(g_2)} \Omega_a(g_3, g_4) ^{U_M(g_1)} \Omega_a(g_2, g_3 g_4)
  \Omega_a(g_1, g_2 g_3 g_4) \label{asdf4} 
\end{align}
Comparing \eqnref{asdf4} with \eqnref{asdf5} we see that $\omega$ must obey the
3-cocycle condition
\begin{equation}
  \omega(g_1 g_2, g_3, g_4) \omega(g_1, g_2, g_3 g_4) = \omega(g_1, g_2, g_3)
  \omega(g_1, g_2 g_3, g_4) \omega(g_2, g_3, g_4).
\end{equation}

Next we prove independence from the choice of restriction $U(g) \to U_M(g)$. Indeed,
consider two restrictions $U_M(g)$ and $\widetilde{U}_M(g) = \Sigma(g)
U_M(g)$.
where $\Sigma(g)$ is a local unitary acting near $\partial M = \{ a, b \}$.
Then we find that
\begin{equation}
  \widetilde{U}_M(g) \widetilde{U}_M(g^{\prime}) =
\widetilde{\Omega}(g,g^{\prime}) \widetilde{U}_M(g g^{\prime}),
\end{equation}
where
\begin{equation}
\widetilde{\Omega}(g,g^{\prime}) = \Sigma(g) ^{U_M(g)} \Sigma(g^{\prime})
\Omega(g,g^{\prime}) \Sigma(gg^{\prime})^{-1}.
\end{equation}
It is obvious that the equivalence class of the 3-cocycle is independent of
the choice of restriction $\Omega \to \Omega_a$, so we are free to choose a
restriction of $\widetilde{\Omega}$ such that
\begin{equation}
  \label{bacardia}
\widetilde{\Omega}_a(g,g^{\prime}) = \Sigma_a(g) ^{U_M(g)} \Sigma_a(g^{\prime})
\Omega_a(g,g^{\prime}) \Sigma_a(gg^{\prime})^{-1},
\end{equation}
where $\Sigma_a(g)$ is the restriction of $\Sigma(g)$ to the point $a$.
Now we calculate
\begin{align}
  & \widetilde{\Omega}_a(g_1, g_2) \widetilde{\Omega}_a(g_1 g_2, g_3) \Sigma_a(g_1 g_2
g_3) \\
&= \widetilde{\Omega}_a(g_1, g_2) \Sigma_a(g_1 g_2) ^{U_M(g_1 g_2)} \Sigma_a(g_3)
\Omega(g_1 g_2, g_3) \\
&= \Sigma_a(g_1) ^{U_M(g_1)} \Sigma_a(g_2) \Omega_a(g_1, g_2) ^{U_M(g_1 g_2)}
\Sigma_a(g_3) \Omega_a(g_1, g_2)
\Omega_a(g_1 g_2, g_3) \\
&= \Sigma_a(g_1) ^{U_M(g_1)} \Sigma_a(g_2) ^{\Omega_a(g_1, g_2) U_M(g_1 g_2)}
\Sigma_a(g_3) \Omega_a(g_1, g_2) \Omega_a(g_1 g_2, g_3) \label{foobar1} \\
&= \Sigma_a(g_1) ^{U_M(g_1)} \Sigma_a(g_2) ^{\Omega(g_1, g_2) U_M(g_1 g_2)}
\Sigma_a(g_3) \label{foobar2} 
\Omega_a(g_1, g_2) \Omega_a(g_1 g_2, g_3) \\
&= \Sigma_a(g_1) ^{U_M(g_1)} \Sigma_a(g_2) ^{U_M(g_1) U_M(g_2)} \Sigma_a(g_3)
\Omega_a(g_1, g_2) \Omega_a(g_1 g_2, g_3), \label{humphrey}
\end{align}
where we applied \eqnref{bacardia} several times, and we also used, again,
\eqnref{xavier} to go from \eqnref{foobar1} to \eqnref{foobar2}. The final line
used the definition of $\omega$, \eqnref{hacker}.
On the other hand:
\begin{align}
  & ^{\widetilde{U}_M(g_1)}\widetilde{\Omega}_a(g_2, g_3)
  \widetilde{\Omega}_a(g_1, g_2 g_3) \Sigma_a(g_1 g_2
  g_3) \\
  &= \Sigma_a(g_1) ^{U_M(g_1)}\widetilde{\Omega}_a(g_2, g_3) \Sigma_a(g_1)^{-1}
  \widetilde{\Omega}_a(g_1, g_2 g_3) \Sigma(g_1 g_2 g_3) \\
  &= \Sigma_a(g_1) ^{U_M(g_1)}\widetilde{\Omega}_a(g_2, g_3) ^{U_M(g_1)}
  \Sigma_a(g_2 g_3)\Omega_a(g_1, g_2 g_3)\\
  &= \Sigma_a(g_1) \; ^{U_M(g_1)}\left\{\widetilde{\Omega}_a(g_2, g_3)
\Sigma_a(g_2 g_3) \right \} \Omega_a(g_1, g_2 g_3)  \\
  &= \Sigma_a(g_1) \; ^{U_M(g_1)}\left\{\Sigma_a(g_2) ^{U_M(g_2)} \Sigma_a(g_3)
  \Omega_a(g_2, g_3) \right \} \Omega_a(g_1, g_2 g_3)  \\
  &= \Sigma_a(g_1) \; ^{U_M(g_1)} \Sigma_a(g_2) ^{U_M(g_1) U_M(g_2)}
  \Sigma_a(g_3) ^{U_M(g_1)} \Omega_a(g_2, g_3) \Omega_a(g_1, g_2 g_3) \\
  &= \omega(g_1, g_2, g_3) \Sigma_a(g_1) \; ^{U_M(g_1)} \Sigma_a(g_2) ^{U_M(g_1) U_M(g_2)}
  \Sigma_a(g_3) \Omega_a(g_1, g_2) \Omega_a(g_1 g_2, g_3)
  \label{hubert}
\end{align}
Comparing Eqs.\ (\ref{hubert}) and (\ref{humphrey}), we find that
\begin{equation}
  ^{\widetilde{U}_M(g_1)}\widetilde{\Omega}_a(g_2, g_3)
  \widetilde{\Omega}_a(g_1, g_2 g_3) = \omega(g_1, g_2, g_3)
  \widetilde{\Omega}_a(g_1, g_2) \widetilde{\Omega}_a(g_1 g_2, g_3).  \label{bernard}
\end{equation}
On the other hand, by definition, $\omega$ also satisfies \eqnref{bernard} 
with $\widetilde{\Omega}_a$ and $\widetilde{U}_M$ replaced by $\Omega_a$ and $U_M$.
Thus, it does not matter whether we use the restriction $U_M$ or
$\widetilde{U}_M$; one obtains the same 3-cocycle $\omega$.

\section{Completing the proof of separation of phases in (2+1)-D}
\label{sec_completing}
In this section, we will fill in the details of the proof outlined in Section
\ref{separation_proof} showing that (2+1)-D SPT phases corresponding to different elements of
the cohomology group $H^{3}(G, \U(1))$ are necessarily separated by a phase
transition. Although we will write the arguments in lemma-proof form, we
emphasize that we do not aim for mathematical rigor in our treatment of
locality; rigorous proofs could potentially be constructed based on the
arguments sketched here, but would require much more careful estimates of how
relevant quantities decay at large distances.

The first result that needed to be proved was
\begin{lemma}
  \label{lem_tolkien}
  Let $\ket{\Psi}$ be the ground state of an SPT phase in $d$ spatial
  dimensions captured by the cohomological classification, and let $U(g)$ be a local unitary representation of a group $G$ which
  leaves $\ket{\Psi}$ invariant. Let $U_A(g)$ be the restriction of $U(g)$ to a
  region $A$. Then there exists a representation $V_{\partial A}(g)$ acting only in a strip
  near the boundary $\partial A$ (and only within the region $A$) such that
  $U_A(g) \ket{\Psi} = V_{\partial A}(g) \ket{\Psi}$.
  \begin{proof}
    We use the fact, discussed in the Appendix \ref{sec_edge_construction} above, that the state
    $\ket{\Psi}$ can be transformed into a product state $\ket{\Phi} =
    \ket{\phi}^{\otimes N}$ through a local unitary $\mathcal{D}$ that commutes with
    $U(g)$. Hence, as in Appendix \ref{sec_edge_construction}, if we define the restriction $\mathcal{D}_A$, it follows that
    $\mathcal{D}_A U_A(g) \mathcal{D}_A^{\dagger} = W_B(g) U_{A \setminus B}(g)$,
    where $W_B(g)$ acts within a strip $B$ near the boundary, and $U_A(g)$ and $U_{A \setminus
    B}(g)$ are the restriction of $U(g)$ to the respective regions (and $A
    \setminus B$ is the region $A$ with the strip $B$ excluded.)
    It is clear (by a similar argument to the one given above in Appendix
    \ref{sec_edge_construction}) that $U_{A \setminus B}(g) \ket{\Phi} = \ket{\Phi}$, and
    hence we find that
    \begin{align}
      U_A(g) \ket{\Psi} &= \mathcal{D}_A^{\dagger} \mathcal{D}_A U_A(g) \mathcal{D}_A^{\dagger} \mathcal{D}_A \ket{\Psi} \\
                        &= \mathcal{D}_A^{\dagger} W_B(g) U_{A \setminus B}(g) \mathcal{D}_A \ket{\Psi} \label{barry_lyndon}, \\
                        &= \mathcal{D}_A^{\dagger} W_B(g) \mathcal{D}_A \ket{\Psi} \label{a_space_odyssey}
    \end{align}
    where in going from \eqnref{barry_lyndon} to \eqnref{a_space_odyssey} we
    used the fact that $\mathcal{D}_A \ket{\Psi}$ looks like $\ket{\Phi}$ on $A
    \setminus B$, and therefore $U_{A \setminus B}(g) \mathcal{D}_A \ket{\Psi} =
    \mathcal{D}_A \ket{\Psi}$.
    Hence, defining $V_{\partial A}(g) = \mathcal{D}_A^{\dagger}
    W_B(g) \mathcal{D}_A$, we have the desired result.
  \end{proof}
\end{lemma}

To proceed further we will also require the following lemma, which states that a
``trivial'' state cannot be invariant under an ``anomalous'' symmetry
representation.
\begin{lemma}
  \label{lem_milton}
  Let $\ket{\Psi}$ be some short-ranged entangled state, and let $U(g)$ be some local unitary representation of a symmetry group $G$ on a
  closed 1-dimensional subregion $C$ of the space on which $\ket{\Psi}$ is
  defined, such that $U(g)
  \ket{\Psi} = \ket{\Psi}$. Then the element of the cohomology group $H^3(G,
  \U(1))$ obtained via the reduction procedure of Section \ref{sec_2p1d} is necessarily
  trivial.
  \begin{proof}
    Consider the restriction $U_M(g)$ to a subregion $M$ of $C$ with boundary.
    Then $U_M(g) \ket{\Psi}$ looks like $\ket{\Psi}$ away from the boundary
    $\partial M$. This implies that, for \emph{short-ranged entangled} states
    $\ket{\Psi}$ it must be the case that $U_M(g) \ket{\Psi} =
    \Sigma_{\partial M}(g) \ket{\Psi}$ for some set of local unitaries
    $\Sigma_{\partial M}(g)$ acting near the boundary \cite{levin_gu}. (To show this, one first
    establishes it for a product state, from which one can show that it must
    also apply for any state that can be turned into a product state by a local
    unitary.) However, the restriction $U(g) \to U_M(g)$ was only defined up to
    unitaries at the boundary anyway, so we are free to set $U_M(g) \ket{\Psi} =
    \ket{\Psi}$. Then, defining $\Omega(g_1, g_2)$ via
    \begin{equation}
      U_M(g_1) U_M(g_2) = \Omega(g_1, g_2) U_M(g_1 g_2)
    \end{equation}
    we can deduce that $\Omega(g_1, g_2) \ket{\Psi} = \ket{\Psi}$. Assuming that $\partial M
    = \{ a, b \}$ where $a$ and $b$ are two points, we can
    choose the restriction $\Omega_a(g_1, g_2)$ such that $\Omega_a(g_1, g_2)
    \ket{\Psi} = \ket{\Psi}$. Then, given the definition [\eqnref{hacker} in
    Section \ref{sec_2p1d}] of the 3-cocycle $\omega(g_1, g_2, g_3)$, one finds that
      $\omega(g_1, g_2, g_3) \ket{\Psi} = \ket{\Psi}$,
    and hence $\omega(g_1, g_2, g_3) = 1$.
  \end{proof}
\end{lemma}

\begin{figure}
\begin{tikzpicture}
  \draw [decoration={markings, mark=at position 0 with {\arrow{triangle 60}}},
  postaction={decorate}] (0,0) circle(1.3);
\draw (0,0) node{$A$};
  \draw [decoration={markings, mark=at position 0 with {\arrow{triangle 60
    reversed}}},
  postaction={decorate}] (0,0) circle(1.8);
\draw (0,1.53) node{$K$};
\draw (0,2.5) node{$A^{\prime}$};
\end{tikzpicture}
\caption{\label{twas_brillig} The regions $A$ and $A^{\prime}$ on which we can
prove that the anomalous symmetry on the boundary is classified by the same
element of $H^3(G, \U(1))$. The orientations of the boundaries $\partial A$ and
$\partial A^{\prime}$ are depicted with arrows.}
\end{figure}
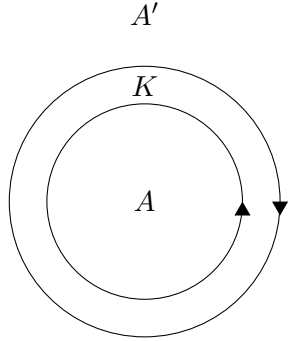
Now, let us consider a (possibly spatially inhomogeneous) ground state $\ket{\Psi}$ in an
SPT phase, invariant under an on-site symmetry representation $U(g)$ of $G$. We choose
two regions $A$ and $A^{\prime}$, separated by a
quasi-one-dimensional buffer region $K$, as depicted in Figure \ref{twas_brillig}. We assume
that the combined region $A \cup K \cup A^{\prime}$ has no boundary. We define
$V_{\partial A}(g)$ and $V_{\partial A^{\prime}}(g)$ according to $U_A(g) \ket{\Psi} =
V_{\partial A} \ket{\Psi}$ as per Lemma \ref{lem_tolkien}.  Denote the corresponding
classes of 3-cocycles as $[\omega], [\omega^{\prime}] \in H^3(G,
\U(1))$. We want to show that $[\omega] = [\omega^{\prime}]$.

First we need to discuss an important subtlety involved in the definitions of
$[\omega]$ and $[\omega^{\prime}]$. Specifically, the reduction procedure of
Section \ref{sec_general} implicitly depends on an orientation for the one-dimensional
space in which the local unitaries act, in order to provide a consistent
convention for reducing from $\partial M = \{ a, b \}$ to a single point $a$.
Opposite orientations give rise to inverse cocycles. We will take the
orientations of $\partial A$ and $\partial A^{\prime}$ to be derived from that of $A$ and
$A^{\prime}$; if we choose $A$ and $A^{\prime}$ to have the same
orientation (e.g. both specified by normal vectors pointing out of the page),
then $\partial A$ and $\partial A^{\prime}$ have \emph{opposite}
orientations, as depicted by the arrows in Figure \ref{twas_brillig}. 

We now observe that $\ket{\Psi}$ is
invariant under $V_{\mathrm{sum}}(g) = U_K(g) V_{\partial A}(g) V_{\partial
A^{\prime}}(g)$. Now, it can readily be verified [using the fact that the three
components of $V_{\mathrm{sum}}(g)$ all commute with each other] that the element of
$H^3(G, \U(1))$ characterizing $V_{\mathrm{sum}}(g)$ is equal
to $[\omega] [\omega^{\prime}]^{-1}$. (The inverse comes from the
fact that we have defined the orientations of $\partial A$ and $\partial
A^{\prime}$ to be opposite.) On
the other hand, by Lemma \ref{lem_milton}, this product must be trivial; hence we have
established that
$[\omega] = [\omega^{\prime}]$.

Finally, let us justify the following claim we made in Section
\ref{separation_proof}:
\begin{lemma}
  Consider two gapped systems $\mathcal{S}$ and $\mathcal{S}^{\prime}$ connected
  without a
  phase transition, and two well-separated regions $A$ and $A^{\prime}$. Then
  one can construct an interpolating system such that the ground state looks
  like that of $\mathcal{S}$ on $A$ and like that of $\mathcal{S}^{\prime}$ on
  $A^{\prime}$.
  \begin{proof}
    Let $\ket{\Psi}$ and $\ket{\Psi^{\prime}}$ be the corresponding ground
    states. Then there must exist a local unitary $\mathcal{D}$ such that
    $\mathcal{D} \ket{\Psi} = \ket{\Psi^{\prime}}$. Define the restriction
    $\mathcal{D}_{\widetilde{A}^{\prime}}$ of $\mathcal{D}$ to some region
    $\widetilde{A^{\prime}}$ that contains $A^{\prime}$ well inside,
  but is also well-separated from $A$. Then applying
  $\mathcal{D}_{\widetilde{A}^{\prime}}$ to $\mathcal{S}$ gives a system with the desired
  properties.
  \end{proof}
\end{lemma}

\section{Proof of Lemma \ref{lemma:central}.}
\label{appendix_universal}
Here we will give a proof of Lemma \ref{lemma:central} which we stated in
Section \ref{nlsm_cochains}.
The proof is based on a result called the universal coefficient theorem
\cite{algebraic_topology}.
Let us first state some definitions.
For a manifold $T$, we define $C^k(T, \U(1))$ to be the group of \emph{closed}
$\U(1)$ $k$-cochains, i.e.\ those cochains $\omega$ for which $d\omega = 0$, and
$B^k(T, \U(1))$ to be the group of \emph{exact} cochains, i.e.\ those which can
be written as $\omega = d\kappa$ for some $\kappa$. We define the
\emph{cohomology group} $H^k(T, \U(1)) \equiv C^k(T, \U(1))/B^k(T, \U(1))$. Similarly,
we define $C_k(T)$  and $M_k(T)$ to be the group of closed (i.e.\ boundaryless)
and exact (can be expressed as a boundary) $k$-chains respectively; and the
homology group $H_k(T) = C_k(T) / M_k(T)$.

We observe that there is a natural
homomorphism $\gamma : H_k(T, \U(1)) \to \mathrm{Hom}(H^k(T), \U(1))$ defined
according to
$[\omega] \mapsto ([\sigma] \mapsto \omega(\sigma))$ (where $[\cdots]$ denotes
equivalence classes in cohomology or homology.)
The universal
coefficient theorem states that $\gamma$ is in fact an isomorphism.
[In general, replacing $\U(1)$ with an arbitrary abelian group $A$, the
universal coefficient theorem states that the homomorphism is surjective and its
kernel is isomorphic to $\mathrm{Ext}(G, A)$. But $\U(1)$ is divisible and it follows
that $\mathrm{Ext}(G, \U(1)) = 0$.]

Hence we can prove
\newtheorem*{lemma:central}{Lemma \ref{lemma:central}}
  \begin{lemma:central}
  A $\U(1)$ $k$-cochain $\omega$ on a manifold $T$ is exact if and only if
  $\omega(C) = 0$ for all closed (i.e.\ boundaryless) $k$-chains $C$.
  \begin{proof}
    If $\omega$ is exact, then $\omega = d\kappa$, and hence $\omega(C) =
    \kappa(\partial C) = 0$ for any closed $k$-chain $C$.

    Conversely, let $\omega$ be some $k$-cochain such that $\omega(C) = 0$ for all closed
    $A$. Then $\omega$ is closed because $(d\omega)(A) = \omega(\partial A) = 0$.
    Also, $\gamma([\omega]) = 0$. Since $\gamma$ is an isomorphism, it follows that
    the equivalence class $[\omega] = 0$. Hence, $\omega$ is exact.
  \end{proof}
  \end{lemma:central}

  \section{Calculating the element of the cohomology group for discrete non-linear sigma models}
  \label{sec_app_discrete}
The action for the field theories of
Ref.~\onlinecite{spto_higher,*spto_higher_prb} is a discrete analogue of the
topological theta term that appeared in the continuous NL$\sigma$M's in Section
\ref{sec_nlsm}. Recall from Section \ref{nlsm_cochains} that the theta term is derived
from a $\U(1)$ cochain defined on the target manifold $T$. The same is true in the
discrete case, except that the target space $T$ is now discrete, and so the
interpretation of the ``chains'' which are the arguments of the cochains needs to be revised.
Specifically, we define a $k$-chain on $T$ to be a formal linear combination
(with integer coefficients) of
``$k$-simplices'', which are simply ordered $k$-tuples $\Delta = (\Delta_1,
\cdots, \Delta_k) \in T^{\times k}$. Then we can define the ``boundary'' operator
$\partial$ acting linearly on $k$-chains by specifying its action on $k$-simplices:
\begin{equation}
  \label{abstract_boundary}
  \partial(\Delta_1, \cdots, \Delta_k) = \sum_{j=1}^k (-1)^{j-1} (\Delta_1, \cdots,
  \Delta_{j-1}, \Delta_{j+1}, \cdots, \Delta_k).
\end{equation}

To construct the discrete topological term corresponding to a $\U(1)$ cochain on
$T$, one considers a triangulation of a $D$-dimensional spacetime manifold $M$; that is, we
build $M$ up out of $D$-simplices. The degrees of freedom of the field theory
will live on the vertices of the simplices. We can represent the
$D$-simplices in spacetime in terms of their vertices $(x_1, \cdots, x_D)$ [the abstract definition of boundary given in 
  \eqnref{abstract_boundary} then agrees
with the geometrical definition]. Thus, we interpret the manifold $M$ as a
formal $D$-chain $M = \sum_{\Delta} (\Delta_1, \cdots, \Delta_D)$. Given a
$\U(1)$ $D$-cochain $\omega$ on $M$, and a function $\alpha$ assigning a value of the
target space to each vertex, we can define the action
\begin{align}
  S_{\mathrm{top}} &= \omega(\alpha(M)) \label{discrete_top} \\
                   &\equiv \sum_{\Delta} \omega(\alpha(\Delta_1), \cdots, \alpha(\Delta_D))
\end{align}
We ensure that this action will vanish for \emph{closed} space-time
manifolds $M$ by requring that $\omega(C) = 0$ for any closed chain $C$. By Lemma
\ref{lemma:central} in Section \ref{nlsm_cochains} above (which holds equally well for discrete cochains),
this is equivalent to requiring that $\omega$ be exact.

If we have an action of a group $G$ on the target space $T$, then for each
symmetric $D$-cochain $\omega$ one can derive an element of the group cohomology group
$H^{D}(G, \U(1))$ by following the
exact same reduction procedure as we did in the continuous case in Section
\ref{nlsm_cochains}.  One might, however, object that the physical significance of this is
not clear unless we specify some way to quantize a field theory defined in discrete time.
 For this reason, we want to reinterpret discrete field
theories like those of \ref{nlsm_cochains} as prescriptions for constructing lattice
models.

Specifically, consider a triangulation of a closed $d$-dimensional
\emph{spatial} manifold $M$ ($d = D - 1$).
At each vertex, we put a quantum particle whose basis states are
labeled by the elements of the target space $T$. (Hence, each basis state of
the whole system is labeled by a function $\alpha$ mapping vertices into $T$.) We
can define a quantum state for the system by ``imaginary-time evolution'' of a
$(d+1)$-dimensional discrete
topological action derived from an exact $\U(1)$ $(d+1)$-cochain on $T$. This
state is given by
\begin{align}
  \ket{\Psi} &= \sum_\alpha \Psi(\alpha) \ket{\alpha}, \\
  \Psi(\alpha) &= \exp\bigl[i \kappa(\alpha(M))\bigr],
\end{align}
where $\kappa$ is a $\U(1)$ $d$-cochain such that $d\kappa = \omega$, and we
define $\kappa(\alpha(M))$ in the analogous way to \eqnref{discrete_top} above.
This wavefunction is invariant under the on-site representation of the symmetry,
\begin{equation}
  \label{sg}
  S(g) = \sum_\alpha \ket{g \alpha} \bra{\alpha}.
\end{equation}
Once we have the wavefunction, it is easy to construct a corresponding
local Hamiltonian for which it is the gapped ground state. For example, if we
let $\mathcal{V} = \sum_{\alpha} \Psi(\alpha) \ket{\alpha} \bra{\alpha}$ be
the local unitary which creates $\ket{\Psi}$ from the trivial product state 
$\ket{\Psi_{\mathrm{prod}}} = \sum_\alpha \ket{\alpha}$, then, starting from a
local Hamiltonian $H_{\mathrm{prod}}$ which has $\ket{\Psi_{\mathrm{prod}}}$ as
its gapped ground state, we can define $H = \mathcal{V} H_{\mathrm{prod}} \mathcal{V}^{\dagger}$.

In Appendix \ref{sec_edge_construction}, we give a general discussion of how to decouple a
bulk theory from its boundary in order to find the form of the edge symmetry.
Applying the method of Appendix \ref{sec_edge_construction} to the situation at hand, one finds that
the edge symmetry takes the form
$U_{\mathrm{edge}}(g) = N(g) S(g)$, where $S(g)$ is as in \eqnref{sg} (but acting
only on the degrees of freedom at the edge), and
\begin{equation}
  N(g) = \sum_\alpha \exp\{i \kappa^{(2)}_g(\alpha(\partial M))\} \ket{\alpha}
  \bra{\alpha},
\end{equation}
where $\kappa^{(2)}_g$ is the $\U(1)$ $d-1$-cochain such that $d \kappa^{(2)}_g =
g\kappa - \kappa$. This is precisely the discrete version of \eqnref{Ng_nlsm}. It is
now easy to see that the general reduction procedure of Section
\ref{sec_higher_dims} will produce the same result as naively applying the method of Section
\ref{nlsm_cochains} for the discrete case.

It turns out that the reduction procedure can actually be done explicitly
starting from an arbitrary exact $\U(1)$ $(d+1)$-cochain $\omega$
on a target space $T$ symmetric under the action of a group $G$. To see this,
choose some arbitrary fixed $t_{*} \in T$, and define
\begin{multline}
  \kappa^{(k)}(g_1, \cdots, g_k)(\Delta_1, \cdots, \Delta_{d-k}) = \omega(
           g_k^{-1} \cdots g_1^{-1} t_{*}, \cdots, g_1^{-1} t_{*}, t_{*}, \, \Delta_1,
  \cdots, \Delta_{d-k}) 
\end{multline}
and $\omega^{(k)} = d \kappa^{(k)}$.
Using the fact that $\omega$ is invariant under the symmetry and $\omega(C) = 0$ for closed chains $C$, one can show
that (a) $\omega^{(0)} = \omega$; and (b) $\delta_k \kappa^{(k)} =
\omega^{(k+1)}$. Thus, we have explicitly constructed the
reduction sequence of Section \ref{nlsm_cochains}, and we find that the resulting element of
the group cohomology group $H^{d+1}(G, \U(1))$ is the equivalence class of the following
$\U(1)$ $(d+1)$-(group cocycle):
\begin{equation}
  \label{hom_inhom}
  \nu(g_1, \cdots, g_{d+1}) = \omega(g_{d+1}^{-1} \cdots g_1^{-1} t_*, \cdots, g_1^{-1}
  t_{*}, t_{*}).
\end{equation}

In particular, following Ref.~\onlinecite{spto_higher,*spto_higher_prb}, we can consider the case where the target
space $T$ is the symmetry group $G$ itself, with $G$ acting on itself by left multiplication.
In that case, it is easy to see that \eqnref{hom_inhom} actually defines a one-to-one mapping between symmetric
exact ``topological'' cochains on the right-hand side and \emph{group} cocycles on the
left-hand side.
Thus, for every element of the
group cohomology group $H^{d+1}(G, \U(1))$, one can construct a discrete topological
term in $d+1$ space-time dimensions via \eqnref{hom_inhom}, and applying our general reduction procedure returns the same
element of $H^{d+1}(G, \U(1))$.

\section{Classification of (2+1)-D fermionic SPT's}
\label{appendix_fermionic}
Here we will implement the ideas discussed in Section \ref{sec_fermions} in order to give a
classification of (2+1)-D fermion SPT's.
Consider a (2+1)-D fermionic SPT with fermionic symmetry group  $G_f$
(represented on-site), including an element $\pi$ corresponding to the fermion
parity. All the symmetries are assumed to be
local, so $\pi$ must commute with all the elements of $G_f$. The fermion parity
plays a such a key role in the following argument that we find it convenient to
write elements of $G_f$ in the form $\varpi(g) \pi^m$, where $m = 0$ or $1$, $g
\in G_b \equiv G_f/ Z_2^f$, and $\varpi$ is an arbitrary identification of
elements of $G_b$ with coset representatives in $G_f$, such that $\varpi(g_1)
\varpi(g_2) = \pi^{\lambda(g_1, g_2)} \varpi(g_1, g_2)$. Here $\lambda(g_1,
g_2)$ takes values 0 or 1, and associativity implies that it must be a
$\mathbb{Z}_2$ 2-cocycle, i.e.\ $\delta \lambda = 0$ [mod 2], where $\delta$ is the
coboundary operator
\begin{equation}
  (\delta \lambda)(g_1, g_2) = \lambda(g_1, g_2) + \lambda(g_1 g_2, g_3) + \lambda(g_2, g_3) + \lambda(g_1, g_2 g_3)
\end{equation}

Now, we assume that the edge of this SPT can be
realized in a strictly (1+1)-D local fermionic system and be invariant under a local unitary
(but not necessarily on-site) representation of $G_f$. 
The fermion parity is still represented as $\Pi \equiv (-1)^{F}$ on the
edge. We write
the fermionic local unitary operator implementing $\varpi(g) \pi^m$ on the edge as $U(g)
\Pi^m$. Then $U(g)$ must satisfy 
\begin{equation}
  \label{fermionic_representation}
U(g_1) U(g_2) = \Pi^{\lambda(g_1, g_2)} U(g_1 g_2).
\end{equation}
If we restrict the symmetry action to an interval $M = [a,b]$, then
the restricted unitaries must
satisfy \eqnref{fermionic_representation} up to a boundary term $\Omega(g_1,
g_2)$:
\begin{equation}
  \label{british_columbia}
  U_M(g_1) U_M(g_2) = \Omega(g_1, g_2) \Pi^{\lambda(g_1, g_2)} U_M(g_1 g_2)
\end{equation}
Using the associativity of the $U_M$'s, combined with $\delta \lambda = 0$ and the fact
that the $U_M$'s commute with $\Pi$, we see that the $\Omega$'s must satisfy an
identical equation to the bosonic case:
\begin{equation}
  \label{quebec}
  \Omega(g_1, g_2) \Omega(g_1 g_2, g_3) = \; ^{U_M({g_1})} \Omega(g_2, g_3)
\Omega({g_1},{g_2}{g_3}),
\end{equation}
As discussed in Section \ref{sec_fermions}, in defining the restriction $\Omega \to \Omega_a$ we might obtain an operator
carrying non-trivial charge under fermion parity. We define the function
$\sigma(g_1, g_2)$ to be 0 if $\Omega_a$ is a fermionic local unitary (no charge
under fermion parity) acting at
the point $a$, and 1 if it is equal to such a local unitary, multiplied by $(c_a
+ c_a^{\dagger})$.

The restriction
$\Omega_a(g_1, g_2)$ must satisfy \eqnref{quebec} up to a phase factor:
\begin{equation}
  \label{fermionic_omega}
  ^{U_M(g_1)} \Omega_a(g_2, g_3) \Omega_a(g_1, g_2 g_3) = \omega(g_1, g_2, g_3)
  \Omega_a(g_1, g_2) \Omega_a(g_1 g_2, g_3),
\end{equation}
where $\omega$ is a $\U(1)$-valued function. The pair of functions $(\omega,
\sigma)$ constitutes the fermionic 3-cocycle. From \eqnref{fermionic_omega} we
immediately see that $\sigma$ must be a $\mathbb{Z}_2$ cocycle, i.e.\ 
\begin{equation}
  \label{sigma_cocycle}
\delta \sigma = 0.
\end{equation}
Following a similar derivation to
the one in Appendix \ref{sec_crossed_module} that gave the bosonic 3-cocycle
condition\footnote{Specifically, the difference from the bosonic case comes from
  in going from \eqnref{asdf1} to
  \eqnref{asdf2} [due to the potential for $\Omega_b(g_1, g_2)$ and $\Omega_a(g_3,
  g_4)$ to anticommute], and in going from \eqnref{asdf2} to \eqnref{asdf3} [due to the
factor of $\Pi^{\lambda(g_1, g_2)}$ appearing in \eqnref{british_columbia}.]},
  we also
also find that $\omega$ must obey
\begin{equation}
  \label{fermionic_omega_cocycle}
  (\delta \omega)(g_1, g_2, g_3, g_4)
  = (-1)^{[\sigma(g_1,
  g_2) + \lambda(g_1, g_2)] \sigma(g_3, g_4)}.
\end{equation}
where 
\begin{equation}
  (\delta \omega)(g_1, g_2, g_3, g_4) = \omega(g_1, g_2, g_3) \omega(g_1 g_2, g_3, g_4)^{-1} \omega(g_1, g_2 g_3,
  g_4) \omega(g_1, g_2, g_3 g_4)^{-1} \omega(g_2, g_3, g_4).
\end{equation}
Eqs.~(\ref{sigma_cocycle}) and (\ref{fermionic_omega_cocycle}) constitute the
condition for ($\omega$,$\sigma$) to be a fermionic 3-cocycle.

Furthermore, the freedom to redefine the restriction $U \to U_M$ and $\Omega \to
\Omega_a$ implies (again following similar arguments as in Appendix
\ref{sec_crossed_module}) that we must identify fermionic 3-cocycles that differ by the
transformation
\begin{align}
  \label{three_coboundary_fermionic}
  \sigma(g_1, g_2) &\to \sigma(g_1, g_2) + (\delta\mu)(g_1, g_2) \quad \mbox{[mod 2]}\\
  \omega(g_1, g_2, g_3) &\to \omega(g_1, g_2, g_3) (-1)^{[\sigma(g_1, g_2) +
  \lambda(g_1, g_2)] \mu(g_3) + \mu(g_1)[\sigma(g_2, g_3) + (d\mu)(g_2, g_3)]}
  (\delta\beta)(g_1, g_2, g_3)
\end{align}
where
\begin{align}
  (\delta \beta)(g_1, g_2, g_3) &= \beta(g_1, g_2) \beta(g_1
  g_2, g_3) \beta(g_2, g_3)^{-1} \beta(g_1, g_2 g_3)^{-1}, \\
  (\delta \mu)(g_1, g_2) &= \mu(g_1) + \mu(g_2) + \mu(g_1 g_2),
\end{align}
and $\beta$ and $\mu$ take values in  $\U(1)$ and $\{ 0, 1\}$ respectively.
[The numbers $\mu(g)$ correspond to the fermion parity of the restriction
$\Sigma_a(g)$ of the operator $\Sigma(g)$ that implements the redefintion
$U_M(g) \to \Sigma(g) U_M(g)$.]
If $(\omega,\sigma)$ is not equivalent to the trivial fermionic 3-cocycle according to
the above equivalence relation, then we expect that the edge must
correspond to the boundary of a non-trivial SPT phase. This is because such an anomalous non-trivial
symmetry precludes a gapped ground state unless the symmetry is spontaneously
broken. (This can be derived in a similar way to
the equivalent bosonic result, Lemma \ref{lem_milton} in Appendix
\ref{sec_completing}.) Similarly, two SPT phases characterized by fermionic
3-cocycles \emph{not} related by the above equivalence relation must be
separated by a phase transition.

We also can define a product rule for fermionic 3-cocycles. Physically, the
product rule corresponds to ``stacking'' two SPT phases on top of each other. If
the edges of the two systems are characterized by $(\sigma,\omega)$ and $(\sigma^{\prime},
\omega^{\prime})$, then one can show that the edge of the combined system will
be described by the fermionic 3-cocycle $(\sigma_{\mathrm{prod}},
\omega_{\mathrm{prod}})$, where
\begin{align}
  \sigma_{\mathrm{prod}} &= \sigma + \sigma^{\prime} \\
  \omega_{\mathrm{prod}}(g_1, g_2, g_3) &= (-1)^{\sigma^{\prime}(g_2, g_3) \sigma(g_1,
  g_2 g_3) + \sigma^{\prime}(g_1, g_2) \sigma(g_1 g_2, g_3)} \omega(g_1, g_2,
  g_3) \omega^{\prime}(g_1, g_2, g_3).
\end{align}

Finally, let us remark that, if we set $\sigma = 0$, then the fermionic
3-cocycles reduce to ordinary 3-cocycles for the ``bosonic'' symmetry group $G_b
= G_f/Z_2^f$. This reflects the fact that bosonic
SPT's can be realized in a fermion system by pairing fermions to form bosons.
However, according to the equivalence relation
\eqnref{three_coboundary_fermionic}, 
when $\lambda \neq 0$ (i.e.\ the fermionic symmetry group is
not simply a direct product $G_f = G_b \times Z_2^f$), a non-trivial bosonic
3-cocycle might still be trivial as a fermionic 3-cocycle.
Thus, there is the possibility that a bosonic SPT
phase could become trivial in the presence of fermions. Examples of this
phenomenon (albeit for symmetry groups including anti-unitary symmetries, which
we have not considered here) can be found in Ref.~\onlinecite{lu_vishwanath}.

%


\begin{thebibliography}{88}%
\makeatletter
\providecommand \@ifxundefined [1]{%
 \@ifx{#1\undefined}
}%
\providecommand \@ifnum [1]{%
 \ifnum #1\expandafter \@firstoftwo
 \else \expandafter \@secondoftwo
 \fi
}%
\providecommand \@ifx [1]{%
 \ifx #1\expandafter \@firstoftwo
 \else \expandafter \@secondoftwo
 \fi
}%
\providecommand \natexlab [1]{#1}%
\providecommand \enquote  [1]{``#1''}%
\providecommand \bibnamefont  [1]{#1}%
\providecommand \bibfnamefont [1]{#1}%
\providecommand \citenamefont [1]{#1}%
\providecommand \href@noop [0]{\@secondoftwo}%
\providecommand \href [0]{\begingroup \@sanitize@url \@href}%
\providecommand \@href[1]{\@@startlink{#1}\@@href}%
\providecommand \@@href[1]{\endgroup#1\@@endlink}%
\providecommand \@sanitize@url [0]{\catcode `\\12\catcode `\$12\catcode
  `\&12\catcode `\#12\catcode `\^12\catcode `\_12\catcode `\%12\relax}%
\providecommand \@@startlink[1]{}%
\providecommand \@@endlink[0]{}%
\providecommand \url  [0]{\begingroup\@sanitize@url \@url }%
\providecommand \@url [1]{\endgroup\@href {#1}{\urlprefix }}%
\providecommand \urlprefix  [0]{URL }%
\providecommand \Eprint [0]{\href }%
\providecommand \doibase [0]{http://dx.doi.org/}%
\providecommand \selectlanguage [0]{\@gobble}%
\providecommand \bibinfo  [0]{\@secondoftwo}%
\providecommand \bibfield  [0]{\@secondoftwo}%
\providecommand \translation [1]{[#1]}%
\providecommand \BibitemOpen [0]{}%
\providecommand \bibitemStop [0]{}%
\providecommand \bibitemNoStop [0]{.\EOS\space}%
\providecommand \EOS [0]{\spacefactor3000\relax}%
\providecommand \BibitemShut  [1]{\csname bibitem#1\endcsname}%
\let\auto@bib@innerbib\@empty
\bibitem [{\citenamefont {Haldane}(1983{\natexlab{a}})}]{haldane1}%
  \BibitemOpen
  \bibfield  {author} {\bibinfo {author} {\bibfnamefont {F.~D.~M.}\
  \bibnamefont {Haldane}},\ }\href {\doibase 10.1016/0375-9601(83)90631-X}
  {\bibfield  {journal} {\bibinfo  {journal} {Phys. Lett. A}\ }\textbf
  {\bibinfo {volume} {93}},\ \bibinfo {pages} {464} (\bibinfo {year}
  {1983}{\natexlab{a}})}\BibitemShut {NoStop}%
\bibitem [{\citenamefont {Haldane}(1983{\natexlab{b}})}]{haldane2}%
  \BibitemOpen
  \bibfield  {author} {\bibinfo {author} {\bibfnamefont {F.~D.~M.}\
  \bibnamefont {Haldane}},\ }\href {\doibase 10.1103/PhysRevLett.50.1153}
  {\bibfield  {journal} {\bibinfo  {journal} {Phys. Rev. Lett.}\ }\textbf
  {\bibinfo {volume} {50}},\ \bibinfo {pages} {1153} (\bibinfo {year}
  {1983}{\natexlab{b}})}\BibitemShut {NoStop}%
\bibitem [{\citenamefont {Affleck}\ \emph {et~al.}(1988)\citenamefont
  {Affleck}, \citenamefont {Kennedy}, \citenamefont {Lieb},\ and\ \citenamefont
  {Tasaki}}]{aklt1}%
  \BibitemOpen
  \bibfield  {author} {\bibinfo {author} {\bibfnamefont {I.}~\bibnamefont
  {Affleck}}, \bibinfo {author} {\bibfnamefont {T.}~\bibnamefont {Kennedy}},
  \bibinfo {author} {\bibfnamefont {E.}~\bibnamefont {Lieb}}, \ and\ \bibinfo
  {author} {\bibfnamefont {H.}~\bibnamefont {Tasaki}},\ }\href {\doibase
  10.1007/BF01218021} {\bibfield  {journal} {\bibinfo  {journal} {Comm. Math.
  Phys.}\ }\textbf {\bibinfo {volume} {115}},\ \bibinfo {pages} {477} (\bibinfo
  {year} {1988})}\BibitemShut {NoStop}%
\bibitem [{\citenamefont {Affleck}\ \emph {et~al.}(1987)\citenamefont
  {Affleck}, \citenamefont {Kennedy}, \citenamefont {Lieb},\ and\ \citenamefont
  {Tasaki}}]{aklt2}%
  \BibitemOpen
  \bibfield  {author} {\bibinfo {author} {\bibfnamefont {I.}~\bibnamefont
  {Affleck}}, \bibinfo {author} {\bibfnamefont {T.}~\bibnamefont {Kennedy}},
  \bibinfo {author} {\bibfnamefont {E.~H.}\ \bibnamefont {Lieb}}, \ and\
  \bibinfo {author} {\bibfnamefont {H.}~\bibnamefont {Tasaki}},\ }\href
  {\doibase 10.1103/PhysRevLett.59.799} {\bibfield  {journal} {\bibinfo
  {journal} {Phys. Rev. Lett.}\ }\textbf {\bibinfo {volume} {59}},\ \bibinfo
  {pages} {799} (\bibinfo {year} {1987})}\BibitemShut {NoStop}%
\bibitem [{\citenamefont {Verstraete}\ \emph {et~al.}(2004)\citenamefont
  {Verstraete}, \citenamefont {Mart\'in-Delgado},\ and\ \citenamefont
  {Cirac}}]{diverging_prl}%
  \BibitemOpen
  \bibfield  {author} {\bibinfo {author} {\bibfnamefont {F.}~\bibnamefont
  {Verstraete}}, \bibinfo {author} {\bibfnamefont {M.~A.}\ \bibnamefont
  {Mart\'in-Delgado}}, \ and\ \bibinfo {author} {\bibfnamefont {J.~I.}\
  \bibnamefont {Cirac}},\ }\href {\doibase 10.1103/PhysRevLett.92.087201}
  {\bibfield  {journal} {\bibinfo  {journal} {Phys. Rev. Lett.}\ }\textbf
  {\bibinfo {volume} {92}},\ \bibinfo {pages} {087201} (\bibinfo {year}
  {2004})},\ \Eprint {http://arxiv.org/abs/0311087} {arXiv:0311087}
  \BibitemShut {NoStop}%
\bibitem [{\citenamefont {Hasan}\ and\ \citenamefont
  {Kane}(2010)}]{hasan_kane_review}%
  \BibitemOpen
  \bibfield  {author} {\bibinfo {author} {\bibfnamefont {M.~Z.}\ \bibnamefont
  {Hasan}}\ and\ \bibinfo {author} {\bibfnamefont {C.~L.}\ \bibnamefont
  {Kane}},\ }\href {\doibase 10.1103/RevModPhys.82.3045} {\bibfield  {journal}
  {\bibinfo  {journal} {Rev. Mod. Phys.}\ }\textbf {\bibinfo {volume} {82}},\
  \bibinfo {pages} {3045} (\bibinfo {year} {2010})},\ \Eprint
  {http://arxiv.org/abs/1002.3895} {arXiv:1002.3895} \BibitemShut {NoStop}%
\bibitem [{\citenamefont {Moore}(2010)}]{moore_birth_2010}%
  \BibitemOpen
  \bibfield  {author} {\bibinfo {author} {\bibfnamefont {J.~E.}\ \bibnamefont
  {Moore}},\ }\href {\doibase 10.1038/nature08916} {\bibfield  {journal}
  {\bibinfo  {journal} {Nature}\ }\textbf {\bibinfo {volume} {464}},\ \bibinfo
  {pages} {194} (\bibinfo {year} {2010})}\BibitemShut {NoStop}%
\bibitem [{\citenamefont {Qi}\ and\ \citenamefont
  {Zhang}(2011)}]{qi_zhang_review}%
  \BibitemOpen
  \bibfield  {author} {\bibinfo {author} {\bibfnamefont {X.-L.}\ \bibnamefont
  {Qi}}\ and\ \bibinfo {author} {\bibfnamefont {S.-C.}\ \bibnamefont {Zhang}},\
  }\href {\doibase 10.1103/RevModPhys.83.1057} {\bibfield  {journal} {\bibinfo
  {journal} {Rev. Mod. Phys.}\ }\textbf {\bibinfo {volume} {83}},\ \bibinfo
  {pages} {1057} (\bibinfo {year} {2011})},\ \Eprint
  {http://arxiv.org/abs/1008.2026} {arXiv:1008.2026} \BibitemShut {NoStop}%
\bibitem [{\citenamefont {Gu}\ and\ \citenamefont {Wen}(2009)}]{gu_wen_2009}%
  \BibitemOpen
  \bibfield  {author} {\bibinfo {author} {\bibfnamefont {Z.-C.}\ \bibnamefont
  {Gu}}\ and\ \bibinfo {author} {\bibfnamefont {X.-G.}\ \bibnamefont {Wen}},\
  }\href {\doibase 10.1103/PhysRevB.80.155131} {\bibfield  {journal} {\bibinfo
  {journal} {Phys. Rev. B}\ }\textbf {\bibinfo {volume} {80}},\ \bibinfo
  {pages} {155131} (\bibinfo {year} {2009})},\ \Eprint
  {http://arxiv.org/abs/0903.1069} {arXiv:0903.1069} \BibitemShut {NoStop}%
\bibitem [{\citenamefont {Chen}\ \emph {et~al.}(2010)\citenamefont {Chen},
  \citenamefont {Gu},\ and\ \citenamefont {Wen}}]{wen_lu}%
  \BibitemOpen
  \bibfield  {author} {\bibinfo {author} {\bibfnamefont {X.}~\bibnamefont
  {Chen}}, \bibinfo {author} {\bibfnamefont {Z.-C.}\ \bibnamefont {Gu}}, \ and\
  \bibinfo {author} {\bibfnamefont {X.-G.}\ \bibnamefont {Wen}},\ }\href
  {\doibase 10.1103/PhysRevB.82.155138} {\bibfield  {journal} {\bibinfo
  {journal} {Phys. Rev. B}\ }\textbf {\bibinfo {volume} {82}},\ \bibinfo
  {pages} {155138} (\bibinfo {year} {2010})},\ \Eprint
  {http://arxiv.org/abs/arXiv:1004.3835} {arXiv:1004.3835} \BibitemShut
  {NoStop}%
\bibitem [{\citenamefont {Chen}\ \emph
  {et~al.}(2011{\natexlab{a}})\citenamefont {Chen}, \citenamefont {Gu},\ and\
  \citenamefont {Wen}}]{chen_gu_wen}%
  \BibitemOpen
  \bibfield  {author} {\bibinfo {author} {\bibfnamefont {X.}~\bibnamefont
  {Chen}}, \bibinfo {author} {\bibfnamefont {Z.-C.}\ \bibnamefont {Gu}}, \ and\
  \bibinfo {author} {\bibfnamefont {X.-G.}\ \bibnamefont {Wen}},\ }\href
  {\doibase 10.1103/PhysRevB.83.035107} {\bibfield  {journal} {\bibinfo
  {journal} {Phys. Rev. B}\ }\textbf {\bibinfo {volume} {83}},\ \bibinfo
  {pages} {035107} (\bibinfo {year} {2011}{\natexlab{a}})},\ \Eprint
  {http://arxiv.org/abs/arXiv:1008.3745} {arXiv:1008.3745} \BibitemShut
  {NoStop}%
\bibitem [{\citenamefont {Schuch}\ \emph {et~al.}(2011)\citenamefont {Schuch},
  \citenamefont {P{\'e}rez-Garc{\'i}a},\ and\ \citenamefont {Cirac}}]{schuch}%
  \BibitemOpen
  \bibfield  {author} {\bibinfo {author} {\bibfnamefont {N.}~\bibnamefont
  {Schuch}}, \bibinfo {author} {\bibfnamefont {D.}~\bibnamefont
  {P{\'e}rez-Garc{\'i}a}}, \ and\ \bibinfo {author} {\bibfnamefont
  {I.}~\bibnamefont {Cirac}},\ }\href {\doibase 10.1103/PhysRevB.84.165139}
  {\bibfield  {journal} {\bibinfo  {journal} {Phys. Rev. B}\ }\textbf {\bibinfo
  {volume} {84}},\ \bibinfo {pages} {165139} (\bibinfo {year} {2011})},\
  \Eprint {http://arxiv.org/abs/1010.3732} {arXiv:1010.3732} \BibitemShut
  {NoStop}%
\bibitem [{\citenamefont {Pollmann}\ \emph {et~al.}(2012)\citenamefont
  {Pollmann}, \citenamefont {Berg}, \citenamefont {Turner},\ and\ \citenamefont
  {Oshikawa}}]{pollmann-arxiv-2009}%
  \BibitemOpen
  \bibfield  {author} {\bibinfo {author} {\bibfnamefont {F.}~\bibnamefont
  {Pollmann}}, \bibinfo {author} {\bibfnamefont {E.}~\bibnamefont {Berg}},
  \bibinfo {author} {\bibfnamefont {A.~M.}\ \bibnamefont {Turner}}, \ and\
  \bibinfo {author} {\bibfnamefont {M.}~\bibnamefont {Oshikawa}},\ }\href
  {\doibase 10.1103/PhysRevB.85.075125} {\bibfield  {journal} {\bibinfo
  {journal} {Phys. Rev. B}\ }\textbf {\bibinfo {volume} {85}},\ \bibinfo
  {pages} {075125} (\bibinfo {year} {2012})},\ \Eprint
  {http://arxiv.org/abs/0909.4059} {arXiv:0909.4059} \BibitemShut {NoStop}%
\bibitem [{\citenamefont {Pollmann}\ \emph {et~al.}(2010)\citenamefont
  {Pollmann}, \citenamefont {Turner}, \citenamefont {Berg},\ and\ \citenamefont
  {Oshikawa}}]{pollmann-prb-2010}%
  \BibitemOpen
  \bibfield  {author} {\bibinfo {author} {\bibfnamefont {F.}~\bibnamefont
  {Pollmann}}, \bibinfo {author} {\bibfnamefont {A.~M.}\ \bibnamefont
  {Turner}}, \bibinfo {author} {\bibfnamefont {E.}~\bibnamefont {Berg}}, \ and\
  \bibinfo {author} {\bibfnamefont {M.}~\bibnamefont {Oshikawa}},\ }\href
  {\doibase 10.1103/PhysRevB.81.064439} {\bibfield  {journal} {\bibinfo
  {journal} {Phys. Rev. B}\ }\textbf {\bibinfo {volume} {81}},\ \bibinfo
  {pages} {064439} (\bibinfo {year} {2010})},\ \Eprint
  {http://arxiv.org/abs/arXiv:0910.1811} {arXiv:0910.1811} \BibitemShut
  {NoStop}%
\bibitem [{\citenamefont {Chen}\ \emph {et~al.}(2012)\citenamefont {Chen},
  \citenamefont {Gu}, \citenamefont {Liu},\ and\ \citenamefont
  {Wen}}]{spto_higher}%
  \BibitemOpen
  \bibfield  {author} {\bibinfo {author} {\bibfnamefont {X.}~\bibnamefont
  {Chen}}, \bibinfo {author} {\bibfnamefont {Z.-C.}\ \bibnamefont {Gu}},
  \bibinfo {author} {\bibfnamefont {Z.-X.}\ \bibnamefont {Liu}}, \ and\
  \bibinfo {author} {\bibfnamefont {X.-G.}\ \bibnamefont {Wen}},\ }\href
  {\doibase 10.1126/science.1227224} {\bibfield  {journal} {\bibinfo  {journal}
  {Science}\ }\textbf {\bibinfo {volume} {338}},\ \bibinfo {pages} {1604}
  (\bibinfo {year} {2012})},\ \Eprint {http://arxiv.org/abs/1301.0861}
  {arXiv:1301.0861} \BibitemShut {NoStop}%
\bibitem [{\citenamefont {Chen}\ \emph
  {et~al.}(2013{\natexlab{a}})\citenamefont {Chen}, \citenamefont {Gu},
  \citenamefont {Liu},\ and\ \citenamefont {Wen}}]{spto_higher_prb}%
  \BibitemOpen
  \bibfield  {author} {\bibinfo {author} {\bibfnamefont {X.}~\bibnamefont
  {Chen}}, \bibinfo {author} {\bibfnamefont {Z.-C.}\ \bibnamefont {Gu}},
  \bibinfo {author} {\bibfnamefont {Z.-X.}\ \bibnamefont {Liu}}, \ and\
  \bibinfo {author} {\bibfnamefont {X.-G.}\ \bibnamefont {Wen}},\ }\href
  {\doibase 10.1103/PhysRevB.87.155114} {\bibfield  {journal} {\bibinfo
  {journal} {Phys. Rev. B}\ }\textbf {\bibinfo {volume} {87}},\ \bibinfo
  {pages} {155114} (\bibinfo {year} {2013}{\natexlab{a}})},\ \Eprint
  {http://arxiv.org/abs/1106.4772} {arXiv:1106.4772} \BibitemShut {NoStop}%
\bibitem [{\citenamefont {Xu}(2013)}]{cenke_neel_order}%
  \BibitemOpen
  \bibfield  {author} {\bibinfo {author} {\bibfnamefont {C.}~\bibnamefont
  {Xu}},\ }\href {\doibase 10.1103/PhysRevB.87.144421} {\bibfield  {journal}
  {\bibinfo  {journal} {Phys. Rev. B}\ }\textbf {\bibinfo {volume} {87}},\
  \bibinfo {pages} {144421} (\bibinfo {year} {2013})},\ \Eprint
  {http://arxiv.org/abs/1209.4399} {arXiv:1209.4399} \BibitemShut {NoStop}%
\bibitem [{\citenamefont {Vishwanath}\ and\ \citenamefont
  {Senthil}(2013)}]{vishwanath_senthil}%
  \BibitemOpen
  \bibfield  {author} {\bibinfo {author} {\bibfnamefont {A.}~\bibnamefont
  {Vishwanath}}\ and\ \bibinfo {author} {\bibfnamefont {T.}~\bibnamefont
  {Senthil}},\ }\href {\doibase 10.1103/PhysRevX.3.011016} {\bibfield
  {journal} {\bibinfo  {journal} {Phys. Rev. X}\ ,\ \bibinfo {pages} {011016}}
  (\bibinfo {year} {2013})},\ \Eprint {http://arxiv.org/abs/1209.3058}
  {arXiv:1209.3058} \BibitemShut {NoStop}%
\bibitem [{\citenamefont {Lu}\ and\ \citenamefont
  {Vishwanath}(2012)}]{lu_vishwanath}%
  \BibitemOpen
  \bibfield  {author} {\bibinfo {author} {\bibfnamefont {Y.-M.}\ \bibnamefont
  {Lu}}\ and\ \bibinfo {author} {\bibfnamefont {A.}~\bibnamefont
  {Vishwanath}},\ }\href {\doibase 10.1103/PhysRevB.86.125119} {\bibfield
  {journal} {\bibinfo  {journal} {Phys. Rev. B}\ }\textbf {\bibinfo {volume}
  {86}},\ \bibinfo {pages} {125119} (\bibinfo {year} {2012})},\ \Eprint
  {http://arxiv.org/abs/1205.3156} {arXiv:1205.3156} \BibitemShut {NoStop}%
\bibitem [{\citenamefont {Xu}\ and\ \citenamefont
  {Senthil}(2013)}]{cenke_wave_functions}%
  \BibitemOpen
  \bibfield  {author} {\bibinfo {author} {\bibfnamefont {C.}~\bibnamefont
  {Xu}}\ and\ \bibinfo {author} {\bibfnamefont {T.}~\bibnamefont {Senthil}},\
  }\href {\doibase 10.1103/PhysRevB.87.174412} {\bibfield  {journal} {\bibinfo
  {journal} {Phys. Rev. B}\ }\textbf {\bibinfo {volume} {87}},\ \bibinfo
  {pages} {174412} (\bibinfo {year} {2013})},\ \Eprint
  {http://arxiv.org/abs/1301.6172} {arXiv:1301.6172} \BibitemShut {NoStop}%
\bibitem [{\citenamefont {Ye}\ and\ \citenamefont
  {Wen}(2013)}]{wen_projective_construction}%
  \BibitemOpen
  \bibfield  {author} {\bibinfo {author} {\bibfnamefont {P.}~\bibnamefont
  {Ye}}\ and\ \bibinfo {author} {\bibfnamefont {X.-G.}\ \bibnamefont {Wen}},\
  }\href {\doibase 10.1103/PhysRevB.87.195128} {\bibfield  {journal} {\bibinfo
  {journal} {Phys. Rev. B}\ }\textbf {\bibinfo {volume} {87}},\ \bibinfo
  {pages} {195128} (\bibinfo {year} {2013})}\BibitemShut {NoStop}%
\bibitem [{\citenamefont {Liu}\ and\ \citenamefont
  {Wen}(2013)}]{wen_spin_hall}%
  \BibitemOpen
  \bibfield  {author} {\bibinfo {author} {\bibfnamefont {Z.-X.}\ \bibnamefont
  {Liu}}\ and\ \bibinfo {author} {\bibfnamefont {X.-G.}\ \bibnamefont {Wen}},\
  }\href {\doibase 10.1103/PhysRevLett.110.067205} {\bibfield  {journal}
  {\bibinfo  {journal} {Phys. Rev. Lett.}\ }\textbf {\bibinfo {volume} {110}},\
  \bibinfo {pages} {067205} (\bibinfo {year} {2013})}\BibitemShut {NoStop}%
\bibitem [{\citenamefont {Santos}\ and\ \citenamefont
  {Wang}(2014)}]{santos_wang_2014}%
  \BibitemOpen
  \bibfield  {author} {\bibinfo {author} {\bibfnamefont {L.~H.}\ \bibnamefont
  {Santos}}\ and\ \bibinfo {author} {\bibfnamefont {J.}~\bibnamefont {Wang}},\
  }\href {\doibase 10.1103/PhysRevB.89.195122} {\bibfield  {journal} {\bibinfo
  {journal} {Phys. Rev. B}\ }\textbf {\bibinfo {volume} {89}},\ \bibinfo
  {pages} {195122} (\bibinfo {year} {2014})},\ \Eprint
  {http://arxiv.org/abs/1310.8291} {arXiv:1310.8291} \BibitemShut {NoStop}%
\bibitem [{\citenamefont {Wang}\ \emph {et~al.}({\natexlab{a}})\citenamefont
  {Wang}, \citenamefont {Santos},\ and\ \citenamefont
  {Wen}}]{wang_santos_wen_2014}%
  \BibitemOpen
  \bibfield  {author} {\bibinfo {author} {\bibfnamefont {J.}~\bibnamefont
  {Wang}}, \bibinfo {author} {\bibfnamefont {L.~H.}\ \bibnamefont {Santos}}, \
  and\ \bibinfo {author} {\bibfnamefont {X.-G.}\ \bibnamefont {Wen}},\
  }\href@noop {} {} ({\natexlab{a}}),\ \Eprint {http://arxiv.org/abs/1403.5256}
  {arXiv:1403.5256} \BibitemShut {NoStop}%
\bibitem [{\citenamefont {Senthil}\ and\ \citenamefont
  {Levin}(2013)}]{senthil_iqhe_bosons}%
  \BibitemOpen
  \bibfield  {author} {\bibinfo {author} {\bibfnamefont {T.}~\bibnamefont
  {Senthil}}\ and\ \bibinfo {author} {\bibfnamefont {M.}~\bibnamefont
  {Levin}},\ }\href {\doibase 10.1103/PhysRevLett.110.046801} {\bibfield
  {journal} {\bibinfo  {journal} {Phys. Rev. Lett.}\ }\textbf {\bibinfo
  {volume} {110}},\ \bibinfo {pages} {046801} (\bibinfo {year} {2013})},\
  \Eprint {http://arxiv.org/abs/1206.1604} {arXiv:1206.1604} \BibitemShut
  {NoStop}%
\bibitem [{\citenamefont {Metlitski}\ \emph {et~al.}(2013)\citenamefont
  {Metlitski}, \citenamefont {Kane},\ and\ \citenamefont
  {Fisher}}]{boson_top_witten}%
  \BibitemOpen
  \bibfield  {author} {\bibinfo {author} {\bibfnamefont {M.~A.}\ \bibnamefont
  {Metlitski}}, \bibinfo {author} {\bibfnamefont {C.~L.}\ \bibnamefont {Kane}},
  \ and\ \bibinfo {author} {\bibfnamefont {M.~P.~A.}\ \bibnamefont {Fisher}},\
  }\href {\doibase 10.1103/PhysRevB.88.035131} {\bibfield  {journal} {\bibinfo
  {journal} {Phys. Rev. B}\ }\textbf {\bibinfo {volume} {88}},\ \bibinfo
  {pages} {035131} (\bibinfo {year} {2013})},\ \Eprint
  {http://arxiv.org/abs/1302.6535} {arXiv:1302.6535} \BibitemShut {NoStop}%
\bibitem [{\citenamefont {Regnault}\ and\ \citenamefont
  {Senthil}(2013)}]{microscopic_iqhe}%
  \BibitemOpen
  \bibfield  {author} {\bibinfo {author} {\bibfnamefont {N.}~\bibnamefont
  {Regnault}}\ and\ \bibinfo {author} {\bibfnamefont {T.}~\bibnamefont
  {Senthil}},\ }\href {\doibase 10.1103/PhysRevB.88.161106} {\bibfield
  {journal} {\bibinfo  {journal} {Phys. Rev. B}\ }\textbf {\bibinfo {volume}
  {88}},\ \bibinfo {pages} {161106} (\bibinfo {year} {2013})}\BibitemShut
  {NoStop}%
\bibitem [{\citenamefont {Levin}\ and\ \citenamefont {Gu}(2012)}]{levin_gu}%
  \BibitemOpen
  \bibfield  {author} {\bibinfo {author} {\bibfnamefont {M.}~\bibnamefont
  {Levin}}\ and\ \bibinfo {author} {\bibfnamefont {Z.-C.}\ \bibnamefont {Gu}},\
  }\href {\doibase 10.1103/PhysRevB.86.115109} {\bibfield  {journal} {\bibinfo
  {journal} {Phys. Rev. B}\ }\textbf {\bibinfo {volume} {86}},\ \bibinfo
  {pages} {115109} (\bibinfo {year} {2012})},\ \Eprint
  {http://arxiv.org/abs/1202.3120} {arXiv:1202.3120} \BibitemShut {NoStop}%
\bibitem [{\citenamefont {Wang}\ and\ \citenamefont {Senthil}(2013)}]{window}%
  \BibitemOpen
  \bibfield  {author} {\bibinfo {author} {\bibfnamefont {C.}~\bibnamefont
  {Wang}}\ and\ \bibinfo {author} {\bibfnamefont {T.}~\bibnamefont {Senthil}},\
  }\href {\doibase 10.1103/PhysRevB.87.235122} {\bibfield  {journal} {\bibinfo
  {journal} {Phys. Rev. B}\ }\textbf {\bibinfo {volume} {87}},\ \bibinfo
  {pages} {235122} (\bibinfo {year} {2013})},\ \Eprint
  {http://arxiv.org/abs/1302.6234} {arXiv:1302.6234} \BibitemShut {NoStop}%
\bibitem [{\citenamefont {Gu}\ and\ \citenamefont {Wen}()}]{supercohomology}%
  \BibitemOpen
  \bibfield  {author} {\bibinfo {author} {\bibfnamefont {Z.-C.}\ \bibnamefont
  {Gu}}\ and\ \bibinfo {author} {\bibfnamefont {X.-G.}\ \bibnamefont {Wen}},\
  }\href@noop {} {}\Eprint {http://arxiv.org/abs/1201.2648} {arXiv:1201.2648}
  \BibitemShut {NoStop}%
\bibitem [{\citenamefont {Kapustin}()}]{kapustin_cobordism}%
  \BibitemOpen
  \bibfield  {author} {\bibinfo {author} {\bibfnamefont {A.}~\bibnamefont
  {Kapustin}},\ }\href@noop {} {}\Eprint {http://arxiv.org/abs/1403.1467}
  {arXiv:1403.1467} \BibitemShut {NoStop}%
\bibitem [{\citenamefont {Sule}\ \emph {et~al.}(2013)\citenamefont {Sule},
  \citenamefont {Chen},\ and\ \citenamefont {Ryu}}]{orbifolds}%
  \BibitemOpen
  \bibfield  {author} {\bibinfo {author} {\bibfnamefont {O.~M.}\ \bibnamefont
  {Sule}}, \bibinfo {author} {\bibfnamefont {X.}~\bibnamefont {Chen}}, \ and\
  \bibinfo {author} {\bibfnamefont {S.}~\bibnamefont {Ryu}},\ }\href {\doibase
  10.1103/PhysRevB.88.075125} {\bibfield  {journal} {\bibinfo  {journal} {Phys.
  Rev. B}\ }\textbf {\bibinfo {volume} {88}},\ \bibinfo {pages} {075125}
  (\bibinfo {year} {2013})},\ \Eprint {http://arxiv.org/abs/1305.0700}
  {arXiv:1305.0700} \BibitemShut {NoStop}%
\bibitem [{\citenamefont {Wang}\ and\ \citenamefont
  {Levin}(2014)}]{wang_levin}%
  \BibitemOpen
  \bibfield  {author} {\bibinfo {author} {\bibfnamefont {C.}~\bibnamefont
  {Wang}}\ and\ \bibinfo {author} {\bibfnamefont {M.}~\bibnamefont {Levin}},\
  }\href@noop {} {} (\bibinfo {year} {2014}),\ \Eprint
  {http://arxiv.org/abs/1403.7437} {arXiv:1403.7437} \BibitemShut {NoStop}%
\bibitem [{\citenamefont {Schnyder}\ \emph {et~al.}(2008)\citenamefont
  {Schnyder}, \citenamefont {Ryu}, \citenamefont {Furusaki},\ and\
  \citenamefont {Ludwig}}]{schnyder_classification}%
  \BibitemOpen
  \bibfield  {author} {\bibinfo {author} {\bibfnamefont {A.~P.}\ \bibnamefont
  {Schnyder}}, \bibinfo {author} {\bibfnamefont {S.}~\bibnamefont {Ryu}},
  \bibinfo {author} {\bibfnamefont {A.}~\bibnamefont {Furusaki}}, \ and\
  \bibinfo {author} {\bibfnamefont {A.~W.~W.}\ \bibnamefont {Ludwig}},\ }\href
  {\doibase 10.1103/PhysRevB.78.195125} {\bibfield  {journal} {\bibinfo
  {journal} {Phys. Rev. B}\ }\textbf {\bibinfo {volume} {78}},\ \bibinfo
  {pages} {195125} (\bibinfo {year} {2008})},\ \Eprint
  {http://arxiv.org/abs/0803.2786} {arXiv:0803.2786} \BibitemShut {NoStop}%
\bibitem [{\citenamefont {Kitaev}(2009)}]{kitaev_periodic}%
  \BibitemOpen
  \bibfield  {author} {\bibinfo {author} {\bibfnamefont {A.}~\bibnamefont
  {Kitaev}},\ }\href@noop {} {} (\bibinfo {year} {2009}),\ \Eprint
  {http://arxiv.org/abs/0901.2686} {arXiv:0901.2686} \BibitemShut {NoStop}%
\bibitem [{\citenamefont {Ryu}\ \emph {et~al.}(2010)\citenamefont {Ryu},
  \citenamefont {Schnyder}, \citenamefont {Furusaki},\ and\ \citenamefont
  {Ludwig}}]{tenfold_way}%
  \BibitemOpen
  \bibfield  {author} {\bibinfo {author} {\bibfnamefont {S.}~\bibnamefont
  {Ryu}}, \bibinfo {author} {\bibfnamefont {A.~P.}\ \bibnamefont {Schnyder}},
  \bibinfo {author} {\bibfnamefont {A.}~\bibnamefont {Furusaki}}, \ and\
  \bibinfo {author} {\bibfnamefont {A.~W.~W.}\ \bibnamefont {Ludwig}},\ }\href
  {\doibase 10.1088/1367-2630/12/6/065010} {\bibfield  {journal} {\bibinfo
  {journal} {New J. Phys.}\ }\textbf {\bibinfo {volume} {12}},\ \bibinfo
  {pages} {065010} (\bibinfo {year} {2010})}\BibitemShut {NoStop}%
\bibitem [{\citenamefont {Wen}(2012)}]{wen_fermions}%
  \BibitemOpen
  \bibfield  {author} {\bibinfo {author} {\bibfnamefont {X.-G.}\ \bibnamefont
  {Wen}},\ }\href {\doibase 10.1103/PhysRevB.85.085103} {\bibfield  {journal}
  {\bibinfo  {journal} {Phys. Rev. B}\ }\textbf {\bibinfo {volume} {85}},\
  \bibinfo {pages} {085103} (\bibinfo {year} {2012})},\ \Eprint
  {http://arxiv.org/abs/1111.6341} {arXiv:1111.6341} \BibitemShut {NoStop}%
\bibitem [{\citenamefont {Tang}\ and\ \citenamefont
  {Wen}(2012)}]{tang_fermions}%
  \BibitemOpen
  \bibfield  {author} {\bibinfo {author} {\bibfnamefont {E.}~\bibnamefont
  {Tang}}\ and\ \bibinfo {author} {\bibfnamefont {X.-G.}\ \bibnamefont {Wen}},\
  }\href {\doibase 10.1103/PhysRevLett.109.096403} {\bibfield  {journal}
  {\bibinfo  {journal} {Phys. Rev. Lett.}\ }\textbf {\bibinfo {volume} {109}},\
  \bibinfo {pages} {096403} (\bibinfo {year} {2012})}\BibitemShut {NoStop}%
\bibitem [{\citenamefont {Fidkowski}\ and\ \citenamefont
  {Kitaev}(2011)}]{fidkowski_fermions}%
  \BibitemOpen
  \bibfield  {author} {\bibinfo {author} {\bibfnamefont {L.}~\bibnamefont
  {Fidkowski}}\ and\ \bibinfo {author} {\bibfnamefont {A.}~\bibnamefont
  {Kitaev}},\ }\href {\doibase 10.1103/PhysRevB.83.075103} {\bibfield
  {journal} {\bibinfo  {journal} {Phys. Rev. B}\ }\textbf {\bibinfo {volume}
  {83}},\ \bibinfo {pages} {075103} (\bibinfo {year} {2011})},\ \Eprint
  {http://arxiv.org/abs/1008.4138} {1008.4138} \BibitemShut {NoStop}%
\bibitem [{\citenamefont {Wang}\ \emph {et~al.}(2014)\citenamefont {Wang},
  \citenamefont {Potter},\ and\ \citenamefont {Senthil}}]{wang_potter_senthil}%
  \BibitemOpen
  \bibfield  {author} {\bibinfo {author} {\bibfnamefont {C.}~\bibnamefont
  {Wang}}, \bibinfo {author} {\bibfnamefont {A.~C.}\ \bibnamefont {Potter}}, \
  and\ \bibinfo {author} {\bibfnamefont {T.}~\bibnamefont {Senthil}},\ }\href
  {\doibase 10.1126/science.1243326} {\bibfield  {journal} {\bibinfo  {journal}
  {Science}\ }\textbf {\bibinfo {volume} {343}},\ \bibinfo {pages} {629}
  (\bibinfo {year} {2014})},\ \Eprint {http://arxiv.org/abs/1306.3238}
  {arXiv:1306.3238} \BibitemShut {NoStop}%
\bibitem [{\citenamefont {Qi}(2013)}]{qi_fermions}%
  \BibitemOpen
  \bibfield  {author} {\bibinfo {author} {\bibfnamefont {X.-L.}\ \bibnamefont
  {Qi}},\ }\href {\doibase 10.1088/1367-2630/15/6/065002} {\bibfield  {journal}
  {\bibinfo  {journal} {New J. Phys.}\ }\textbf {\bibinfo {volume} {15}},\
  \bibinfo {pages} {065002} (\bibinfo {year} {2013})},\ \Eprint
  {http://arxiv.org/abs/1202.3983} {arXiv:1202.3983} \BibitemShut {NoStop}%
\bibitem [{\citenamefont {Yao}\ and\ \citenamefont {Ryu}(2013)}]{yao_fermions}%
  \BibitemOpen
  \bibfield  {author} {\bibinfo {author} {\bibfnamefont {H.}~\bibnamefont
  {Yao}}\ and\ \bibinfo {author} {\bibfnamefont {S.}~\bibnamefont {Ryu}},\
  }\href {\doibase 10.1103/PhysRevB.88.064507} {\bibfield  {journal} {\bibinfo
  {journal} {Phys. Rev. B}\ }\textbf {\bibinfo {volume} {88}},\ \bibinfo
  {pages} {064507} (\bibinfo {year} {2013})},\ \Eprint
  {http://arxiv.org/abs/1201.5805} {arXiv:1201.5805} \BibitemShut {NoStop}%
\bibitem [{\citenamefont {Gu}\ and\ \citenamefont
  {Levin}(2014)}]{gu_levin_fermions}%
  \BibitemOpen
  \bibfield  {author} {\bibinfo {author} {\bibfnamefont {Z.-C.}\ \bibnamefont
  {Gu}}\ and\ \bibinfo {author} {\bibfnamefont {M.}~\bibnamefont {Levin}},\
  }\href {\doibase 10.1103/PhysRevB.89.201113} {\bibfield  {journal} {\bibinfo
  {journal} {Phys. Rev. B}\ }\textbf {\bibinfo {volume} {89}},\ \bibinfo
  {pages} {201113} (\bibinfo {year} {2014})},\ \Eprint
  {http://arxiv.org/abs/1304.4569} {arXiv:1304.4569} \BibitemShut {NoStop}%
\bibitem [{\citenamefont {Bi}\ \emph {et~al.}({\natexlab{a}})\citenamefont
  {Bi}, \citenamefont {Cheng},\ and\ \citenamefont {Gu}}]{cheng_gu_fermions}%
  \BibitemOpen
  \bibfield  {author} {\bibinfo {author} {\bibfnamefont {Z.}~\bibnamefont
  {Bi}}, \bibinfo {author} {\bibfnamefont {M.}~\bibnamefont {Cheng}}, \ and\
  \bibinfo {author} {\bibfnamefont {Z.-C.}\ \bibnamefont {Gu}},\ }\href@noop {}
  {} ({\natexlab{a}}),\ \bibinfo {note} {in preparation}\BibitemShut {NoStop}%
\bibitem [{\citenamefont {Neupert}\ \emph {et~al.}()\citenamefont {Neupert},
  \citenamefont {Chamon}, \citenamefont {Mudry},\ and\ \citenamefont
  {Thomale}}]{thomale}%
  \BibitemOpen
  \bibfield  {author} {\bibinfo {author} {\bibfnamefont {T.}~\bibnamefont
  {Neupert}}, \bibinfo {author} {\bibfnamefont {C.}~\bibnamefont {Chamon}},
  \bibinfo {author} {\bibfnamefont {C.}~\bibnamefont {Mudry}}, \ and\ \bibinfo
  {author} {\bibfnamefont {R.}~\bibnamefont {Thomale}},\ }\href@noop {}
  {}\Eprint {http://arxiv.org/abs/1403.0953} {arXiv:1403.0953} \BibitemShut
  {NoStop}%
\bibitem [{\citenamefont {Kapustin}\ \emph {et~al.}()\citenamefont {Kapustin},
  \citenamefont {Thorngren}, \citenamefont {Turzillo},\ and\ \citenamefont
  {Wang}}]{kapustin_fermions}%
  \BibitemOpen
  \bibfield  {author} {\bibinfo {author} {\bibfnamefont {A.}~\bibnamefont
  {Kapustin}}, \bibinfo {author} {\bibfnamefont {R.}~\bibnamefont {Thorngren}},
  \bibinfo {author} {\bibfnamefont {A.}~\bibnamefont {Turzillo}}, \ and\
  \bibinfo {author} {\bibfnamefont {Z.}~\bibnamefont {Wang}},\ }\href@noop {}
  {}\Eprint {http://arxiv.org/abs/1406.7329} {arXiv:1406.7329} \BibitemShut
  {NoStop}%
\bibitem [{\citenamefont {Chen}\ \emph
  {et~al.}(2011{\natexlab{b}})\citenamefont {Chen}, \citenamefont {Liu},\ and\
  \citenamefont {Wen}}]{spto_2d}%
  \BibitemOpen
  \bibfield  {author} {\bibinfo {author} {\bibfnamefont {X.}~\bibnamefont
  {Chen}}, \bibinfo {author} {\bibfnamefont {Z.-X.}\ \bibnamefont {Liu}}, \
  and\ \bibinfo {author} {\bibfnamefont {X.-G.}\ \bibnamefont {Wen}},\ }\href
  {\doibase 10.1103/PhysRevB.84.235141} {\bibfield  {journal} {\bibinfo
  {journal} {Phys. Rev. B}\ }\textbf {\bibinfo {volume} {84}},\ \bibinfo
  {pages} {235141} (\bibinfo {year} {2011}{\natexlab{b}})},\ \Eprint
  {http://arxiv.org/abs/1106.4752} {arXiv:1106.4752} \BibitemShut {NoStop}%
\bibitem [{\citenamefont {{Wen}}(2013)}]{Wen13}%
  \BibitemOpen
  \bibfield  {author} {\bibinfo {author} {\bibfnamefont {X.-G.}\ \bibnamefont
  {{Wen}}},\ }\href {\doibase 10.1103/PhysRevD.88.045013} {\bibfield  {journal}
  {\bibinfo  {journal} {\prd}\ }\textbf {\bibinfo {volume} {88}},\ \bibinfo
  {eid} {045013} (\bibinfo {year} {2013})},\ \Eprint
  {http://arxiv.org/abs/1303.1803} {arXiv:1303.1803} \BibitemShut {NoStop}%
\bibitem [{\citenamefont {Kapustin}\ and\ \citenamefont
  {Thorngren}(2014)}]{kapustin_anomalies}%
  \BibitemOpen
  \bibfield  {author} {\bibinfo {author} {\bibfnamefont {A.}~\bibnamefont
  {Kapustin}}\ and\ \bibinfo {author} {\bibfnamefont {R.}~\bibnamefont
  {Thorngren}},\ }\href {\doibase 10.1103/PhysRevLett.112.231602} {\bibfield
  {journal} {\bibinfo  {journal} {Phys. Rev. Lett.}\ }\textbf {\bibinfo
  {volume} {112}},\ \bibinfo {pages} {231602} (\bibinfo {year} {2014})},\
  \Eprint {http://arxiv.org/abs/1403.0617} {arXiv:1403.0617} \BibitemShut
  {NoStop}%
\bibitem [{\citenamefont {Kapustin}\ and\ \citenamefont
  {Thorngren}()}]{kapustin_anomalies_2}%
  \BibitemOpen
  \bibfield  {author} {\bibinfo {author} {\bibfnamefont {A.}~\bibnamefont
  {Kapustin}}\ and\ \bibinfo {author} {\bibfnamefont {R.}~\bibnamefont
  {Thorngren}},\ }\href@noop {} {}\Eprint {http://arxiv.org/abs/1404.3230}
  {arXiv:1404.3230} \BibitemShut {NoStop}%
\bibitem [{\citenamefont {Wen}(1991)}]{wen_gapless}%
  \BibitemOpen
  \bibfield  {author} {\bibinfo {author} {\bibfnamefont {X.~G.}\ \bibnamefont
  {Wen}},\ }\href {\doibase 10.1103/PhysRevB.43.11025} {\bibfield  {journal}
  {\bibinfo  {journal} {Phys. Rev. B}\ }\textbf {\bibinfo {volume} {43}},\
  \bibinfo {pages} {11025} (\bibinfo {year} {1991})}\BibitemShut {NoStop}%
\bibitem [{\citenamefont {Chen}\ and\ \citenamefont
  {Vishwanath}()}]{gauging_time_reversal}%
  \BibitemOpen
  \bibfield  {author} {\bibinfo {author} {\bibfnamefont {X.}~\bibnamefont
  {Chen}}\ and\ \bibinfo {author} {\bibfnamefont {A.}~\bibnamefont
  {Vishwanath}},\ }\href@noop {} {}\Eprint {http://arxiv.org/abs/1401.3736}
  {arXiv:1401.3736} \BibitemShut {NoStop}%
\bibitem [{\citenamefont {Chen}\ and\ \citenamefont
  {Wen}(2012)}]{chen_wen_chiral}%
  \BibitemOpen
  \bibfield  {author} {\bibinfo {author} {\bibfnamefont {X.}~\bibnamefont
  {Chen}}\ and\ \bibinfo {author} {\bibfnamefont {X.-G.}\ \bibnamefont {Wen}},\
  }\href {\doibase 10.1103/PhysRevB.86.235135} {\bibfield  {journal} {\bibinfo
  {journal} {Phys. Rev. B}\ }\textbf {\bibinfo {volume} {86}},\ \bibinfo
  {pages} {235135} (\bibinfo {year} {2012})},\ \Eprint
  {http://arxiv.org/abs/1206.3117} {arXiv:1206.3117} \BibitemShut {NoStop}%
\bibitem [{\citenamefont {Kitaev}\ and\ \citenamefont
  {Preskill}(2006)}]{top_ent_entr}%
  \BibitemOpen
  \bibfield  {author} {\bibinfo {author} {\bibfnamefont {A.}~\bibnamefont
  {Kitaev}}\ and\ \bibinfo {author} {\bibfnamefont {J.}~\bibnamefont
  {Preskill}},\ }\href {\doibase 10.1103/PhysRevLett.96.110404} {\bibfield
  {journal} {\bibinfo  {journal} {Phys. Rev. Lett.}\ }\textbf {\bibinfo
  {volume} {96}},\ \bibinfo {pages} {110404} (\bibinfo {year} {2006})},\
  \Eprint {http://arxiv.org/abs/hep-th/0510092} {arXiv:hep-th/0510092}
  \BibitemShut {NoStop}%
\bibitem [{\citenamefont {Senthil}\ and\ \citenamefont {Levin}()}]{boson_iqhe}%
  \BibitemOpen
  \bibfield  {author} {\bibinfo {author} {\bibfnamefont {T.}~\bibnamefont
  {Senthil}}\ and\ \bibinfo {author} {\bibfnamefont {M.}~\bibnamefont
  {Levin}},\ }\href@noop {} {}\Eprint {http://arxiv.org/abs/1206.1604}
  {arXiv:1206.1604} \BibitemShut {NoStop}%
\bibitem [{\citenamefont {Bi}\ \emph {et~al.}({\natexlab{b}})\citenamefont
  {Bi}, \citenamefont {Rasmussen},\ and\ \citenamefont
  {Xu}}]{adr_classification}%
  \BibitemOpen
  \bibfield  {author} {\bibinfo {author} {\bibfnamefont {Z.}~\bibnamefont
  {Bi}}, \bibinfo {author} {\bibfnamefont {A.}~\bibnamefont {Rasmussen}}, \
  and\ \bibinfo {author} {\bibfnamefont {C.}~\bibnamefont {Xu}},\ }\href@noop
  {} {} ({\natexlab{b}}),\ \Eprint {http://arxiv.org/abs/1309.0515}
  {arXiv:1309.0515} \BibitemShut {NoStop}%
\bibitem [{\citenamefont {Burnell}\ \emph {et~al.}()\citenamefont {Burnell},
  \citenamefont {Chen}, \citenamefont {Fidkowski},\ and\ \citenamefont
  {Vishwanath}}]{beyond_cohomology_walker_wang}%
  \BibitemOpen
  \bibfield  {author} {\bibinfo {author} {\bibfnamefont {F.~J.}\ \bibnamefont
  {Burnell}}, \bibinfo {author} {\bibfnamefont {X.}~\bibnamefont {Chen}},
  \bibinfo {author} {\bibfnamefont {L.}~\bibnamefont {Fidkowski}}, \ and\
  \bibinfo {author} {\bibfnamefont {A.}~\bibnamefont {Vishwanath}},\
  }\href@noop {} {}\Eprint {http://arxiv.org/abs/1302.7072} {arXiv:1302.7072}
  \BibitemShut {NoStop}%
\bibitem [{\citenamefont {Thorngren}()}]{thorngren_cobordism}%
  \BibitemOpen
  \bibfield  {author} {\bibinfo {author} {\bibfnamefont {R.}~\bibnamefont
  {Thorngren}},\ }\href@noop {} {}\Eprint {http://arxiv.org/abs/1404.4385}
  {arXiv:1404.4385} \BibitemShut {NoStop}%
\bibitem [{\citenamefont {Chen}\ \emph
  {et~al.}(2013{\natexlab{b}})\citenamefont {Chen}, \citenamefont {Wang},
  \citenamefont {Lu},\ and\ \citenamefont {Lee}}]{Chen2013}%
  \BibitemOpen
  \bibfield  {author} {\bibinfo {author} {\bibfnamefont {X.}~\bibnamefont
  {Chen}}, \bibinfo {author} {\bibfnamefont {F.}~\bibnamefont {Wang}}, \bibinfo
  {author} {\bibfnamefont {Y.-M.}\ \bibnamefont {Lu}}, \ and\ \bibinfo {author}
  {\bibfnamefont {D.-H.}\ \bibnamefont {Lee}},\ }\href {\doibase
  http://dx.doi.org/10.1016/j.nuclphysb.2013.04.015} {\bibfield  {journal}
  {\bibinfo  {journal} {Nucl. Phys. B}\ }\textbf {\bibinfo {volume} {873}},\
  \bibinfo {pages} {248 } (\bibinfo {year} {2013}{\natexlab{b}})},\ \Eprint
  {http://arxiv.org/abs/1302.3121} {1302.3121} \BibitemShut {NoStop}%
\bibitem [{\citenamefont {Gross}\ \emph {et~al.}(2012)\citenamefont {Gross},
  \citenamefont {Nesme}, \citenamefont {Vogts},\ and\ \citenamefont
  {Werner}}]{gross_index}%
  \BibitemOpen
  \bibfield  {author} {\bibinfo {author} {\bibfnamefont {D.}~\bibnamefont
  {Gross}}, \bibinfo {author} {\bibfnamefont {V.}~\bibnamefont {Nesme}},
  \bibinfo {author} {\bibfnamefont {H.}~\bibnamefont {Vogts}}, \ and\ \bibinfo
  {author} {\bibfnamefont {R.}~\bibnamefont {Werner}},\ }\href {\doibase
  10.1007/s00220-012-1423-1} {\bibfield  {journal} {\bibinfo  {journal} {Comm.
  Math. Phys.}\ }\textbf {\bibinfo {volume} {310}},\ \bibinfo {pages} {419}
  (\bibinfo {year} {2012})}\BibitemShut {NoStop}%
\bibitem [{\citenamefont {Hastings}()}]{hastings_torus_trick}%
  \BibitemOpen
  \bibfield  {author} {\bibinfo {author} {\bibfnamefont {M.~B.}\ \bibnamefont
  {Hastings}},\ }\href@noop {} {}\Eprint {http://arxiv.org/abs/1305.6625}
  {arXiv:1305.6625} \BibitemShut {NoStop}%
\bibitem [{\citenamefont {Ryu}\ and\ \citenamefont
  {Zhang}(2012)}]{ryu_zhang_fermions}%
  \BibitemOpen
  \bibfield  {author} {\bibinfo {author} {\bibfnamefont {S.}~\bibnamefont
  {Ryu}}\ and\ \bibinfo {author} {\bibfnamefont {S.-C.}\ \bibnamefont
  {Zhang}},\ }\href {\doibase 10.1103/PhysRevB.85.245132} {\bibfield  {journal}
  {\bibinfo  {journal} {Phys. Rev. B}\ }\textbf {\bibinfo {volume} {85}},\
  \bibinfo {pages} {245132} (\bibinfo {year} {2012})},\ \Eprint
  {http://arxiv.org/abs/1202.4484} {arXiv:1202.4484} \BibitemShut {NoStop}%
\bibitem [{\citenamefont {Etingof}\ \emph {et~al.}(2010)\citenamefont
  {Etingof}, \citenamefont {Nikshych},\ and\ \citenamefont
  {Ostrik}}]{Etingof10}%
  \BibitemOpen
  \bibfield  {author} {\bibinfo {author} {\bibfnamefont {P.}~\bibnamefont
  {Etingof}}, \bibinfo {author} {\bibfnamefont {D.}~\bibnamefont {Nikshych}}, \
  and\ \bibinfo {author} {\bibfnamefont {V.}~\bibnamefont {Ostrik}},\ }\href
  {\doibase 10.4171/QT/6} {\bibfield  {journal} {\bibinfo  {journal} {Quantum
  Topology}\ }\textbf {\bibinfo {volume} {1}},\ \bibinfo {pages} {209}
  (\bibinfo {year} {2010})}\BibitemShut {NoStop}%
\bibitem [{\citenamefont {{Chen}}\ \emph {et~al.}()\citenamefont {{Chen}},
  \citenamefont {{Burnell}}, \citenamefont {{Vishwanath}},\ and\ \citenamefont
  {{Fidkowski}}}]{Chen14}%
  \BibitemOpen
  \bibfield  {author} {\bibinfo {author} {\bibfnamefont {X.}~\bibnamefont
  {{Chen}}}, \bibinfo {author} {\bibfnamefont {F.~J.}\ \bibnamefont
  {{Burnell}}}, \bibinfo {author} {\bibfnamefont {A.}~\bibnamefont
  {{Vishwanath}}}, \ and\ \bibinfo {author} {\bibfnamefont {L.}~\bibnamefont
  {{Fidkowski}}},\ }\href@noop {} {\enquote {\bibinfo {title} {{Anomalous
  Symmetry Fractionalization and Surface Topological Order}},}\ }\bibinfo
  {note} {ArXiv:1403.6491}\BibitemShut {NoStop}%
\bibitem [{\citenamefont {{Ryu}}\ \emph {et~al.}(2012)\citenamefont {{Ryu}},
  \citenamefont {{Moore}},\ and\ \citenamefont {{Ludwig}}}]{Ryu12}%
  \BibitemOpen
  \bibfield  {author} {\bibinfo {author} {\bibfnamefont {S.}~\bibnamefont
  {{Ryu}}}, \bibinfo {author} {\bibfnamefont {J.~E.}\ \bibnamefont {{Moore}}},
  \ and\ \bibinfo {author} {\bibfnamefont {A.~W.~W.}\ \bibnamefont
  {{Ludwig}}},\ }\href {\doibase 10.1103/PhysRevB.85.045104} {\bibfield
  {journal} {\bibinfo  {journal} {\prb}\ }\textbf {\bibinfo {volume} {85}},\
  \bibinfo {eid} {045104} (\bibinfo {year} {2012})},\ \Eprint
  {http://arxiv.org/abs/1010.0936} {arXiv:1010.0936} \BibitemShut {NoStop}%
\bibitem [{\citenamefont {Wang}\ \emph {et~al.}({\natexlab{b}})\citenamefont
  {Wang}, \citenamefont {Gu},\ and\ \citenamefont {Wen}}]{for_dummies}%
  \BibitemOpen
  \bibfield  {author} {\bibinfo {author} {\bibfnamefont {J.}~\bibnamefont
  {Wang}}, \bibinfo {author} {\bibfnamefont {Z.-C.}\ \bibnamefont {Gu}}, \ and\
  \bibinfo {author} {\bibfnamefont {X.-G.}\ \bibnamefont {Wen}},\ }\href@noop
  {} {} ({\natexlab{b}}),\ \Eprint {http://arxiv.org/abs/1405.7689}
  {arXiv:1405.7689} \BibitemShut {NoStop}%
\bibitem [{\citenamefont {Laughlin}(1981)}]{Laughlin81}%
  \BibitemOpen
  \bibfield  {author} {\bibinfo {author} {\bibfnamefont {R.~B.}\ \bibnamefont
  {Laughlin}},\ }\href {\doibase 10.1103/PhysRevB.23.5632} {\bibfield
  {journal} {\bibinfo  {journal} {Phys. Rev. B}\ }\textbf {\bibinfo {volume}
  {23}},\ \bibinfo {pages} {5632} (\bibinfo {year} {1981})}\BibitemShut
  {NoStop}%
\bibitem [{Note1()}]{Note1}%
  \BibitemOpen
  \bibinfo {note} {By restricting the rotation to an interval, we cannot
  introduce a net twist, but we can separate equal and opposite twists and
  focus on the vicinity of just one of them.}\BibitemShut {Stop}%
\bibitem [{Note2()}]{Note2}%
  \BibitemOpen
  \bibinfo {note} {The precise relation is slightly subtle, however, because
  $\Omega _a$ measures the charge to the left of the endpoint $a$, the very
  place at which charge is accumulating due to the Hall effect. The precise
  manner in which $\Omega _a$ takes into account the charge located at the
  endpoint itself depends on the restriction $U \to U_M$. However, as long as
  we use the same restriction both to define $\Omega _a$ and to implement the
  winding, the value of the measured ``charge'' is independent of the choice of
  restriction. Perhaps the easiest case to interpret is where $U_M$ is defined
  so that the winding of the phase interpolates linearly over a transition
  region between no winding outside the interval and the desired winding in the
  interior of the interval. Then $\Omega _a$ measures the charge to the left of
  a point $x$, averaged over all $x$ in the transition region. This is $nm$,
  rather than the charge $2nm$ measured to the left of a point in the interior
  of the interval.}\BibitemShut {Stop}%
\bibitem [{\citenamefont {Turner}\ \emph {et~al.}(2010)\citenamefont {Turner},
  \citenamefont {Zhang},\ and\ \citenamefont {Vishwanath}}]{Turner09}%
  \BibitemOpen
  \bibfield  {author} {\bibinfo {author} {\bibfnamefont {A.}~\bibnamefont
  {Turner}}, \bibinfo {author} {\bibfnamefont {Y.}~\bibnamefont {Zhang}}, \
  and\ \bibinfo {author} {\bibfnamefont {A.}~\bibnamefont {Vishwanath}},\
  }\href {\doibase 10.1103/PhysRevB.82.241102} {\bibfield  {journal} {\bibinfo
  {journal} {Phys. Rev. B}\ }\textbf {\bibinfo {volume} {82}},\ \bibinfo
  {pages} {241102} (\bibinfo {year} {2010})},\ \Eprint
  {http://arxiv.org/abs/0909.3119} {arXiv:0909.3119} \BibitemShut {NoStop}%
\bibitem [{\citenamefont {{Hughes}}\ \emph {et~al.}(2011)\citenamefont
  {{Hughes}}, \citenamefont {{Prodan}},\ and\ \citenamefont
  {{Bernevig}}}]{Hughes11}%
  \BibitemOpen
  \bibfield  {author} {\bibinfo {author} {\bibfnamefont {T.~L.}\ \bibnamefont
  {{Hughes}}}, \bibinfo {author} {\bibfnamefont {E.}~\bibnamefont {{Prodan}}},
  \ and\ \bibinfo {author} {\bibfnamefont {B.~A.}\ \bibnamefont {{Bernevig}}},\
  }\href {\doibase 10.1103/PhysRevB.83.245132} {\bibfield  {journal} {\bibinfo
  {journal} {\prb}\ }\textbf {\bibinfo {volume} {83}},\ \bibinfo {eid} {245132}
  (\bibinfo {year} {2011})},\ \Eprint {http://arxiv.org/abs/1010.4508}
  {arXiv:1010.4508} \BibitemShut {NoStop}%
\bibitem [{\citenamefont {{Fu}}(2011)}]{Fu11}%
  \BibitemOpen
  \bibfield  {author} {\bibinfo {author} {\bibfnamefont {L.}~\bibnamefont
  {{Fu}}},\ }\href {\doibase 10.1103/PhysRevLett.106.106802} {\bibfield
  {journal} {\bibinfo  {journal} {Phys. Rev. Lett.}\ }\textbf {\bibinfo
  {volume} {106}},\ \bibinfo {eid} {106802} (\bibinfo {year} {2011})},\ \Eprint
  {http://arxiv.org/abs/1010.1802} {arXiv:1010.1802} \BibitemShut {NoStop}%
\bibitem [{\citenamefont {{Jadaun}}\ \emph {et~al.}(2013)\citenamefont
  {{Jadaun}}, \citenamefont {{Xiao}}, \citenamefont {{Niu}},\ and\
  \citenamefont {{Banerjee}}}]{Jadaun13}%
  \BibitemOpen
  \bibfield  {author} {\bibinfo {author} {\bibfnamefont {P.}~\bibnamefont
  {{Jadaun}}}, \bibinfo {author} {\bibfnamefont {D.}~\bibnamefont {{Xiao}}},
  \bibinfo {author} {\bibfnamefont {Q.}~\bibnamefont {{Niu}}}, \ and\ \bibinfo
  {author} {\bibfnamefont {S.~K.}\ \bibnamefont {{Banerjee}}},\ }\href
  {\doibase 10.1103/PhysRevB.88.085110} {\bibfield  {journal} {\bibinfo
  {journal} {\prb}\ }\textbf {\bibinfo {volume} {88}},\ \bibinfo {eid} {085110}
  (\bibinfo {year} {2013})},\ \Eprint {http://arxiv.org/abs/1208.1472}
  {arXiv:1208.1472} \BibitemShut {NoStop}%
\bibitem [{\citenamefont {Moore}\ and\ \citenamefont
  {Balents}(2007)}]{Moore07}%
  \BibitemOpen
  \bibfield  {author} {\bibinfo {author} {\bibfnamefont {J.~E.}\ \bibnamefont
  {Moore}}\ and\ \bibinfo {author} {\bibfnamefont {L.}~\bibnamefont
  {Balents}},\ }\href@noop {} {\bibfield  {journal} {\bibinfo  {journal} {Phys.
  Rev. B}\ }\textbf {\bibinfo {volume} {75}},\ \bibinfo {pages} {121306(R)}
  (\bibinfo {year} {2007})},\ \Eprint {http://arxiv.org/abs/cond-mat/0607314}
  {cond-mat/0607314} \BibitemShut {NoStop}%
\bibitem [{\citenamefont {Fu}\ \emph {et~al.}(2007)\citenamefont {Fu},
  \citenamefont {Kane},\ and\ \citenamefont {Mele}}]{Fu07}%
  \BibitemOpen
  \bibfield  {author} {\bibinfo {author} {\bibfnamefont {L.}~\bibnamefont
  {Fu}}, \bibinfo {author} {\bibfnamefont {C.~L.}\ \bibnamefont {Kane}}, \ and\
  \bibinfo {author} {\bibfnamefont {E.~J.}\ \bibnamefont {Mele}},\ }\href@noop
  {} {\bibfield  {journal} {\bibinfo  {journal} {Phys. Rev. Lett.}\ }\textbf
  {\bibinfo {volume} {98}},\ \bibinfo {pages} {106803} (\bibinfo {year}
  {2007})},\ \Eprint {http://arxiv.org/abs/cond-mat/0607699} {cond-mat/0607699}
  \BibitemShut {NoStop}%
\bibitem [{\citenamefont {Roy}(2009)}]{Roy09}%
  \BibitemOpen
  \bibfield  {author} {\bibinfo {author} {\bibfnamefont {R.}~\bibnamefont
  {Roy}},\ }\href@noop {} {\bibfield  {journal} {\bibinfo  {journal} {Phys.
  Rev. B}\ }\textbf {\bibinfo {volume} {79}},\ \bibinfo {pages} {195322}
  (\bibinfo {year} {2009})},\ \Eprint {http://arxiv.org/abs/cond-mat/0607531}
  {cond-mat/0607531} \BibitemShut {NoStop}%
\bibitem [{\citenamefont {Qi}\ \emph {et~al.}(2008)\citenamefont {Qi},
  \citenamefont {Hughes},\ and\ \citenamefont
  {Zhang}}]{topfieldtheory_insulators}%
  \BibitemOpen
  \bibfield  {author} {\bibinfo {author} {\bibfnamefont {X.-L.}\ \bibnamefont
  {Qi}}, \bibinfo {author} {\bibfnamefont {T.~L.}\ \bibnamefont {Hughes}}, \
  and\ \bibinfo {author} {\bibfnamefont {S.-C.}\ \bibnamefont {Zhang}},\ }\href
  {\doibase 10.1103/PhysRevB.78.195424} {\bibfield  {journal} {\bibinfo
  {journal} {Phys. Rev. B}\ }\textbf {\bibinfo {volume} {78}},\ \bibinfo
  {pages} {195424} (\bibinfo {year} {2008})},\ \Eprint
  {http://arxiv.org/abs/0802.3537} {arXiv:0802.3537} \BibitemShut {NoStop}%
\bibitem [{\citenamefont {{Mesaros}}\ and\ \citenamefont
  {{Ran}}(2013)}]{Mesaros13}%
  \BibitemOpen
  \bibfield  {author} {\bibinfo {author} {\bibfnamefont {A.}~\bibnamefont
  {{Mesaros}}}\ and\ \bibinfo {author} {\bibfnamefont {Y.}~\bibnamefont
  {{Ran}}},\ }\href {\doibase 10.1103/PhysRevB.87.155115} {\bibfield  {journal}
  {\bibinfo  {journal} {\prb}\ }\textbf {\bibinfo {volume} {87}},\ \bibinfo
  {eid} {155115} (\bibinfo {year} {2013})},\ \Eprint
  {http://arxiv.org/abs/1212.0835} {arXiv:1212.0835} \BibitemShut {NoStop}%
\bibitem [{\citenamefont {{Hung}}\ and\ \citenamefont {{Wen}}(2013)}]{Hung13}%
  \BibitemOpen
  \bibfield  {author} {\bibinfo {author} {\bibfnamefont {L.-Y.}\ \bibnamefont
  {{Hung}}}\ and\ \bibinfo {author} {\bibfnamefont {X.-G.}\ \bibnamefont
  {{Wen}}},\ }\href {\doibase 10.1103/PhysRevB.87.165107} {\bibfield  {journal}
  {\bibinfo  {journal} {\prb}\ }\textbf {\bibinfo {volume} {87}},\ \bibinfo
  {eid} {165107} (\bibinfo {year} {2013})},\ \Eprint
  {http://arxiv.org/abs/1212.1827} {arXiv:1212.1827} \BibitemShut {NoStop}%
\bibitem [{\citenamefont {{Lu}}\ and\ \citenamefont
  {{Vishwanath}}(2013)}]{Lu13}%
  \BibitemOpen
  \bibfield  {author} {\bibinfo {author} {\bibfnamefont {Y.-M.}\ \bibnamefont
  {{Lu}}}\ and\ \bibinfo {author} {\bibfnamefont {A.}~\bibnamefont
  {{Vishwanath}}},\ }\href@noop {} {} (\bibinfo {year} {2013}),\ \Eprint
  {http://arxiv.org/abs/1302.2634} {arXiv:1302.2634} \BibitemShut {NoStop}%
\bibitem [{\citenamefont {{Xu}}(2013)}]{Xu13}%
  \BibitemOpen
  \bibfield  {author} {\bibinfo {author} {\bibfnamefont {C.}~\bibnamefont
  {{Xu}}},\ }\href {\doibase 10.1103/PhysRevB.88.205137} {\bibfield  {journal}
  {\bibinfo  {journal} {\prb}\ }\textbf {\bibinfo {volume} {88}},\ \bibinfo
  {eid} {205137} (\bibinfo {year} {2013})},\ \Eprint
  {http://arxiv.org/abs/1307.8131} {arXiv:1307.8131} \BibitemShut {NoStop}%
\bibitem [{\citenamefont {{Essin}}\ and\ \citenamefont
  {{Hermele}}(2013)}]{Essin13}%
  \BibitemOpen
  \bibfield  {author} {\bibinfo {author} {\bibfnamefont {A.~M.}\ \bibnamefont
  {{Essin}}}\ and\ \bibinfo {author} {\bibfnamefont {M.}~\bibnamefont
  {{Hermele}}},\ }\href {\doibase 10.1103/PhysRevB.87.104406} {\bibfield
  {journal} {\bibinfo  {journal} {\prb}\ }\textbf {\bibinfo {volume} {87}},\
  \bibinfo {eid} {104406} (\bibinfo {year} {2013})},\ \Eprint
  {http://arxiv.org/abs/1212.0593} {arXiv:1212.0593} \BibitemShut {NoStop}%
\bibitem [{\citenamefont {Holt}(1979)}]{holt_cohomology}%
  \BibitemOpen
  \bibfield  {author} {\bibinfo {author} {\bibfnamefont {D.}~\bibnamefont
  {Holt}},\ }\href {\doibase 10.1016/0021-8693(79)90084-X} {\bibfield
  {journal} {\bibinfo  {journal} {Journal of Algebra}\ }\textbf {\bibinfo
  {volume} {60}},\ \bibinfo {pages} {307} (\bibinfo {year} {1979})}\BibitemShut
  {NoStop}%
\bibitem [{\citenamefont {Huebschmann}(1980)}]{huebschmann_cohomology}%
  \BibitemOpen
  \bibfield  {author} {\bibinfo {author} {\bibfnamefont {J.}~\bibnamefont
  {Huebschmann}},\ }\href {\doibase 10.1007/BF02566688} {\bibfield  {journal}
  {\bibinfo  {journal} {Commentarii mathematici Helvetici}\ }\textbf {\bibinfo
  {volume} {55}},\ \bibinfo {pages} {302} (\bibinfo {year} {1980})}\BibitemShut
  {NoStop}%
\bibitem [{\citenamefont {Brown}(1982)}]{brown_cohomology}%
  \BibitemOpen
  \bibfield  {author} {\bibinfo {author} {\bibfnamefont {K.~S.}\ \bibnamefont
  {Brown}},\ }\href@noop {} {\emph {\bibinfo {title} {Cohomology of groups}}}\
  (\bibinfo  {publisher} {Springer},\ \bibinfo {year} {1982})\BibitemShut
  {NoStop}%
\bibitem [{\citenamefont {Thomas}()}]{crossed_module_extensions}%
  \BibitemOpen
  \bibfield  {author} {\bibinfo {author} {\bibfnamefont {S.}~\bibnamefont
  {Thomas}},\ }\href@noop {} {}\Eprint {http://arxiv.org/abs/0911.2861}
  {arXiv:0911.2861} \BibitemShut {NoStop}%
\bibitem [{\citenamefont {Dodson}\ and\ \citenamefont
  {Parker}(1997)}]{algebraic_topology}%
  \BibitemOpen
  \bibfield  {author} {\bibinfo {author} {\bibfnamefont {C.~T.}\ \bibnamefont
  {Dodson}}\ and\ \bibinfo {author} {\bibfnamefont {P.~E.}\ \bibnamefont
  {Parker}},\ }\href@noop {} {\emph {\bibinfo {title} {A User's Guide to
  Algebraic Topology}}}\ (\bibinfo  {publisher} {Kluwer Academic Publishers},\
  \bibinfo {year} {1997})\BibitemShut {NoStop}%
\bibitem [{Note3()}]{Note3}%
  \BibitemOpen
  \bibinfo {note} {Specifically, the difference from the bosonic case comes
  from in going from Eq.\ (\ref {asdf1}) to Eq.\ (\ref {asdf2}) [due to the
  potential for $\Omega _b(g_1, g_2)$ and $\Omega _a(g_3, g_4)$ to
  anticommute], and in going from Eq.\ (\ref {asdf2}) to Eq.\ (\ref {asdf3})
  [due to the factor of $\Pi ^{\lambda (g_1, g_2)}$ appearing in Eq.\ (\ref
  {british_columbia}).]}\BibitemShut {NoStop}%
\end{thebibliography}
\end{document}